\documentclass[]{aa}
\usepackage[varg]{txfonts}

\usepackage{graphicx}
\usepackage{color}
\usepackage{gensymb}

\usepackage{natbib}
\bibpunct{(}{)}{;}{a}{}{,} 

\begin{document}

\title{The \textit{ROSAT} RASTER SURVEY IN THE NORTH-ECLIPTIC POLE FIELD: X--RAY CATALOGUE AND OPTICAL IDENTIFICATIONS}

\author{G. Hasinger\inst{1,2}
\and M. Freyberg\inst{3}
\and E. M. Hu\inst{2}
\and C. Z. Waters\inst{4}
\and P. Capak\inst{5}
\and A. Moneti\inst{6}
\and H. J. McCracken\inst{6}
}

\institute{European Space Astronomy Centre (ESA/ESAC), E-28691
Villanueva de la Cañada, Madrid, Spain\\
\email{guenther.hasinger@esa.int}
\label{inst1}
\and Institute for Astronomy, University of Hawaii, 2680 Woodlawn Drive, Honolulu, HI 96822, USA
\label{inst2}
\and Max-Planck-Institut für extraterrestrische Physik, Giessenbachstrasse 1,
85748 Garching, Germany
\label{inst3}
\and Department of Astrophysical Sciences, 4 Ivy Lane,
Princeton University, Princeton, NJ 08544, USA
\label{inst4}
\and Infrared Processing and Analysis Center (IPAC), 1200 East California Boulevard, Pasadena, California 91125, USA; California Institute of Technology, 1200 East California Boulevard, Pasadena, California 91125, USA
\label{inst5}
\and Institut d’Astrophysique de Paris, CNRS (UMR7095), 98 bis Boulevard Arago, F-75014, Paris, France
\label{inst6}
}

\date{Received Nameofmonth dd, yyyy; accepted Nameofmonth dd, yyyy}

\abstract
    {The North-Ecliptic Pole (NEP) is an important region for extragalactic surveys. Deep/wide contiguous surveys are being performed by several space observatories, most currently with the \textit{eROSITA} telescope. Several more are planned for the near future.
    We analyse all the \textit{ROSAT} pointed and survey observations in a region of 40 deg$^2$ around the NEP, restricting the \textit{ROSAT} field-of-view to the inner 30\arcmin~radius. We obtain an X--ray catalogue of 805 sources with 0.5--2 keV fluxes >2.9$\times$10$^{-15}$ erg cm$^{-2}$ s$^{-1}$, about a factor of three deeper than the \textit{ROSAT} All-Sky Survey in this field. The sensitivity and angular resolution of our data are comparable to the \textit{eROSITA} All-Sky Survey expectations. The 50\% position error radius of the sample of X--ray sources is $\sim$10\arcsec.
    We use \textit{HEROES} optical and near-infrared imaging photometry from the \textit{Subaru} and \textit{Canada/France/Hawaii} telescopes together with \textit{GALEX}, \textit{SDSS}, \textit{Pan-STARRS} and \textit{WISE} catalogues, as well as images from a new deep and wide \textit{Spitzer} survey in the field to statistically identify the X--ray sources and to calculate photometric redshifts for the candidate counterparts. In particular we utilize mid-infrared colours to identify AGN X--ray counterparts. Despite the relatively large error circles and often faint counterparts, together with confusion issues and systematic errors, we obtain a rather reliable catalogue of 766 high-quality optical counterparts, corresponding redshifts and optical classifications.
    The quality of the dataset is sufficient to look at ensemble properties of X--ray source classes. In particular we find a new population of luminous absorbed X--ray AGN at large redshifts, identified through their mid-IR colours. This populous group of AGN has not been recognized in previous X--ray surveys, but could be identified in our work due to the unique combination of survey solid angle, X--ray sensitivity and quality of the multiwavelength  photometry. We also use the \textit{WISE} and \textit{Spitzer} photometry to identify a sample of 185 AGN selected purely through their mid-infrared colours, most of which are not detected by \textit{ROSAT}. Their redshifts and upper limits to X--ray luminosity and X--ray to optical flux ratios are even higher than for the new class of X--ray selected luminous AGN2; they are probably a natural extension of this sample.
    This unique dataset is important as a reference sample for future deep surveys in the NEP region, in particular for \textit{eROSITA}, but also for \textit{Euclid} and \textit{SPHEREX}. We predict that most of the absorbed distant AGN should be readily picked up by eROSITA, but they require sensitive mid-IR imaging to be recognized as optical counterparts.}

\keywords{galaxies: active -- galaxies: evolution -- large-scale
structure of Universe -- quasars: general -- surveys}

\titlerunning{\textit{ROSAT} NEP Raster Survey}

\maketitle

\section{Introduction}

The North-Ecliptic Pole (NEP) region around the coordinates $\alpha$(2000)=18$^h$00$^m$00$^s$, $\delta$(2000)=+66$\degree$33$\arcmin$39$\arcsec$ is an important area for space-based extragalactic surveys. Quite a number of spacecraft are powered by fixed solar arrays, which need to face towards the sun. This gives them a degree of freedom to point in any direction roughly perpendicular to the sun. The two ecliptic poles, both the NEP, but also the South Ecliptic Pole (SEP) are therefore always accessible during the mission and thus prime targets for surveys and performance verification or calibration targets. Spacecrafts performing all-sky surveys by continuously scanning the sky perpendicular to the sun accumulate particularly large amounts of exposure time around the ecliptic poles.

The SEP is close to the Small and Large Magellanic Cloud limiting our visibility to the extragalactic sky, and is thus less suitable, but the NEP is perfectly situated for unbiased deep and wide extragalactic surveys. The \textit{ROSAT} X--ray observatory performed an all-sky survey \citep{1982AdSpR...2..241T} perpendicular to the sun-Earth direction and executed a particularly deep and wide survey at the NEP \citep{2006ApJS..162..304H}, hereafter H06, as well as several deep pointings for operational reasons \citep{1991A&A...246L...2H,1996MNRAS.281...59B}. The \textit{AKARI} infrared satellite performed a deep NEP survey \citep{2006PASJ...58..673M,2017PKAS...32..225G}, which was later followed up by the far-infrared observatory \textit{Herschel} \citep{2019PASJ...71...13P}. 

\begin{figure*}[t]
\begin{center}
    \includegraphics[width=.49\textwidth]{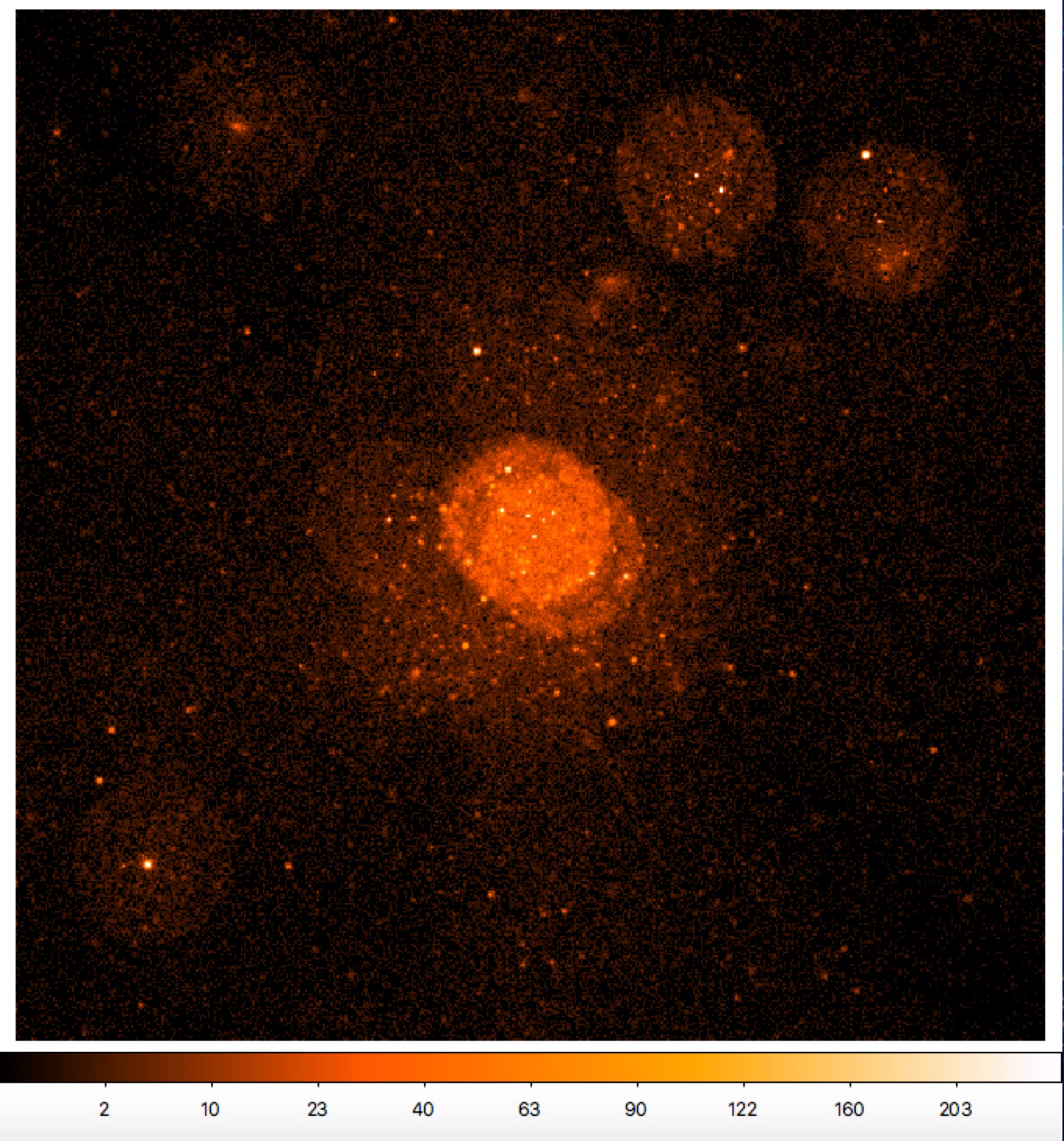}
    \includegraphics[width=.49\textwidth]{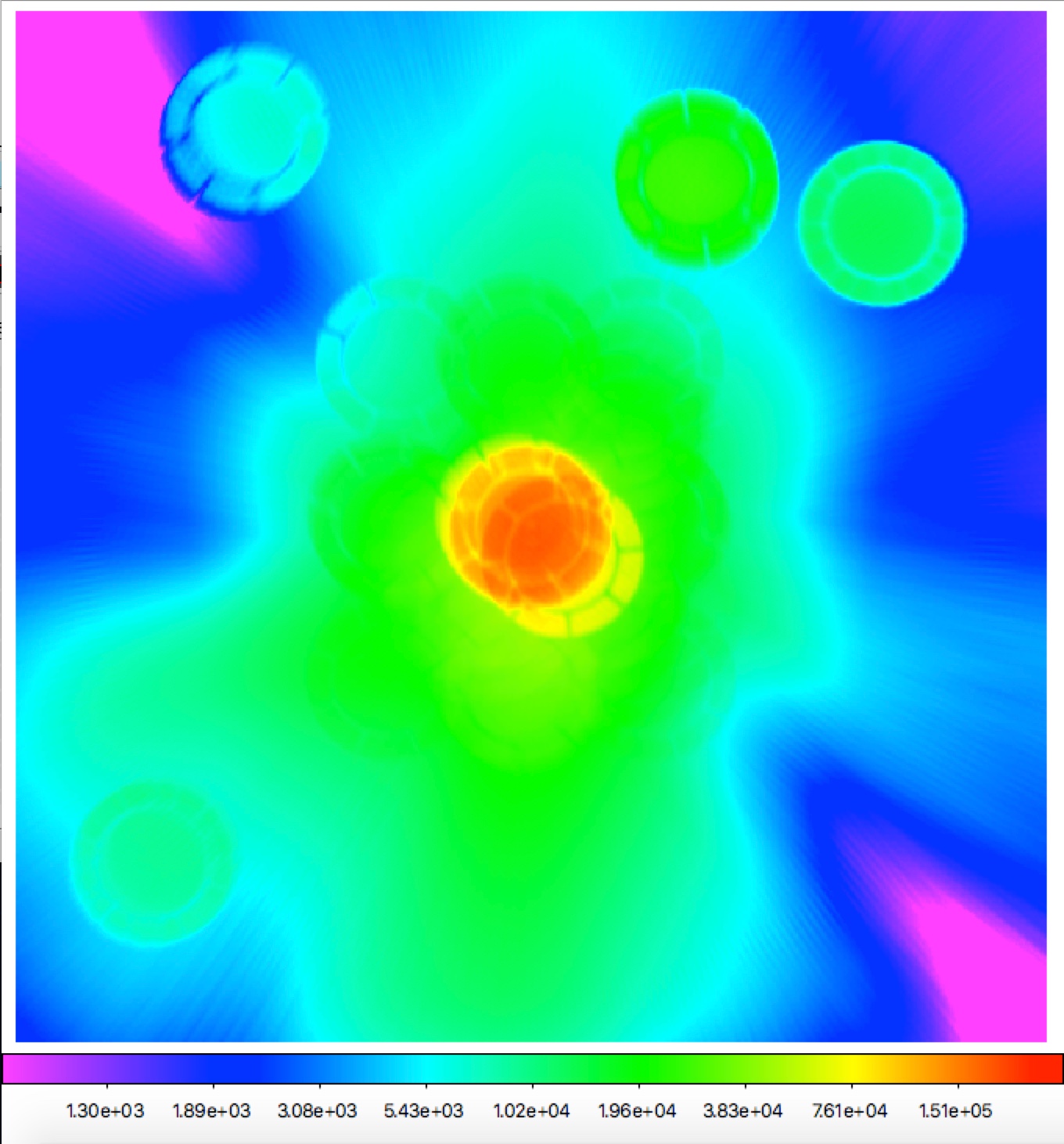}
    \includegraphics[width=.49\textwidth]{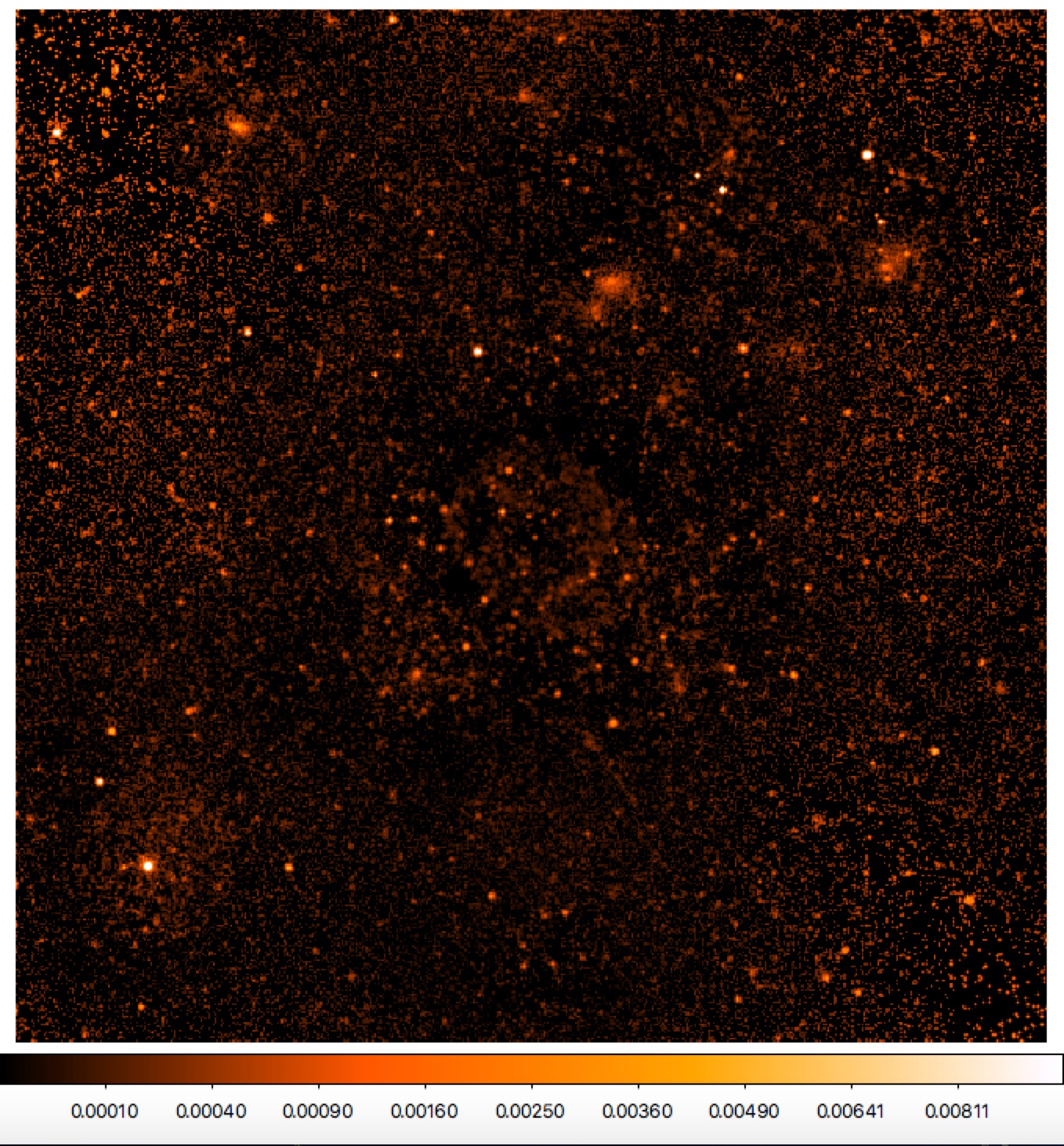}
    \includegraphics[width=.49\textwidth]{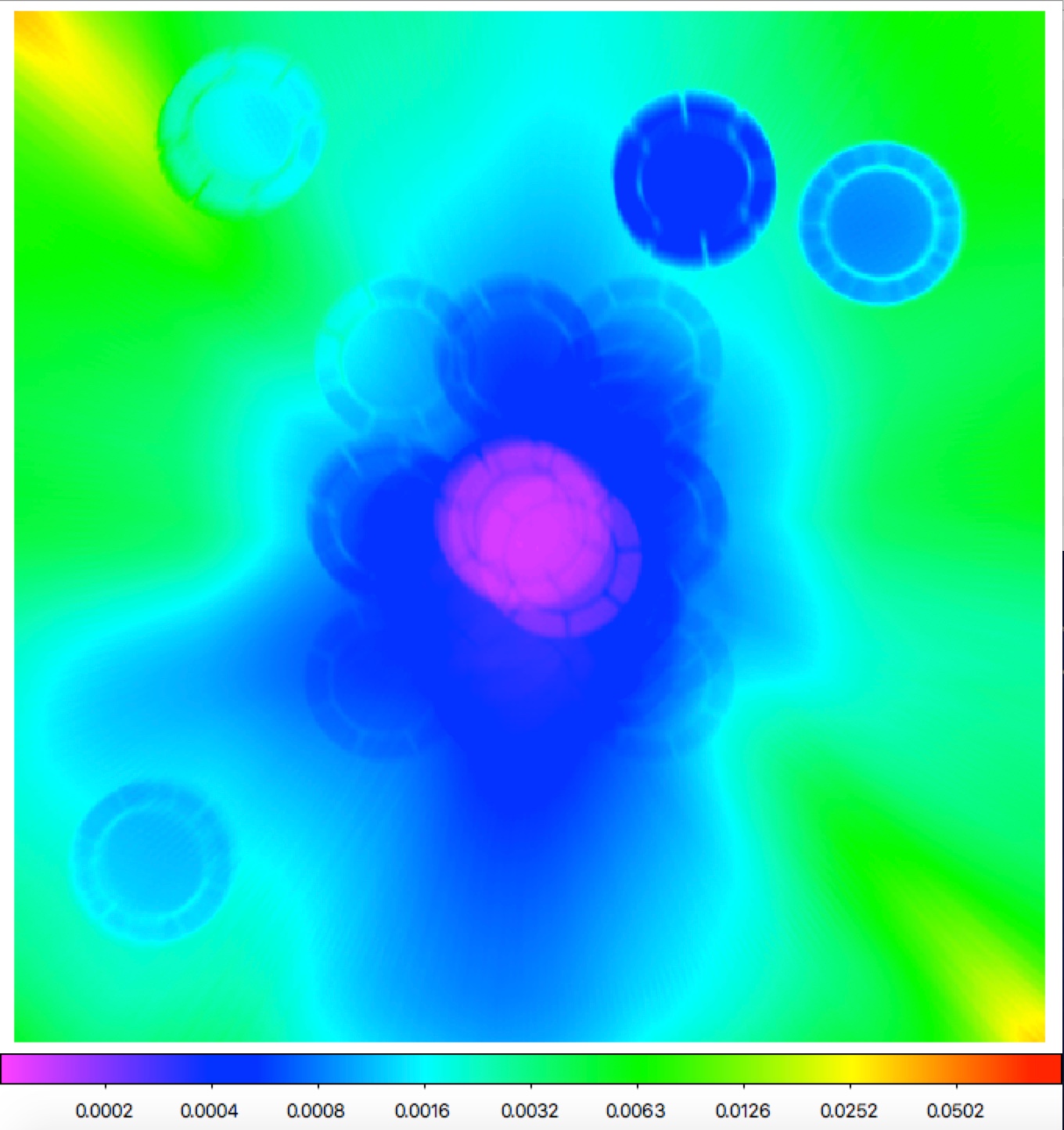}
\end{center}
\caption{Raw hard-band (0.5--2 keV) image of the NEP Raster scan (upper left, scale in raw counts); exposure map (upper right, scale in seconds); exposure-corrected image (lower left, scale in counts/sec); count rate sensitivity map (lower right, scale in  counts/sec). The image size is  $6.4\degree\times6.4\degree$, centered on the NEP. The pixel size is 45 \arcsec. 
}
\label{fig:images}
\end{figure*}

Future missions will also have an important focus on the NEP. The \textit{eROSITA} \citep{2012arXiv1209.3114M,2020arXiv201003477P} and ART-XC \citep{2018SPIE10699E..1YP} telescopes on board of the recently launched \textit{Spektr-RG} mission \citep{2009SPIE.7437E..08P} are currently producing an X--ray all-sky survey more than an order of magnitude deeper than \textit{ROSAT}, which again will have particularly deep and wide coverage at the ecliptic poles \citep[see][]{2020NatAs.tmp..123M}. The future NASA Medium Explorer mission \textit{SPHEREx} \citep{2018SPIE10698E..1UK,2018arXiv180505489D} will perform an all-sky spectroscopic survey in the near-infrared, again using the sun-perpendicular scanning scheme with particularly deep-wide ecliptic pole surveys. The ESA dark energy survey mission \textit{Euclid} has selected three Deep Fields, one of which is also centered on the NEP. The \textit{James Webb Space Telescope} will perform a long-term time-domain survey in its continuous viewing zone field close to the NEP \citep{2018PASP..130l4001J}, which is also currently monitored with \textit{Chandra} \citep{2019AAS...23336315M}. In preparation of these future surveys we have embarked on the wide-deep \textit{UgrizyJ} imaging survey \textit{HEROES}\footnote{Hawaii eROSITA Ecliptic Pole Survey.} with the \textit{Subaru} and \textit{CFHT} telescopes on Maunakea, covering about 40 deg$^2$ centered on the NEP \citep[see e.g.][]{2018ApJ...859...91S}, as well as a deep \textit{Spitzer} coverage of the \textit{Euclid} NEP deep field (Moneti et al. 2021, in prep.).

In addition to the deep coverage in the all-sky survey and several serendipitous pointings centered on the NEP, \textit{ROSAT} has also performed a large number of raster-scan pointings around the NEP, as well as several pointed observations on particular interesting targets. The motivation for this work has been to analyse all the \textit{ROSAT} survey and pointing data in the \textit{HEROES} area in a systematic and coherent fashion. In order to do so, we have restricted the off-axis angle of the \textit{ROSAT} observations to <30\arcmin~ in order to avoid the outer portions of the field-of-view, where the point spread function degrades significantly. As we will see, the soft X--ray sensitivity limit and the resulting angular resolution of this NEP raster-scan survey can be compared to the expected parameters of the \textit{eROSITA} all-sky survey and thus provide a real-sky approximation and reference field for this ongoing survey.  Throughout this work we adopt a $\Lambda$-cosmology with $\Omega_M$=0.3 and $\Omega_{\Lambda}$=0.7, and H$_0$=70 km s$^{-1}$ Mpc$^{-1}$ \citep{2003ApJS..148..175S}, and all magnitudes are given in the AB system.  

\section{The \textit{ROSAT} Data Preparation}
\label{sec:prep}

Table \ref{Table:log} in the appendix shows the observing log of the \textit{ROSAT} observations within $\sim$3.5 deg from the NEP used for our analysis. Apart from the \textit{ROSAT} All-Sky Survey (RASS) there were about 35 "Raster" pointings distributed around the NEP. In addition to the very long \textit{ROSAT} Wide-Field-Camera "WFC background" observations in the beginning of the mission, there were also a number of deep "Idle" or "Dummy" pointings close to the NEP (filling gaps in the observation timeline), plus pointings towards several interesting individual targets in the same region. Please note, that all observations in 1990 have been performed with the primary \textit{ROSAT} Position Sensitive Proportional Counter (PSPC-C) \citep{1987SPIE..733..519P}, while after an incident in which the PSPC-C looked at the sun and was unfortunately destroyed, the redundant PSPC-B was used. This whole data set has never been analysed jointly and in a coherent way. For our analysis we used a version of the interactive scientific analysis \textit{EXSAS} \citep{1993euge.book.....Z} system available on a computer at MPE Garching. In general we followed the analysis procedure in H06, which is based on the standard \textit{ROSAT} data reduction described in detail in \cite{1999A&A...349..389V}. However, we applied a number of non-standard selections and corrections to the data before the actual source detection. We first downloaded all the data listed in Table \ref{Table:log} and projected the X--ray events and the attitude files to a common celestial reference frame centered on the NEP. 

Next we choose an optimum cutoff radius for the detector field of view. The PSPC has a circular field of view with a radius 57\arcmin. The PSPC entrance window has a rib support structure with an inner ring at a radius corresponding to 20\arcmin~\citep{1987SPIE..733..519P,1997astro.ph.12341H}. Both the \textit{ROSAT} telescope angular resolution and its vignetting function are roughly constant within the inner 20\arcmin ring, but degrade significantly towards larger off-axis angles. The combined detector and telescope point-spread functions (PSFs) are described in detail in \citet{2000A&AS..141..507B}. To first order, the PSF at each off-axis angle can be approximated by a Gaussian function with a half-power radius (HPR) of 13, 22, 52, 93, 130, and 180\arcsec, at off-axis angles of 0, 12, 24, 36, 48, and 57\arcmin, respectively (at 1 keV). The vignetting function at 1 keV drops almost linearly to about 50\% at an off-axis angle of 50\arcmin. Taking into account all these effects, the half-power radius of the overall RASS point-spread function is 84\arcsec~ \citep{2000A&AS..141..507B}. This means that the classical confusion limit (40 beams per source) is reached at a source density of about 15 sources/deg$^2$, which is well exceeded in the high-exposure areas of our survey. In addition, we need to optimally discriminate between extended and point-like X--ray sources, calling for as high an angular resolution as possible. We therefore have to reduce the detector FOV. The sharpest imaging is achieved within the inner 20\arcmin~of the PSPC FOV, corresponding to the inner ring-like rib of the PSPC support structure (see Figure \ref{fig:images}). However, there is a trade-off between image sharpness and the number of photons required for detection and image characterization. In particular in the outer areas of our survey, where the RASS exposure times drop significantly, a 20\arcmin~FOV radius does not provide sufficient exposure time. Taking into account the various competing factors in this trade-off we made a few tests varying the FOV cutoff radius, and finally decided on an optimum FOV radius of 30\arcmin. The PSPC detector coordinates have a pixel size of 0.934\arcsec. We thus removed all X--ray events from the data set, which are further than 1925 pixels from the PSPC center pixel coordinate [4119,3929]. A similar cut had to be applied to the modified PSPC instrument map (MOIMP), which is later used for the construction of the survey exposure map. 

Using the information in \citet{2000A&AS..141..507B}, we calculated an effective survey PSF by appropriately weighting the corresponding functions at 0, 12, and 24\arcmin; this function has an HPR of 34\arcsec, i.e. about 2.5 times sharper than the standard RASS PSF, but still about 30\% larger than the expected eROSITA survey PSF \citep{2012arXiv1209.3114M}. It has to be noted, that our survey contains both survey and raster-scan data, for which the PSF approximation averaged over the whole field of view is appropriate. But the survey also contains a number of deep pointed observations of specific targets, for which a local PSF model would be more appropriate. To a large degree the maximum likelihood algorithm discussed below takes care of these differences. 

The next step of the analysis is the astrometric correction of the X--ray events to a common reference frame. For this purpose we analysed all the pointed observations in Table \ref{Table:log} separately and compared the detected X--ray positions to a catalogue of optical reference point sources from the \textit{Gaia} DR2 catalogue \citep{2018A&A...616A...1G}. We detected and corrected an offset of 10\arcsec~for all pointed observations performed with the PSPC-C, while all PSPC-B observations are compatible - within the errors - with the \textit{Gaia} reference frame. For the RASS part of the data we cut the survey into 10 roughly equal time intervals and detected the X--ray sources in individual images of the central NEP region. All the individual detections are consistent with the average survey image and the \textit{Gaia} reference frame, so that we did not need to apply an astrometric correction to the survey data.

\begin{figure*}[htbp]
    \includegraphics[width=.54\textwidth]{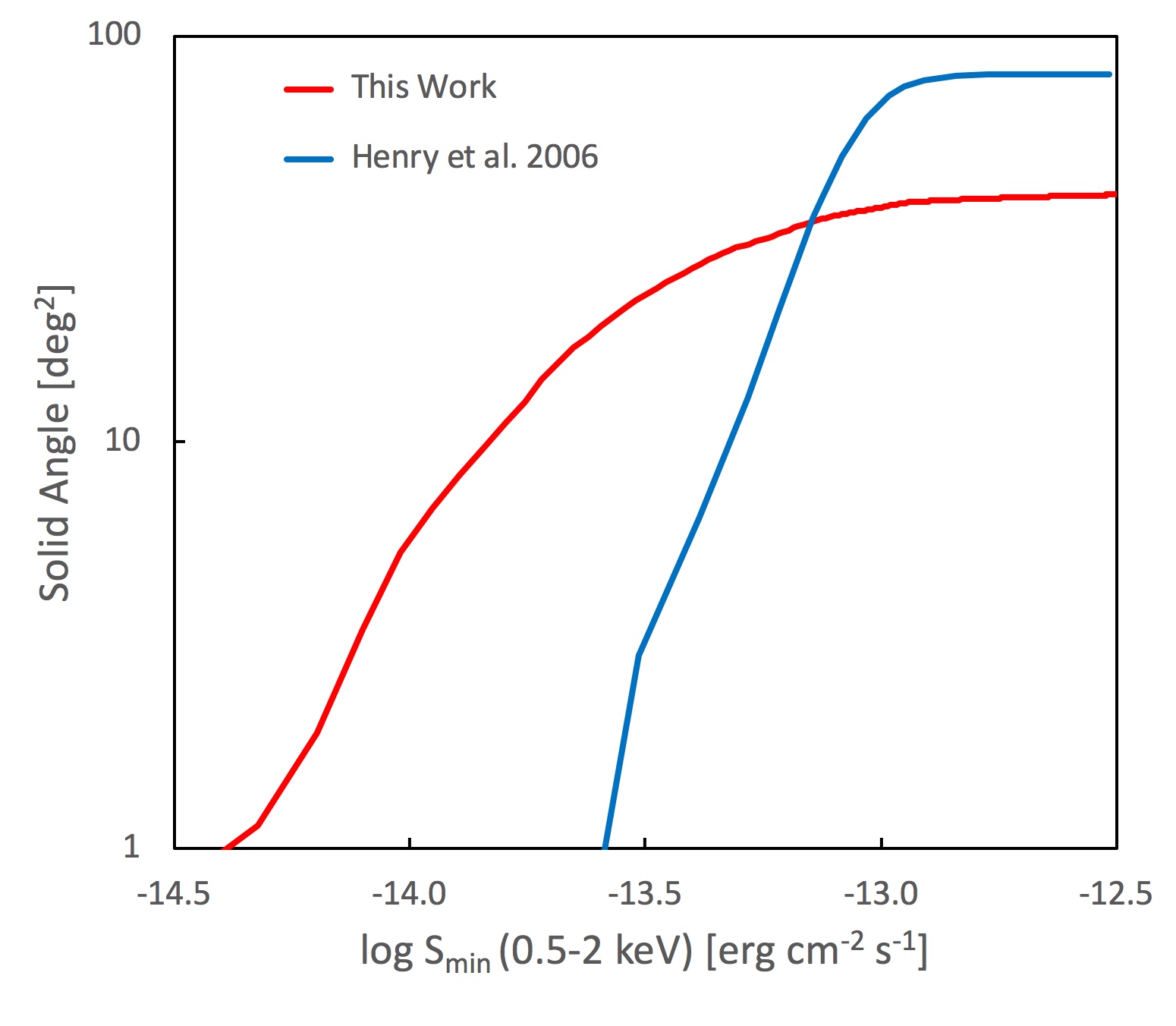}
    \includegraphics[width=.48\textwidth]{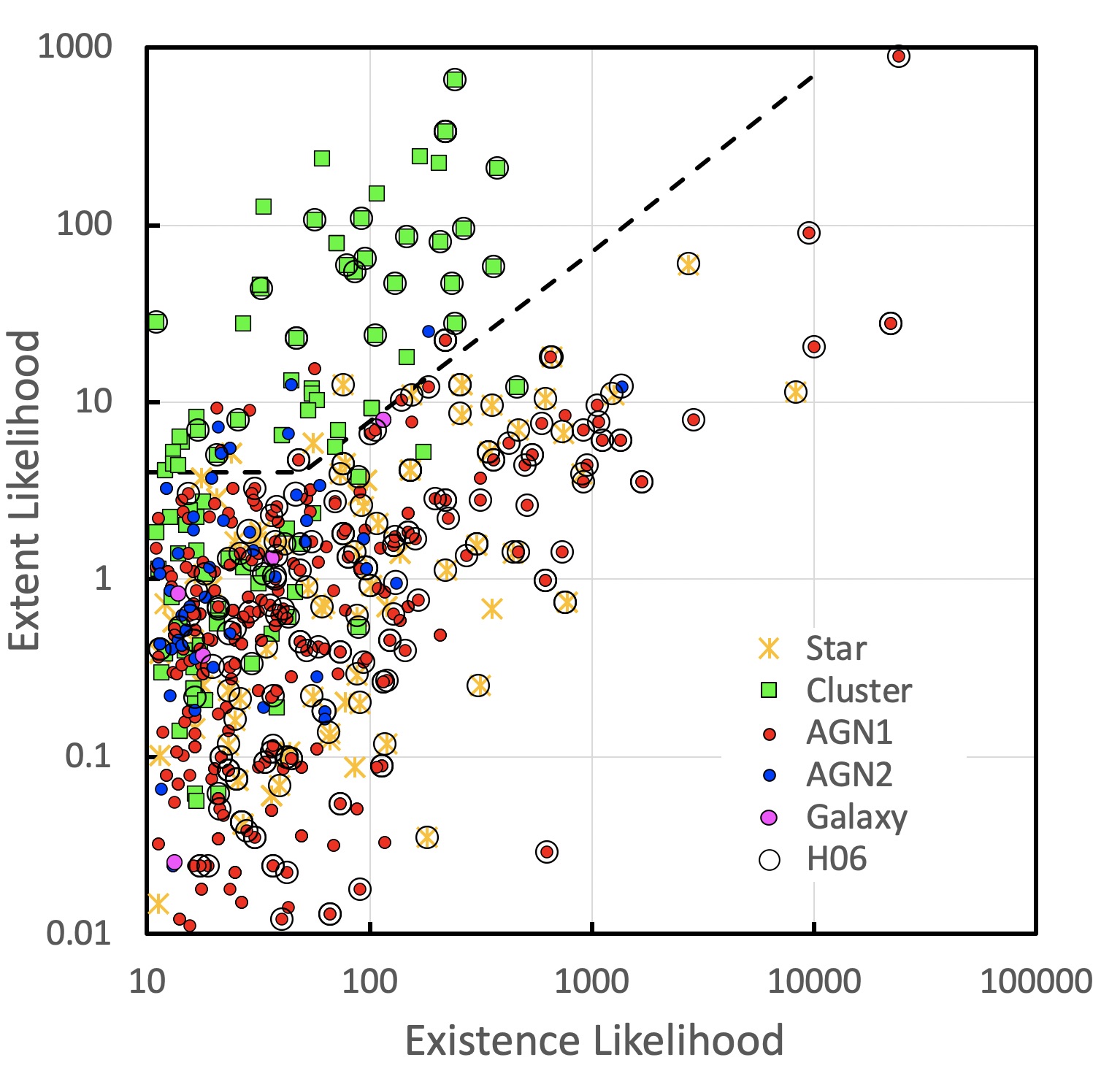}
    \caption{Left: \textit{ROSAT} NEP Raster Survey sensitivity function (red) compared the the RASS NEP survey of H06 (blue). Right: Characterization of extended sources in the diagram of existence versus extent likelihood. Red circles correspond to type1 AGN, blue circles to type2 AGN, pink circles to normal galaxies, green squares to clusters of galaxies, and yellow asterisks to stars. Black circles around the symbols indicate sources from the H06 catalogue. This figure, as also the corresponding following figures are only shown for the unique optical counterparts (see section \ref{sec:optids}). Bright point sources (AGN, stars) appear as extended because of the imperfect description of the point spread function in the ML algorithm. The dashed line shows an empirical separation between true and spurious extension.}
    \label{fig:solid_extent}
\end{figure*}

\section{X--ray Analysis}

\subsection{Source Detection}

We have restricted the basic source detection to the standard "hard" \textit{ROSAT} PSPC energy band (0.5--2 keV), corresponding to PSPC PHA channels 52--201. For extragalactic work this is the most sensitive band, least affected by interstellar absorption and also with the best PSPC angular resolution. The interstellar neutral hydrogen column density in our region of interest varies from $N_H$= 2.5$\times$10$^{20}$ to 8.3$\times$10$^{20}$ cm$^{-2}$ with a median of 4.1$\times$10$^{20}$ cm$^{-2}$ \citep{1994ApJS...95..413E}. In the hard band this leads to negligible absorption differences, while in the soft (0.1--0.4 keV) band the variation in transmission would be a factor of 7 across the field. 

We accumulated an image in the 0.5--2 keV band centered on the NEP with $512\times512$ pixels of 45\arcsec each. Figure \ref{fig:images} (upper left) shows the resulting raw counts image, where the different exposure times are clearly visible. The image in the upper right of the figure shows the exposure map, calculated from the \textit{ROSAT} attitude files and the PSPC modified instrument map (MOIMP) representing the detailed geometry of the PSPC support window and the telescope vignetting. Some of the deeper pointings have been carried out in a static (i.e. non "wobbled") observing mode. For these, the structure of the PSPC support window is clearly visible in the exposure map, while it is partly smeared out in the "wobbled" exposures. The detailed effect of the PSPC support structure and the Wobble-Mode on the ROSAT deep surveys has been discussed in \citet{1997astro.ph.12341H}. The image on the lower left is the exposure-corrected count rate map. Apart from some enhanced noise artifacts due to the relatively low exposure in the upper left and lower right corners, the image appears quite homogeneous, despite the large variations in exposure time, confirming the quality of the exposure map. Some bright diffuse X--ray sources can be readily identified.

For the source detection we followed the RASS procedure described in \citet{1999A&A...349..389V} and H06. However, we apply the detection only to the hard band image and use the SExtractor \citep{1996A&AS..117..393B} detection technique instead of the standard map detection method MDETECT before the final Maximum Likelihood (ML) source characterization. The first step is the local detection algorithm LDETECT, using a $3\times3$ pixel detect cell, and a local measurement of the background in the surrounding 16 pixels. After a first pass at full resolution, the detect cell and corresponding background area are successively doubled in several more passes, in order to also detect larger extended sources. The LDETECT step is applied to the raw image (see Figure \ref{fig:images} upper left), uncorrected for exposure time. 

A more elaborate background map is used for the next detection step. Around every X--ray source detected by LDETECT (using a higher likelihood threshold of $L\ge10$) a circular region with a radius about the size of the detect cell is cut out from the raw image (the "Swiss cheese map"). The remaining image is then divided by the exposure map and binned into coarser pixels. This cleaned, exposure-corrected image is fit by a smooth two-dimensional spline. After the spline fit, a $4\sigma$ cut is applied to pixels with count rates above the determined background, in order to remove artefacts from large diffuse sources and bright source haloes. This procedure is repeated until no more excess coarse pixels are found. Finally a background map is produced by applying the spline parameters to all pixels in the original image, and multiplying by the exposure map. Figure \ref{fig:images} (lower right) shows the count rate sensitivity map, which is a combination of the background map and the exposure map (see below). 

The standard second stage of the source detection using the map detection algorithm MDETECT on the raw image with the same detection cells as LDETECT, but taking the background map estimate instead of the local background, turned out not to be appropriate for our complex set of observations. On one hand, we have high-sensitivity confusion-limited areas in the image, which are not treated well by the standard LDETECT and MDETECT algorithms. On the other hand, large exposure gradients and diffuse emission features produce a number of artefacts posing difficulties for the standard detection procedures. We therefore decided to apply the SExtractor algorithm well known from optical astronomy \citep{1996A&AS..117..393B} as the interim detection step offering possible "regions of interest" to the final ML source detection and characterization. This procedure works reasonably well for isolated point sources, and even moderately confused sources, as well as bright diffuse sources. It, however, breaks down at the low exposure regions, where Gaussian statistics are no longer appropriate and have to be replaced by Poissonian estimates. This is the reason, why we performed a detailed visual inspection and screening of the SExtractor source list selecting 1200 regions of interest to be finally offered to the ML procedure. In this step we also manually inserted a number of source regions of interest, which were missed by the algorithms; these are marked in the final catalogue.

\setlength{\tabcolsep}{9pt}

\begin{table*}
\caption{ \textit{ROSAT} NEP Raster X--ray Catalogue.}
\vskip -0.5truecm
\label{Table:XID}
\begin{center}
\begin{tabular}{rllrrrrrrrr}
\hline\hline
(1) & (2)    & (3)     & (4) & (5)& (6) & (7)  & (8) & (9)& (10) & (11)\\ 
XID & RA$_X$ & DEC$_X$ & err & $L_{exi}$ & CTS & eCTS & $L_{ext}$ & Exposure & F$_X$(0.5-2)& m\\
    &   J2000     &     J2000    & [\arcsec] &    &     &      &     & [s] & [$10^{-14}$ cgs]\\
\hline
1&261.4722856&68.15910084&9.9&13.3&8.2&3.1&0.0&1803&5.42\\
2&261.7538598&69.44866306&11.0&30.5&12.3&3.6&0.0&1341&11.00\\
3&261.9778949&67.81116425&14.1&29.1&16.9&4.3&9.0&2261&8.92\\
4&262.1400312&67.54254828&9.8&21.1&14.4&4.2&0.7&2270&7.60\\
5&262.4181202&68.79506796&9.2&26.9&14.4&4.0&0.0&1448&11.91\\
6&262.4862510&66.86526820&11.5&12.2&8.9&3.4&0.0&1741&6.10\\
7&262.5572915&65.73547731&9.5&17.4&10.8&3.6&0.0&3546&3.65\\
8&262.5800581&68.16710175&8.5&17.3&8.6&3.1&0.0&1839&5.62\\
9&262.9844661&65.37719281&13.5&19.3&14.0&4.1&0.0&3379&4.95\\
10&263.0151617&67.80792566&9.7&13.6&9.8&3.5&0.1&2370&4.95\\

...\\
796&277.4381113&67.81986356&5.4&68.4&25.3&5.3&0.0&1637&18.49&1\\
797&277.4522820&64.58551610&8.4&24.8&18.5&4.8&0.2&4657&4.75\\	
798&277.5094136&66.75661056&6.0&53.0&22.9&5.0&0.0&2095&13.04\\	
799&278.0683716&68.54311384&9.8&21.1&10.5&3.5&0.1&1091&11.55&1\\
800&278.1405809&68.61500503&8.1&42.7&17.1&4.2&0.1&1044&19.62\\	
801&278.1552805&68.80281717&6.9&364.3&127.9&11.5&58.1&783&195.44&1\\
802&278.1773278&69.07251242&13.6&14.7&6.3&2.7&0.0&438&17.22\\	
803&278.4506109&69.36011661&13.7&18.7&9.8&3.3&0.0&323&36.12\\	
804&278.6549948&69.52945061&11.7&155.2&14.9&3.8&10.9&320&55.50\\	
805&278.6965904&69.41463855&23.1&46.9&5.0&2.2&0.9&306&19.50\\	
\hline
\end{tabular}
\end{center}
\vskip -0.2truecm
Column explanation: (1) internal XID identification; (2) and (3) X--ray source coordinates in J2000.0; (4) position error (including systematics); (5) existence likelihood $L_{exi}$; (6) and (7) detected number of net counts and statistical error; (8) extent likelihood $L_{ext}$ (the actual extent determination is discussed in the text); (9) exposure time; (10) 0.5--2 keV source fluxes in units of 10$^{-14}$ erg cm$^{-2}$ s$^{-1}$; (11) flag for manual ML input.
\end{table*}

\subsection{The \textit{ROSAT} NEP Raster Catalogue}
\label{subsec:XIDcatalog}

The third and last detection step uses a ML algorithm \citep{1988ESOC...28..177C,2001A&A...370..649B} applied to unbinned individual photons to both detect the sources and measure their final parameters. X--ray events in a circle of 3\arcmin~radius around every entry in the regions of interest list were selected. For a subset of 34 significantly extended sources (marked in the final catalogue) we increased the event selection radius to 6\arcmin. The ML fit takes into account each photon with the appropriate PSF corresponding to the off-axis angle and energy at which it was detected. An effectively smaller extraction radius and therefore higher weight is given to photons detected near the center of the field, compared to those at the outskirts with the worse PSF. The ML analysis yields a number of parameters for each source. Most important is the source existence likelihood $L_{exi} = -ln(P_0)$, where $P_0$ is the probability that the source count rate is zero \citep[see][]{2001A&A...370..649B}. The threshold for this parameter has been set to $L_{exi}\ge11$ to define the final content of our survey. The likelihood analysis also determines the best parameters for the source position and a corresponding position error, the net detected counts of the source and its error, and an estimate of the angular extent of the source and its likelihood to be extended. To estimate this extent the ML algorithm fits a Gaussian model added in quadrature to the PSF.

Table \ref{Table:XID} gives the final catalogue of 805 detected X--ray sources in an abbreviated form. (The complete catalogue is available in the online publication.) For all sources the parameters of the ML analysis are quoted. A threshold of $L_{exi}$ $\ge$11 has been applied throughout (see below). For a true estimate of source extent a combination of both existence likelihood and extent likelihood have to be considered (see discussion below). In order to convert the source count rates to fluxes, following previous work, we assume an extragalactic point source with a photon index --2 and an interstellar column density of $N_H$=4.1$\times$10$^{20}$ cm$^{-2}$, folded through the PSPC response matrix. A source with a 0.5--2 keV flux of $10^{-11}$ erg cm$^{-2}$ s$^{-1}$ yields a PSPC count rate of 0.815 cts s$^{-1}$ in the 0.5--2 keV band. Because we restrict the source detection to the hard band, variations of $N_H$ across the field can be neglected. Therefore the extragalactic point source fluxes can be readily converted into other spectral model fluxes. For 34 sources significantly extended in the first ML analysis or by visual inspection, the extent characterization of the ML algorithm may be insufficient, among other reasons because of the limited photon extraction radius, and therefore the detected count rate could be significantly underestimated. For these sources we determined the ML parameters by doubling the event extraction radius and marked them down with the flag (\textit{m}) in column (11) of the catalogue. Another ten non-extended sources, which were originally not included in the source region of interest list, were manually fed through the normal ML detection procedure and are also marked in column (11).

As a final quality check of the source detection we compared our catalogue to previous X--ray information in the HEROES field, predominantly with the Second \textit{ROSAT} all-sky survey (2RXS) source catalogue \citep{2016A&A...588A.103B} and the Second \textit{XMM-Newton} Slew Survey (XMMSL2)\footnote{https://www.cosmos.esa.int/web/xmm-newton/xmmsl2-ug}
\citep[see also][]{2018MNRAS.473.4937S}, as well as with the fourth-generation \textit{XMM-Newton} serendipitous source catalogue  \citep[4XMM-DR9,][]{2020A&A...641A.136W}. A total of 477 of our 805 X--ray sources have counterparts in the literature. In particular there are 431 matches with 2RXS, 83 with 4XMM-DR9, and 40 with XMMSL2. About half of these (228) are also present in the H06 catalogue. As expected, the 2RXS analysis has a larger difficulty in determining source  extents, and resolving confusion, but yields better accuracy for isolated sources in the outskirts of our survey. As we will see below, the literature data (in particular the better \textit{XMM-Newton} positions) confirmed almost all of our optical counterparts in the overlapping areas, but for 17 cases identified high-quality optical counterparts that would have been missed in our analysis.

\subsection{Sensitivity Limits}

An important consideration is the setting of the source detection threshold $L_{exi}$ and thus the corresponding sensitivity function for the survey. It has to be chosen to maximise the number of true sources in the survey, on the other hand to minimise the number of spurious sources. The expected number of spurious sources is the above probability $P_0$ multiplied with the number of statistically independent trials $n_{trial}$ across the field. Given the rather complex setup of our survey, and the intricacies of the ML source detection algorithm, it is not possible to determine $n_{trial}$ analytically. A rough estimate is e.g. the number of statistically independent detection cells in the LDETECT process, which is $(512\times512)/(3\times3)$=29127. Further down, in the context of calculating the survey sensitivity function, we derive the average \textit{effective} extraction radius for our ML analysis, which is 1.35 pixels, or 60.8\arcsec. Using this number for the size of the detection cell, we arrive at 45785 independent trials. Therefore an existence likelihood threshold of $L_{exi} \ge 11$ has been applied to obtain statistically less than one spurious source in the overall survey. This threshold is more conservative than e.g. the limits of $L_{exi} \ge 6.5$, or $L_{exi} \ge 9$ selected for the Second \textit{ROSAT} all-sky survey (2RXS) source catalogue \citep{2016A&A...588A.103B}, which lead to a much larger spurious source fraction of about 30\% and 5\%, respectively. The treatment is, however, rather simplified. In reality systematic effects can increase the number of spurious sources, e.g. the high diffuse surface brightness around bright extended and point sources discussed above, and the effects of confusion. Using the likelihood threshold of $L_{exi} \ge 11$, we thus can expect a handful of spurious sources, consistent with the optical identifications discussed in Subsection \ref{sec:optids}. Together with a careful manual screening of spurious sources and merging of split source components, we arrive at a catalogue of 805 X--ray sources (see Table \ref{Table:XID}). 254 of our sources are matching with entries in the H06 catalogue within 2.5$\sigma$ error circles. These are indicated with black solid circles in the relevant figures throughout this paper. The classical confusion limit in radio astronomy is defined as 40 "beams" (i.e. statistically independent detection cells) per source \citep{1974ApJ...188..279C}. With the number of independent trials estimated above, this corresponds to 1145 sources in the whole survey, or about 28 sources deg$^{-2}$. Our average source density is below this number, but in the deepest pointed areas around the NEP the source density is higher than this classical confusion limit.

In order to use the survey for quantitative statistics, it is important to calculate a survey selection function. This is equivalent to the sky coverage solid angle, within which sources of a particular brightness would have been detected in the survey. For the maximum likelihood algorithm described above, with its intricacy of different effective detection cell radii for individual photons, it is not possible to calculate a sensitivity limit analytically. Therefore some publications resort to extensive Monte Carlo simulations for the determination of the sky coverage function \citep[see e.g.][]{2009A&A...497..635C}. For the statistical applications in this paper, it is appropriate to approximate the sensitivity map by a simple signal to noise ratio calculation, following the procedure described in H06. In the case of significant background in the detection cell, the likelihood can be approximated by a Gaussian probability distribution. A detection likelihood of $L_{exi} \ge 11$ almost exactly corresponds to a Gaussian probability of $3\sigma$, so we base our sensitivity calculation on this limit. In the background limited case, the signal to noise ratio can be calculated as $S/N = S / \sqrt{S+B}$, where $S$ is the net detected counts of a source, and $B$ is the number of background counts in the source detection cell. For a given background brightness per pixel of $b$, the background counts in a circular detect cell of radius $r_b$ is simply $B=b \pi r_b^2$. For the purpose of this analysis $b$ is assumed error-free, because it has been determined from a large background map with high statistical accuracy. 

Compared to the original extraction radius of 3\arcmin~for the ML algorithm, the \textit{effective} extraction radius applicable for the background calculation is smaller, because every photon is treated with its own PSF. The task at hand is now to determine the effective background cell radius $r_b$, which is equivalent to the ML treatment in our survey. For this purpose we chose 71 X--ray sources detected with likelihoods of $10 \le L_{exi} \le 12$ and varied the background cell radius $r_b$ until the distribution of signal to noise ratios determined by the maximum likelihood procedure agreed with that of the Gaussian calculation. This calibration results in an effective background cell radius of $r_b$ = 1.35 pixels = 60.8\arcsec. The survey-integrated point-spread function determined above contains 77\% of the flux within this radius. Multiplying the background cell area with this radius to the background map derived above, applying the above S/N calculation formula, and dividing by the exposure map, we arrive at the count rate sensitivity map shown in Figure \ref{fig:images} on the lower right. 

To convert this map into proper source fluxes we have to sum the count rate sensitivity distribution over all pixels in the $512 \times 512$ pixel images to obtain the corresponding cumulative area in deg$^2$, and to divide the count rates by the PSF correction factor of 0.77 and by the above count rate to flux conversion factor of 8.15$\times$10$^{10}$ cts erg$^{-1}$ cm$^2$. This way we arrive at the final survey selection function shown in red in Figure \ref{fig:solid_extent} (left) in comparison to the corresponding curve of H06 (using their tabulated values for extragalactic point sources). The total solid angle covered in our survey at high fluxes is 40.9 deg$^2$, about half that of H06. On the other hand, our survey is about 0.5 dex deeper than H06. 

\subsection{Extended Source Analysis}

The ML algorithm is very useful for the separation between point sources and clusters. It weights different photons according to the size of their individual PSF model, and therefore is arguably the most sensitive method to detect extended sources in a photon-starved situation, where the extent is not much larger than the PSF. However, because of the necessary approximations in the description of the actual PSF, which depends on the off-axis angle and energy and also has extended scattering wings, the method tends to detect spurious extents in bright X--ray point sources. This is shown in Figure \ref{fig:solid_extent} (right), where the extent likelihood is compared to the existence likelihood for the whole X--ray catalogue. Most clusters from the H06 catalogue (green squares surrounded by black circles) are clearly segregated from the rest of the sample, with extent likelihoods $L_{ext}\ge4-5$. However, above existence likelihoods of $L_{exi}\ge50$, the bright point sources (yellow stars and red AGN1) from H06 are creeping into the significantly extended area. We therefore empirically determined the dashed line to discriminate truly extended sources. There is one star from H06 (\#3970), which appears significantly extended with an extent likelihood $L_{ext}$=12.4. There are actually two bright stars in the same error circle, so that the extent could be due to the double star nature. An additional complexity occurs in the case of AGN residing in clusters of galaxies, which may show X--ray extent despite being identified with a point source. The most interesting case is the AGN1 \#5340 in the H06 catalogue, the object in the top right of Figure \ref{fig:solid_extent} (right). This is the well-known bright HEAO-1 catalogue source H1821+643 \citep{1984ApJ...281..570P} sitting inside the massive cluster CL 1821+643 \citep{1992AJ....103.1047S}, later also detected by \textit{Planck} through its Sunyaev-Zeldovich effect \citep{2014A&A...571A..29P}. An extended X--ray source superposed on the AGN point source extent has been independently measured for this object with the ROSAT HRI \citep{1997AJ....113.1179H}, therefore the extent measured in our analysis is approximately correct and thus the dashed line maybe a bit too conservative.

\section{Multi-band observations}

\subsection{\textit{HEROES} Optical/NIR wide-field imaging}

The ultimate goal of \textit{HEROES} \citep[see e.g.][]{2018ApJ...859...91S} is to cover an area of 120 deg$^2$ around the NEP, where the \textit{eROSITA} X--ray all-sky survey \citep{2012arXiv1209.3114M,2020arXiv201003477P} will have some of its deepest coverage. This survey utilizes the wide-field capability of the Hyper Suprime-Cam instrument \citep[HSC;][]{2012SPIE.8446E..0ZM} on the \textit{Subaru} 8.2 m telescope, as well as the MegaPrime/MegaCam and WIRCam instruments on the 3.6 m \textit{CFHT} to map this large area. \textit{HEROES} comprises \textit{grizy} broad-band images with limiting magnitudes around 26.5--24.5 and  NB816 \& NB921 narrow-band images with limiting magnitudes around 24 from \textit{Subaru}, as well as U and J band images with limiting magnitudes 25.5 and 22.1, respectively, from \textit{CFHT}. So far an area of approximately 40 deg$^2$ has been covered with HSC during the periods 2016-07-01 to 2016-07-10, and 2017-06-21 to 2017-06-28, while the \textit{CFHT} data were taken in the interval 2016-03-18 to 2016-08-20. The HSC observations were aquired in a close packed set of dithered observations described in detail in \citet{2018ApJ...859...91S}.

A formal publication of the \textit{HEROES} catalogue is in preparation. Here we just give some basic information about our data reduction process. Rather than using the HSC standard pipeline, which at the time of the massive data reduction task was not yet available to us, we analysed all HSC images with the \textit{Pan-STARRS} Image Processing Pipeline \citep[IPP;][]{2016arXiv161205240M}, which was well tested and available on a dedicated computer cluster allowing fast processing of the large data volume. The IPP pipeline can be adapted to any wide-field imaging dataset, as long as the instrument calibration specifics are incorporated. For HSC this required mainly the accurate description of the significant differential image distortion across the large field of view.  The details of the \textit{Pan-STARRS} Pixel Processing, i.e. the detrending, warping, and stacking of the images are described in \citet{2016arXiv161205245W}. Each exposure is cleaned from instrumental effects, and photometry and astrometry are performed by comparing the objects detected on the individual images with the \textit{Pan-STARRS} reference catalog. This process also yields the individual image quality in terms of seeing and photometric transmission. The seeing was typically very good during the observations, with a median around 0.7\arcsec~and a large fraction of photometric transparency. For stacking the images into a common pixel grid, we selected exposures with seeing better than 1.36\arcsec~and photometric zero points not fainter than 0.3 mag from the median. Co-adding the images to a certain degree also reduces non-astronomical artefacts like the reflection "ghosts" from bright stars outside the field of view. Faint sources are detected by forced photometry running simultaneously across all filters \citep{2016arXiv161205244M}. If a source is detected with more than $5\sigma$ in any particular band, photometry is forced on all other bands. This enables to run photometric redshifts on a large fraction of all sources. Whenever available, we used Kron magnitudes for this purpose. For each band we also obtain a star/galaxy separation parameter. The original HSC catalogue of objects detected significantly in at least one band contains 23.9 million objects. However, one has to be careful about false positive detections, because a single artefact in any band will produce an entry in the catalogue. Therefore we have selected objects detected in at least two bands for the photometric redshift determination described below.

The \textit{CFHT} WIRCAM J-band data was reduced and stacked with a custom version of the \textit{AstrOmatic} image analysis system \citep{2012ASSP...29...71B}, in particular using the packages SCAMP, SWarp, MissFITS and SExtractor \citep{1996A&AS..117..393B}. The \textit{CFHT} MegaCam U-band data have kindly been reduced and stacked with the excellent MegaPipe imaging pipeline at the Canadian Astronomy Data Center (courtesy Stephen Gwyn). The astrometric calibration was done with \textit{Gaia}, and the photometric calibration with a combination of SDSS data and a nightly zero-point calibration from MegaCam on photometric nights. This photometric calibration was bootstrapped to the few non-photometric nights, so that the zero-point calibration is self-consistent to 0.015 magnitudes. Again we used the Kron magnitudes  for the subsequent analysis.

\begin{figure}[htbp]
    \includegraphics[width=0.48\textwidth]{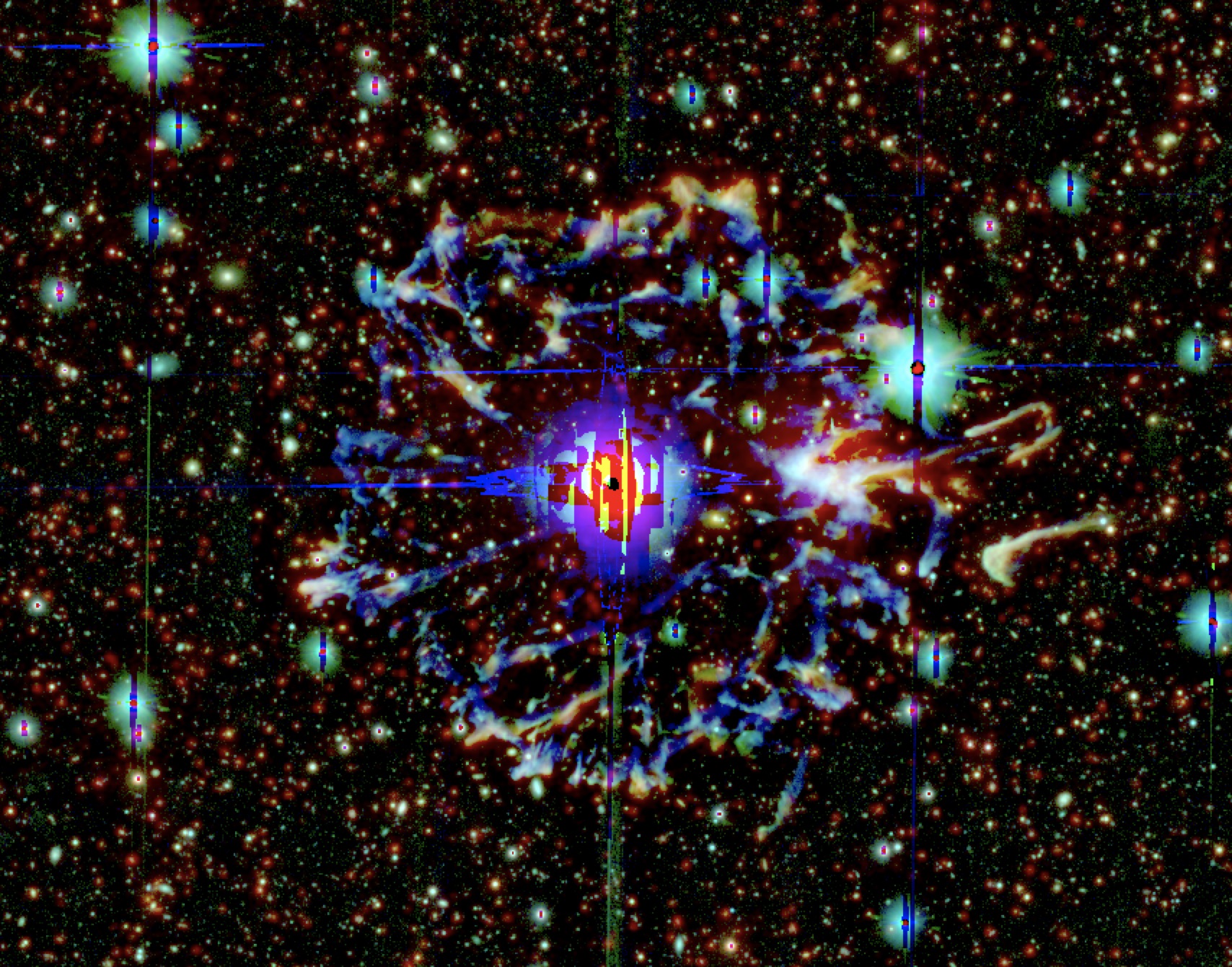}
    \caption{False colour \textit{HEROES} image of the shell around the planetary nebula NGC6543 (Cat's Eye Nebula) close to the NEP. Blue represents the HSC \textit{g}-band image and shows predominantly the green [OIII] line. Green represents the HSC \textit{r}-band and shows predominantly H$\alpha$. Red corresponds to the 4.5$\mu$m {\it Spitzer} image and mainly shows dust emission.}
    \label{fig:PN}
\end{figure}

\begin{figure*}[htbp]
    \includegraphics[width=0.52\textwidth]{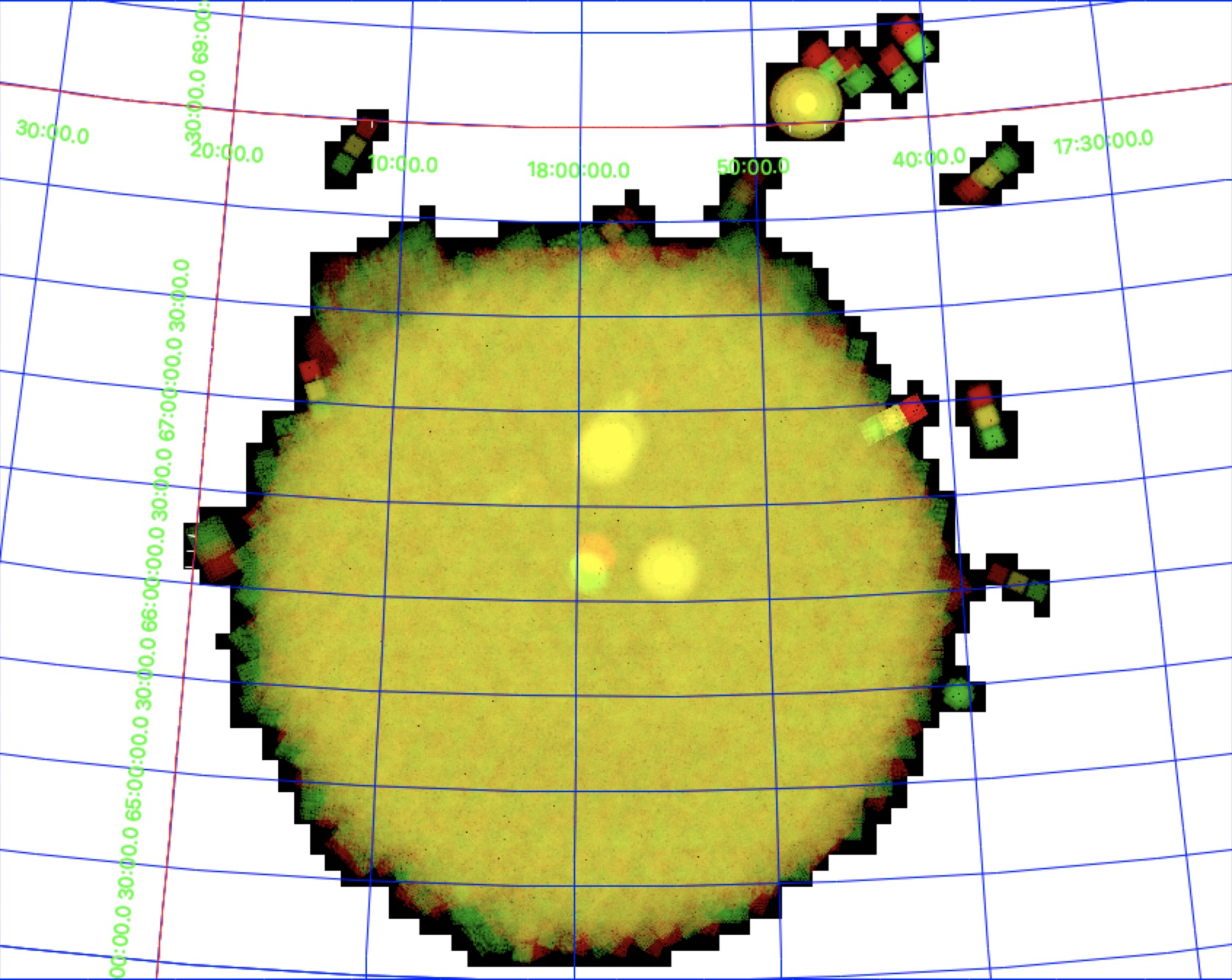}
    \includegraphics[width=0.45\textwidth]{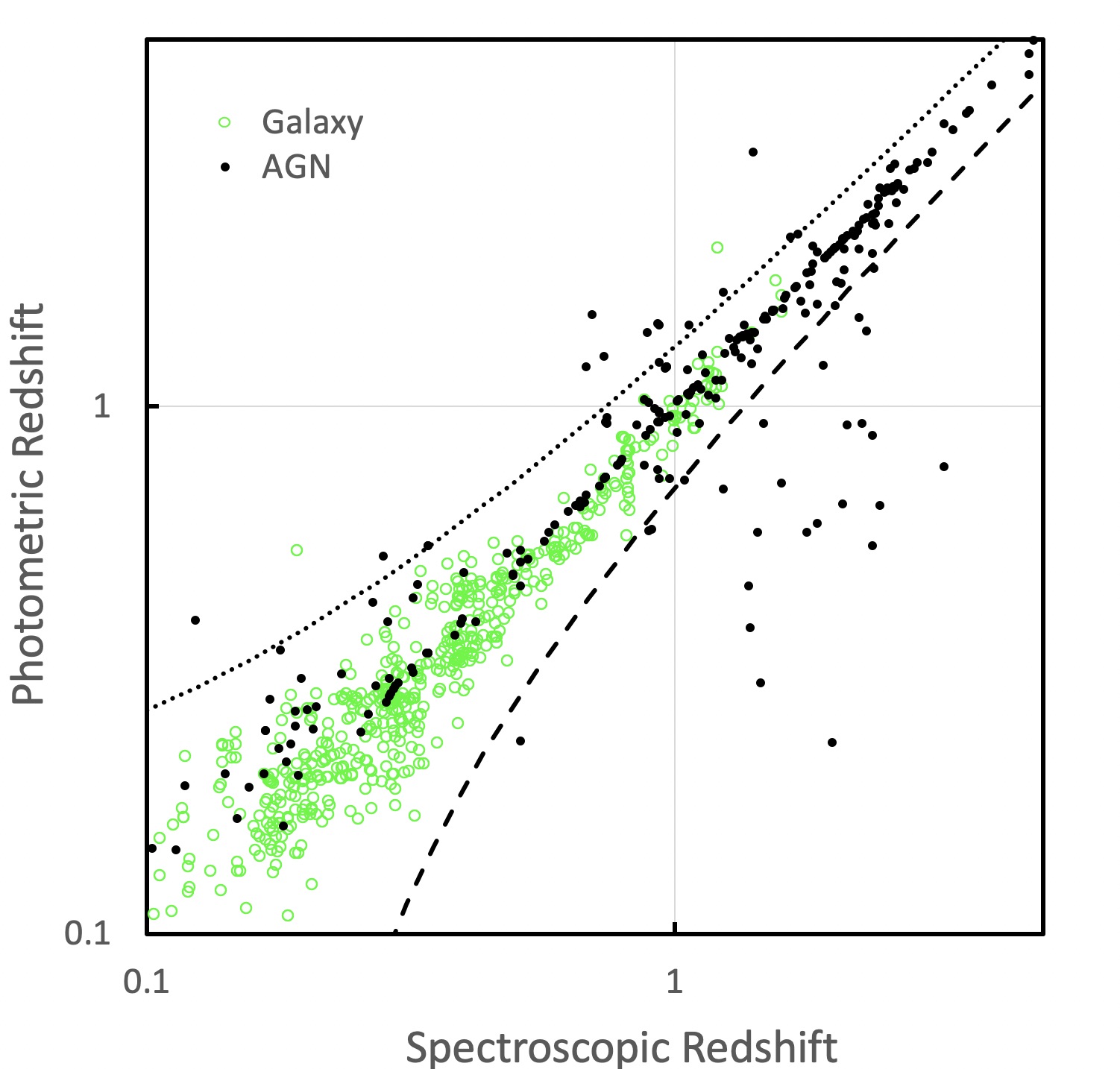}
    \caption{Left: Map of the \textit{Spitzer} IRAC 3.6$\mu$m (green) and 4.5$\mu$m (red) coverage in the \textit{Spitzer Cosmic Dawn Survey} around the NEP (Moneti et al. 2021, in prep.). Right: Comparison between spectroscopic and photometric redshifts in our sample. Green circles are 541 "clean" galaxies with spectroscopic redshifts in the \textit{HEROES} field. Black dots are a sample of 263 AGN with spectroscopic redshifts in the field.The dashed and dotted lines refer to a redshift error of $\Delta$\textit{z}/(1+\textit{z})=$\pm$0.15.}
    \label{fig:SCDS}
\end{figure*}

An example of the excellent \textit{HEROES} image quality, but also the various artefacts produced by bright objects in the field, is given in Figure \ref{fig:PN} showing the shell around the planetary nebula NGC6543 (Cat's Eye Nebula), that was ejected by its predecessor red giant star. The image is a combination of the HSC \textit{g}-band and \textit{r}-band data, predominantly showing the [OIII] and H$\alpha$ emission lines, and the \textit{Spitzer} 4.5$\mu$m data highlighting the dust emission. The actual PN in the center of the image is completely over-exposed. The high density of faint background objects shows the excellent sensitivity of the data. The NEP lies at relatively low Galactic latitudes, and therefore contains a rather large number of over-exposed foreground stars showing up in cyan colours in this image. In order to obtain reliable optical photometry for brighter objects, which are saturated in the \textit{HEROES} HSC images, we also made use of the \textit{SDSS} DR16 \citep{2020ApJS..249....3A}, the \textit{Pan-STARRS} DR1 \& DR2 \citep{2016arXiv161205560C}, and \textit{Gaia} DR2 \citep{2018A&A...616A...1G} catalogues. We also crosscorrelated our samples with the Far-UV and Near-UV photometry from the \textit{GALEX} surveys \citep{2005ApJ...619L...1M}.

\begin{figure*}[htbp]
    \includegraphics[width=\textwidth]{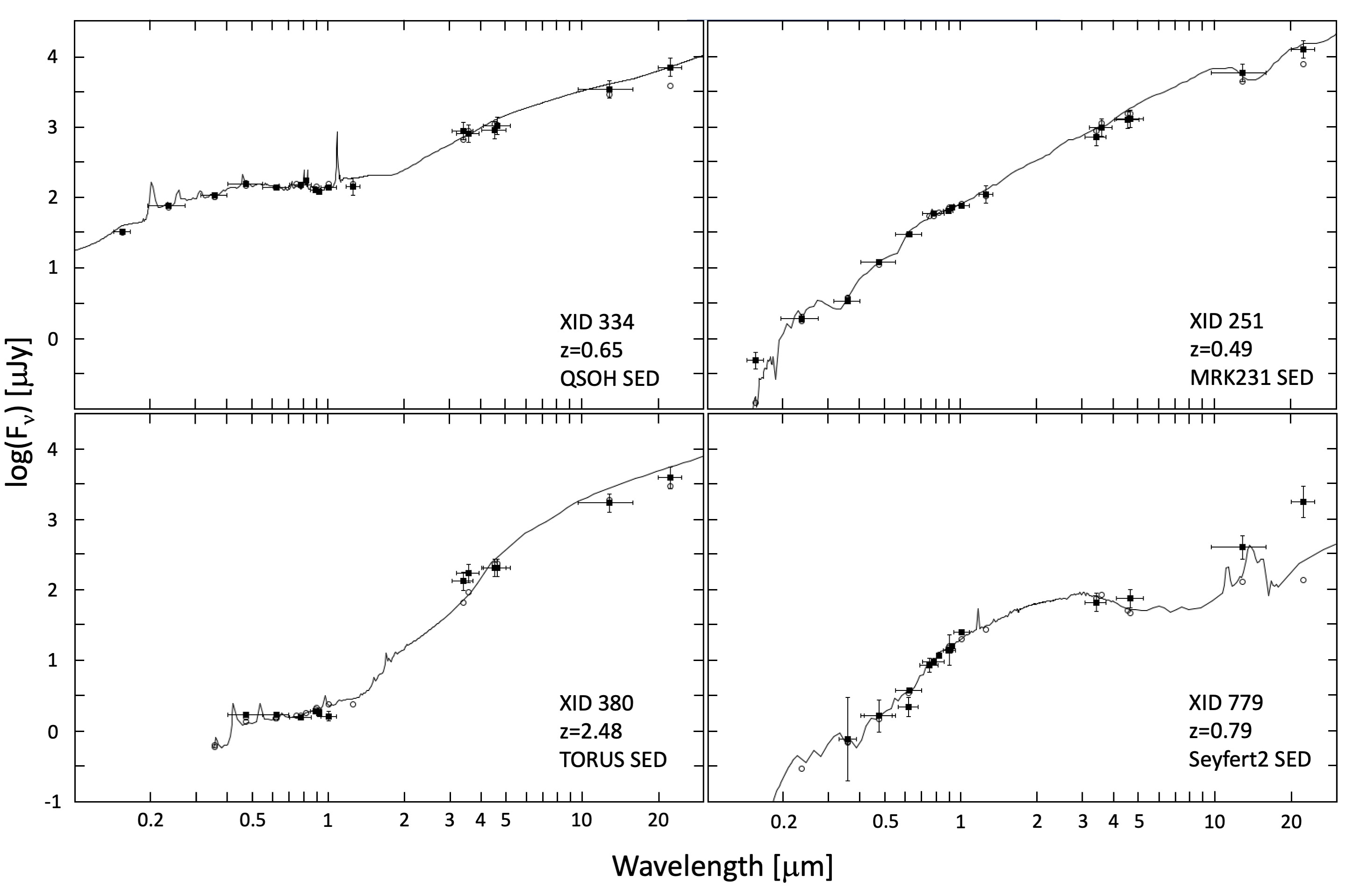}
    \caption{Some high-quality examples of photometric redshift fits to X--ray counterparts, illustrating different AGN SED models. The \texttt{MRK231} and \texttt{TORUS} spectra have the largest mid-infrared fraction of all chosen models.}
    \label{fig:SED}
\end{figure*}

\subsection{Mid-Infrared observations of the \textit{HEROES} field}

For photometric information in the 3--25$\mu$m band across the whole \textit{HEROES} field we used the \textit{WISE} all-sky survey catalogues. For the shorter wavelength W1 (3.4$\mu$m) and W2 (4.6$\mu$m) bands we used the \textit{CatWISE2020} all-sky catalogue \citep{2020ApJS..247...69E} (see also Marocco, F. et al. 2020 in prep.), containing about 1.89 billion objects observed by the Wide-field Infrared Survey Explorer (\textit{WISE} and \textit{NEOWISE}) and having both higher sensitivity and higher angular resolution than the original \textit{ALLWISE} catalogue \citep{2013yCat.2328....0C}. The \textit{CatWISE2020} catalogue in the \textit{HEROES} field contains 2.4 million objects and gives fluxes in the W1 and W2 band, which we converted to the AB magnitude system. For the longer wavelength W3 (12$\mu$m) and W4 (22 $\mu$m) bands we used the original \textit{ALLWISE} catalogue \citep{2013yCat.2328....0C}, again converting to AB magnitudes. At the faintest magnitudes the \textit{CatWISE2020} and \textit{ALLWISE} catalogues are, however, severely confusion limited. In the center of the HEROES field, we therefore also made use of the \textit{Spitzer} Observations of the \textit{Euclid} Deep Field North (Moneti et a., 2021, in prep.).

ESA's \textit{Euclid} mission is making great progress towards its launch, scheduled in 2022. \textit{Euclid's} main goal is to survey a large fraction of the sky and image billions of galaxies to investigate dark energy and dark matter over the history of the Universe. Roughly 10\% of the observing time will be dedicated to the Euclid Deep Fields, repeatedly observing three specific areas  of the sky, covering a total of 40 deg$^2$. These three fields were carefully selected to contain a minimum number of bright stars, low dust emission and zodiacal light. In addition these fields already have substantial multi-wavelength coverage, and will be observed with other space observatories, enabling a large amount of ancillary science. One of those fields encompasses an area of 10 deg$^2$ around the NEP, in the middle of the \textit{HEROES} field. NASA's \textit{Spitzer} telescope has performed a large survey (P.I. P. Capak) of two Euclid Deep Fields, the \textit{Euclid/WFIRST Spitzer Legacy Survey} \citep{2016sptz.prop13058C}. A total of 5286hr of \textit{Spitzer} observing time are distributed over 20 square degrees split between the Chandra Deep Field South and the NEP, with an exposure time of 2 hrs per pixel. The primary goal is to enable definitive studies of reionization, \textit{z}>7 galaxy formation, and the first massive black holes. The data will also enhance the cosmological constraints provided by \textit{Euclid} and \textit{Nancy Grace Roman Space Telescope (WFIRST)}. We are using a preliminary \textit{Spitzer} data product. The final survey is being prepared for publication as part of the \textit{Spitzer Cosmic Dawn Survey} (Moneti et al. 2021, in prep.), covering the three \textit{Euclid} deep fields and several other \textit{Euclid} calibration fields. These authors developed a new IRAC data processing pipeline and used the availability of highly precise astrometry available from \textit{Gaia} to reprocess nearly \textit{all} available Spitzer data (excluding short observations like calibrations on bright stars) in this field, which will be essential for the \textit{Euclid} calibration and for high-redshift legacy science. Figure \ref{fig:SCDS} (left) shows the sky coverage of the two channels 3.6$\mu$m (I1) and 4.5$\mu$m (I2) with the widest deep coverage of the NEP. We used SExtractor \citep{1996A&AS..117..393B} to extract source positions and magnitudes from the IRAC images in these two bands, yielding a catalogue of almost one million sources.

\subsection{Photometric Redshifts}
\label{subsec:photoz}

Based on the \textit{grizy} HSC detections, we joined the \textit{GALEX}, MegaCam, HSC, WIRCAM, \textit{SDSS}, \textit{Pan-STARRS} (partially), \textit{CatWISE2020}, \textit{ALLWISE}, and \textit{Spitzer} IRAC catalogues into a single source list by association through positional matching. For the UV, optical, and NIR images as well as the IRAC catalogue we allowed a maximum distance of 1\arcsec, while for \textit{GALEX} and \textit{WISE} we allowed a maximum of 3\arcsec. This way we obtained a combination of a maximum of 22 photometric bands (\textit{GALEX} FUV\&NUV, \textit{HEROES} \textit{UgrizyJ} plus NB816 \& NB921, \textit{SDSS} \textit{ugriz}, as well as \textit{CatWISE2020} W1\&W2, \textit{ALLWISE} W3\&W4, and IRAC I1 \& I2 bands). Ideally, for the highest accuracy of photometric redshifts, one should use forced aperture photometry in all bands. However, this is not possible in the case of a combination with catalogues from the literature. In some cases of very faint optical counterparts or objects confused with brighter nearby sources, we determined the correct magnitudes through manual aperture photometry. 

A nice review about the application of photometric redshift techniques in modern wide-field surveys is given by \citet{2019NatAs...3..212S}. 
We determined photometric redshifts using the \textit{LePhare} code \citep{1999MNRAS.310..540A,2006A&A...457..841I}. We followed the procedure described in \citet{2009ApJ...690.1236I}, basically fitting three different model families (galaxies, AGN and stars) to the spectral energy distribution. For the galaxy SED templates we used the models from \citet{2016ApJS..224...24L}, including emission lines. For AGN we applied the specific modifications described in \citet{2009ApJ...690.1250S}, wherever possible correcting for time variability between the \textit{SDSS} and \textit{HEROES} data, and taking the different SED templates used for point-like and extended AGN in \cite{2017ApJ...850...66A}. Because we include the \textit{WISE} W3\&W4 mid-infrared bands, we had to extrapolate some of the \cite{2017ApJ...850...66A} hybrid galaxy/QSO SEDs to longer wavelengths. Our AGN photometry also required to include the heavily absorbed \texttt{TORUS} model SED from \cite{2007ApJ...663...81P}. We adopted the Small Magellanic Cloud extinction law from \citet{1984A&A...132..389P} \citep[see][]{2009ApJ...690.1250S} to allow for intrinsic reddening of the sources, exploring E(B-V) values from 0 to 0.35 in steps of 0.05. Before the final fit we made slight adjustments to the zero-points for each band, using the \textit{LePhare} self-calibration procedure with 541 spectroscopic redshifts for clean galaxies (i.e. no AGN contribution, not confused, intermediate magnitudes) and 263 AGN with spectroscopic redshifts observed by \textit{HEROES} in the field. 

Figure \ref{fig:SED} shows four high-quality examples of photometric redshift fits for X--ray detected AGN candidates, illustrating different spectral energy distributions. The \texttt{MRK231} and \texttt{TORUS} SEDs have the largest relative mid-infrared contributions, and sizeable number of X--ray counterparts (16 and 5, respectively), and even more mid-infrared selected AGN (54 and 26, respectively) require these models. They  are reminiscent of the \textit{Spitzer} power law AGN SEDs detected in the \textit{Chandra} deep fields \citep{2007ApJ...660..167D}. The \texttt{TORUS} SED model from \cite{2007ApJ...663...81P} is an addition compared to the work of \citet{2009ApJ...690.1250S} and \cite{2017ApJ...850...66A}.

We checked the quality of the photometric redshifts using the 541 reference galaxies (see the green circles in Figure \ref{fig:SCDS} right). The fraction of catastrophic outliers with $|\Delta$\textit{z}|/(1+\textit{z})$\ge$0.15 is only 4.6\%, and the r.m.s. error of all galaxy photometric redshifts is <|$\Delta$\textit{z}|/(1+\textit{z})>=0.033. The accuracy of the galaxy photometric redshifts is thus quite comparable to other surveys using broad-band photometry, but somewhat worse than e.g. those in the COSMOS field \citep{2009ApJ...690.1236I,2009ApJ...690.1250S,2016ApJS..224...24L}, mainly because COSMOS has a large number of intermediate band filters and much deeper imaging. We also compared the photometric and spectroscopic redshifts for the reference sample of 263 AGN in the field (black dots in Figure \ref{fig:SCDS} right). The fraction of catastrophic outliers is about 22\%, and these are dominated by relatively bright type-1 AGN. The r.m.s. error of all AGN photometric redshifts is <|$\Delta$\textit{z}|/(1+\textit{z})>=0.06. In about 4\% of all cases there is a secondary maximum in the photometric redshift probability distribution fitting better to the spectroscopic redshift. This quality is very similar to the results obtained by \citet{2017ApJ...850...66A} for a sample of similar quality. Broad-band photometric redshifts for type-1 AGN are notoriously difficult for several reasons: their SEDs are practically power laws with superposed emission lines, which are not prominent in broad photometric bands. Time variability or photometric errors can cause spurious spectral features, which the SED fit clings on to. However, despite the potentially erroneous photometric redshifts, the classification as AGN is rather unique. Therefore our photometric redshifts are more than sufficient for a crude optical identification and source classification of the X--ray counterparts.

\subsection{Optical Identifications}

\label{sec:optids}
The first step towards the optical identification of X--ray sources is the astrometric correction, which was already applied as part of the data preparation in section \ref{sec:prep}. Therefore the final output catalogue does not need further astrometric corrections. Optical identification is an iterative procedure, which in case of relatively large error circles with multiple possible counterparts has a significant statistical uncertainty. The results therefore contain a probabilistic element, which is addressed below. The availability of the optical identification catalogue of H06, as well as the existence of a number of additional spectroscopic redshifts from the literature, are important prerequisites for the identification procedure. The NEP field is at a comparatively low Galactic latitude (\textit{b}$\approx$28$\degree$) and therefore contains a large number of stars, many of which may also be faint X--ray sources. The optical position accuracy for bright stars is reduced, partially because they are often saturated in the deep CCD images. Wherever possible, we therefore make use of the \textit{Gaia} DR2 catalogue \citep{2018A&A...616A...1G}. 

But arguably the most important element of reliable identifications is the existence of the mid-infrared catalogues from \textit{WISE} and \textit{Spitzer}. As we will show below, the identification procedure for the 805 X--ray catalogue sources yields 766 high-likelihood optical candidates (identification quality IQ=2), while in 39 cases there is an ambiguity with several possible counterparts (IQ=1). There is another 74 possibly interesting objects in the error circles of high-likelihood counterparts, which are noted in the optical ID catalogue with IQ=0. The following figures show only the 766 high-likelihood (IQ=2) counterparts.

\begin{figure}[htbp]
    \includegraphics[width=.50\textwidth]{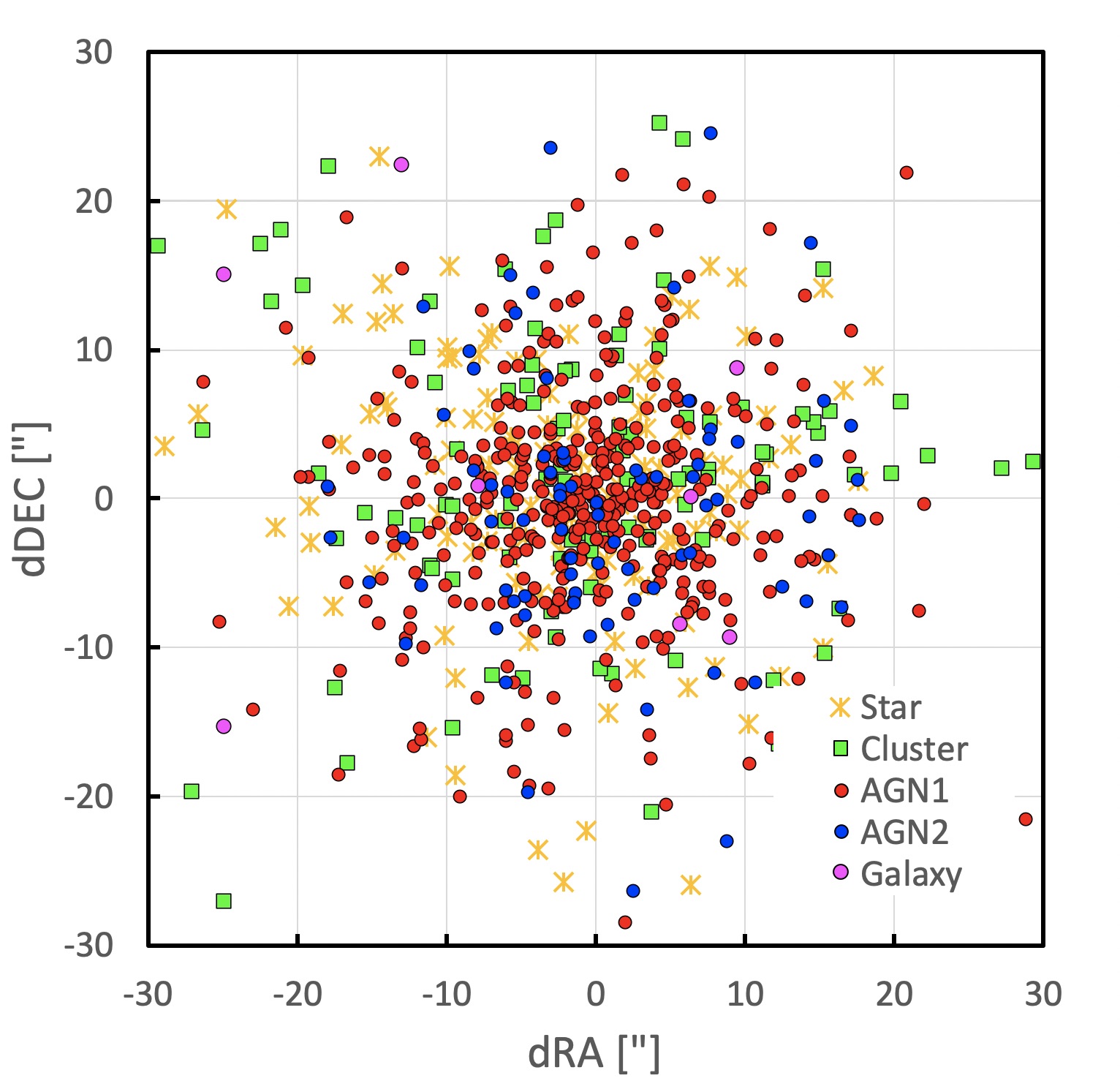}
    \caption{Separation between X--ray and optical/NIR counterparts in arcseconds of Right Ascension and Declination. The symbols are the same as in Figure \ref{fig:solid_extent} (right).}
    \label{fig:offset}
\end{figure}

\begin{figure*}[htbp]
    \includegraphics[width=.49\textwidth]{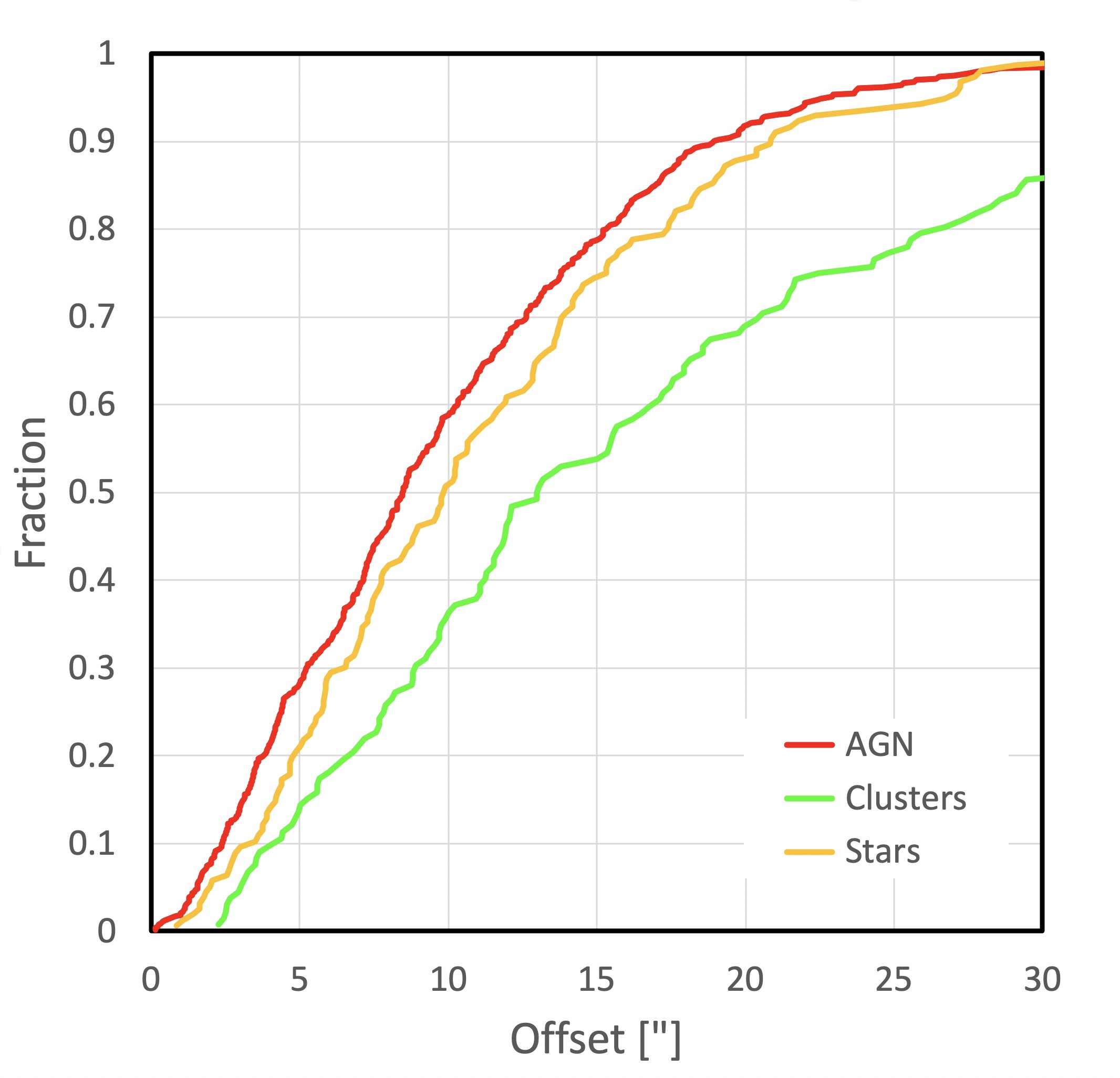}
    \includegraphics[width=.49\textwidth]{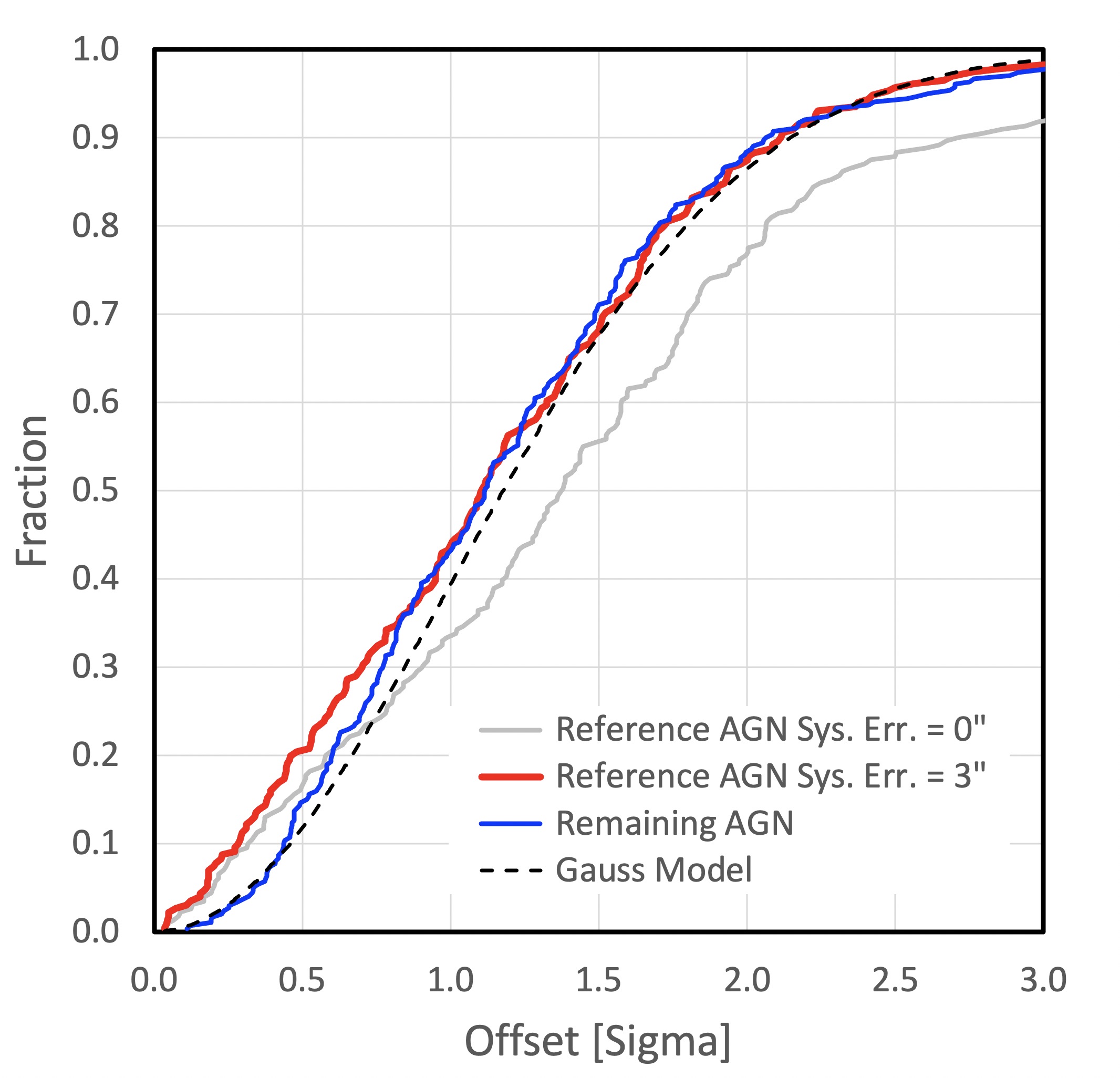}
    \caption{Left: cumulative distribution of position offsets (in arcseconds) for different classes of sources. AGN have a narrow distribution with a half-radius of 8.5\arcsec, while stars have a slightly wider distribution (9.8\arcsec), indicating some possible misidentification among $\sim$15\% of the stars. Clusters and groups have an even wider distribution due to the fact that the X--ray emission is not always centered on the brightest galaxy. Right: Cumulative distribution of position offsets in units of the 1$\sigma$ position errors compared to a Gaussian model (black dashes). The grey line shows the distribution  for 231 high-quality reference AGN (see text) before the application of a systematic position error. The red line shows the same 231 reference AGN after a systematic position error of 3\arcsec~ has been applied to each statistical error. The blue line shows the offset distribution for the remaining 301 AGN applying the same 3\arcsec~ systematic position error.}
    \label{fig:histoffset}
\end{figure*}

Figure \ref{fig:offset} shows the offsets in Right Ascension and Declination between the position of the X--ray source and that of the high-likelihood optical counterpart. Figure \ref{fig:histoffset} (left) shows the cumulative distributions of position offsets (in arcseconds) for AGN (red), stars (yellow), and galaxies, clusters and groups (green), identified below. AGN show the narrowest distribution, with a half-radius of 8.5\arcsec. Stars, on the other hand, have a somewhat wider distribution (half-radius 9.8\arcsec). This indicates that about 15\% of the stellar identifications may be wrongly associated, likely due to the large density of stars in the field. Clusters, groups and individual galaxies have an even wider distribution. This may be due to the fact that the X--ray emission is not always centered on the brightest galaxy. We therefore use only AGN to calibrate the identification procedure. The first step is the determination of possible systematic position errors in the dataset. The maximum-likelihood X--ray detection algorithm gives the statistical position error (see Table \ref{Table:XID}), which can be compared to the distribution of counterpart offsets. For this purpose we selected a reference sample of 231 high-quality AGN identifications, consisting of 141 AGN from H06, 10 other AGN with spectroscopic redshifts from the literature, and 80 high-quality AGN identifications selected from their mid-infrared colours \citep[with \textit{WISE} W1--W2>0.8, see][]{2013ApJ...772...26A}. First we calculate the cumulative distribution of position offsets in units of the 1$\sigma$ statistical position errors, which is shown as a grey line in Figure \ref{fig:histoffset} (right). This is significantly wider than the Gaussian model expectation, shown as black dashed curve. We then iteratively applied a systematic position error in quadrature to the statistical errors, until we found a reasonable match with the Gaussian expectation at a systematic error value of 3\arcsec. The corresponding cumulative distribution for our reference AGN sample is shown as the red curve. We then applied the same systematic error to the remaining 301 AGN identifications in the sample. Their cumulative position offsets are shown as blue line in Figure \ref{fig:histoffset} (right). We also tested the normalized offset distributions separately for AGN1 and AGN2 for both the reference sample and the remaining AGN and did not find significant differences. The fact that all normalized offset distributions almost perfectly match the Gaussian expectation, both for the reference sample, which is typically from brighter X--ray objects, and the fainter remaining AGN, confirms the accuracy of the maximum likelihood errors as well as the systematic errors.  

\begin{figure*}[htbp]
    \includegraphics[width=.49\textwidth]{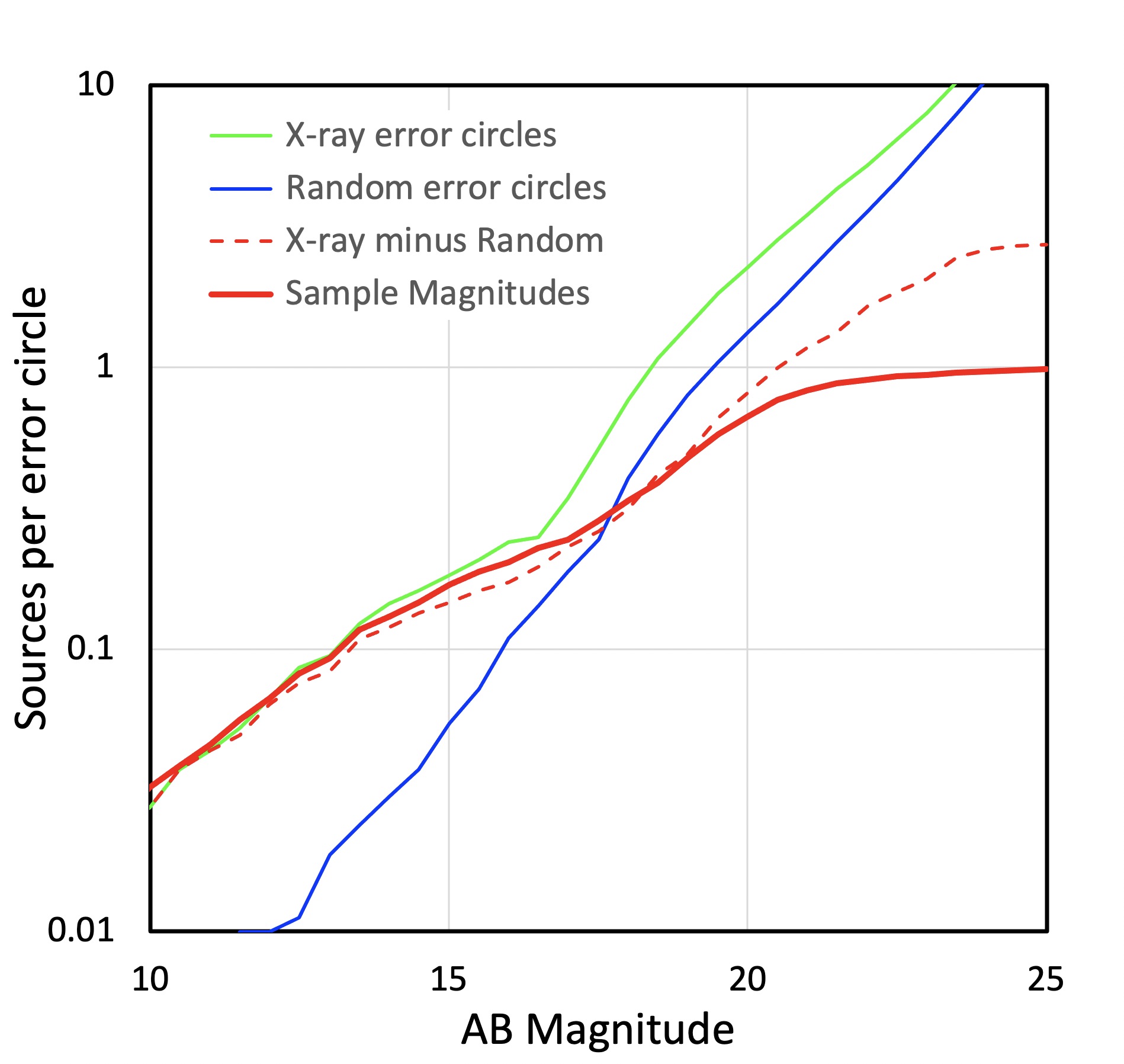}
    \includegraphics[width=.49\textwidth]{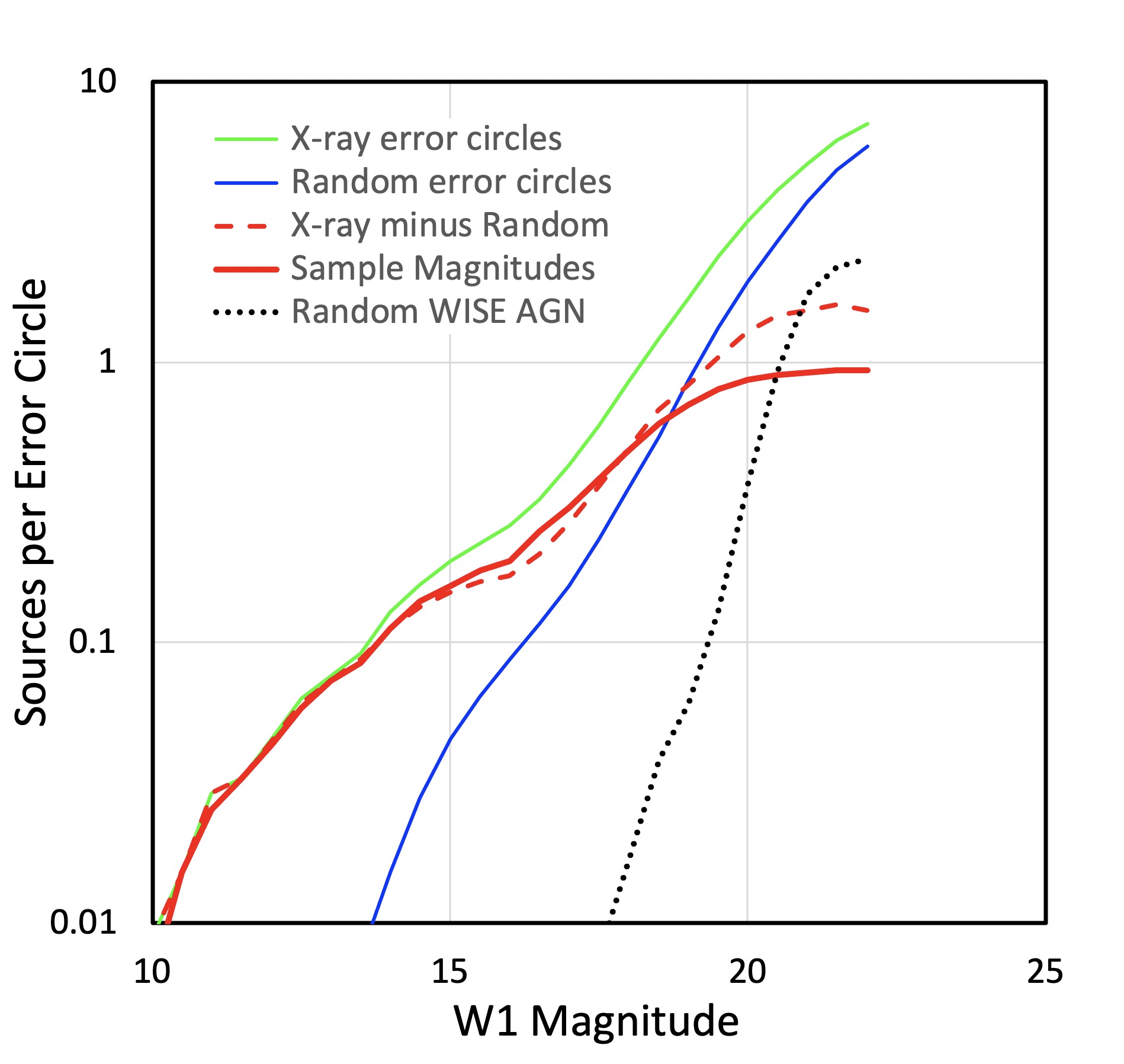}
    \caption{Cumulative magnitude distributions in circles with 2.5$\sigma$ error radii around the 805 X--ray sources (green line), and the same number of randomly chosen positions (blue line). The dashed red line shows the difference between source and random circles, while the thick red line shows the actual magnitude distribution of the X--ray counterparts. Left:\textit{Subaru} HSC \textit{i}-band. Right: \textit{CatWISE2020} W1 band. The dotted black line shows the cumulative distribution of random mid-infrared selected AGN candidates with W1-W2>0.8.}
    \label{fig:histmag}
\end{figure*}

We now can correlate the whole sample both with the optical (\textit{HEROES}) and the mid-infrared (\textit{WISE}, \textit{Spitzer}) catalogues to look at the number and magnitude distribution of the expected counterparts in the X--ray error circles -- both the real counterparts and random field associations. We use a correlation radius of 2.5$\sigma$ around each of the 805 X--ray sources to obtain the cumulative \textit{i}$_{AB}$ and W1 magnitude distributions of objects, respectively, shown by the green lines in Figure \ref{fig:histmag}. The blue lines show the magnitude distribution of field objects within 805 randomly chosen circles of the same 2.5$\sigma$ radius. The dashed red lines show the difference between X--ray error circles and field circles, i.e. the expected cumulative magnitude distribution for all counterparts associated with the X--ray sources. This allows for the possibility to have more than one physical association per error circle, e.g. pairs, groups or clusters of galaxies, merging AGN, or star clusters. In making this subtraction one has to take care of the effect discussed in \citet{2007ApJS..172..353B} and \citet{2013ApJS..209...30N}, namely that the magnitude distribution of field galaxies around bright objects is significantly shallower than around faint objects, leading to negative values in the subtraction. The dashed red lines have therefore been calculated piece-wise in different magnitude intervals before constructing the cumulative distribution. The solid red lines show the actual magnitude distribution of the selected "best" optical counterparts in Table \ref{Table:ID}. For magnitudes <19 the dashed and solid red curves are quite close, indicating that the large majority of the selected "best" counterparts should be correct. At fainter magnitudes the dashed red curves increases above the solid red curve, up to values of 2.7 and 1.5 at \textit{i}$_{AB}$=25 and W1=22, respectively.  This is likely due to some true physical associations, e.g. interacting pairs of galaxies or cluster/group galaxies appearing in the same X--ray error circle. 

\begin{figure*}[htbp]
    \includegraphics[width=\textwidth]{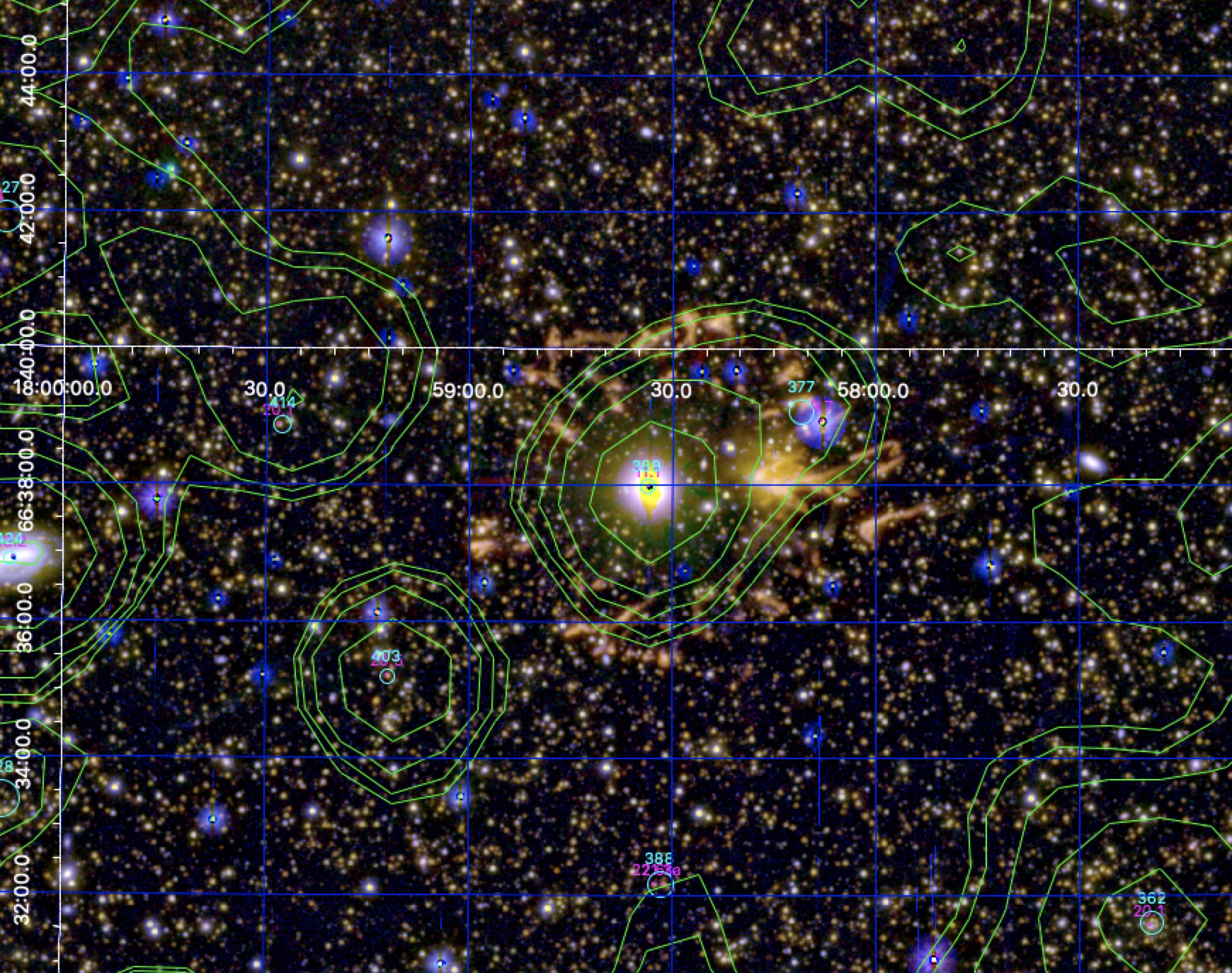}
    \caption{A 18\arcmin$\times$15\arcmin~ cutout of the center of the \textit{HEROES} field. ROSAT X--ray contours (green) are superposed on a false-colour optical/mid-infrared image with the HSC \textit{i}-band in blue and the \textit{Spitzer} I1 and I2 bands in green and red, respectively. Maximum-likelihood X--ray error circles are shown in cyan with the radius of 2$\sigma$. Optical counterpart magnitudes are indicated in magenta.}
    \label{fig:findingPN}
\end{figure*}

At magnitudes \textit{i}$_{AB}$ and W1>19.5 the expected number of field objects per error circle increases above one, reaching values of $\sim$20 and $\sim$6 at \textit{i}$_{AB}$=25 and W=22, respectively. In this situation the optical identification purely by positional association becomes meaningless. We therefore have to introduce additional information and prior expectations into the identification process. As discussed above, this is naturally a probabilistic approach and does no longer yield unique identifications. In principle there is the elaborate Bayesian multi-catalogue matching tool NWAY\footnote{https://github.com/JohannesBuchner/nway}~\citep{2018MNRAS.473.4937S}, where several input catalogs can be cross-matched and a number of priors can be introduced, which also calculates the likelihood for every possible positional coincidence. However, in the case of incomplete catalogue information, and in the presence of significant systematic errors (e.g. non-astronomical false positive detections in the optical catalog, or the presence of source confusion both in the X--ray catalogue and in the \textit{CatWISE2020} catalogue, it is very cumbersome to construct the appropriate prior for this  method. 

\begin{figure*}[htbp]
    \includegraphics[width=.49\textwidth]{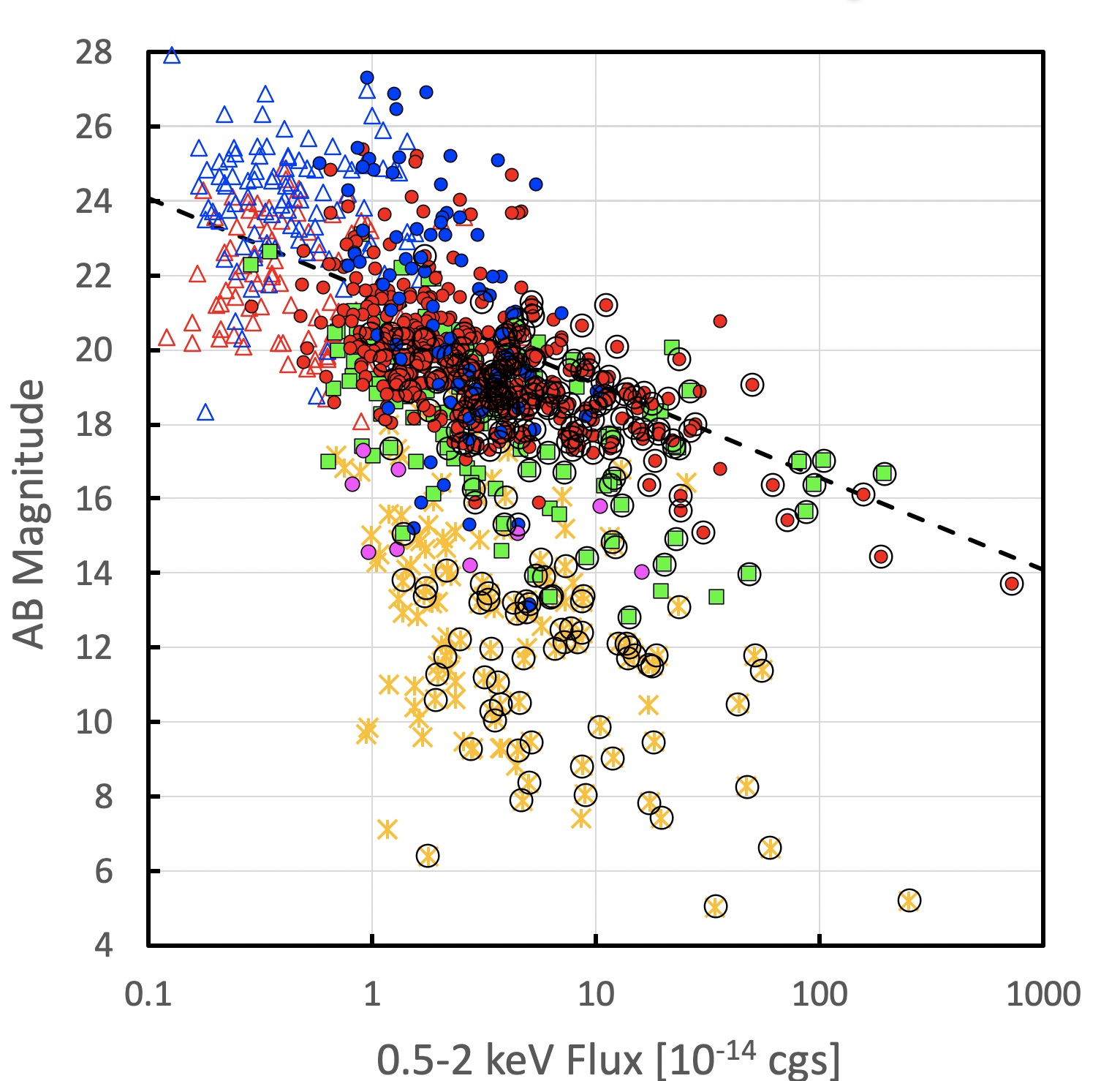}
    \includegraphics[width=.49\textwidth]{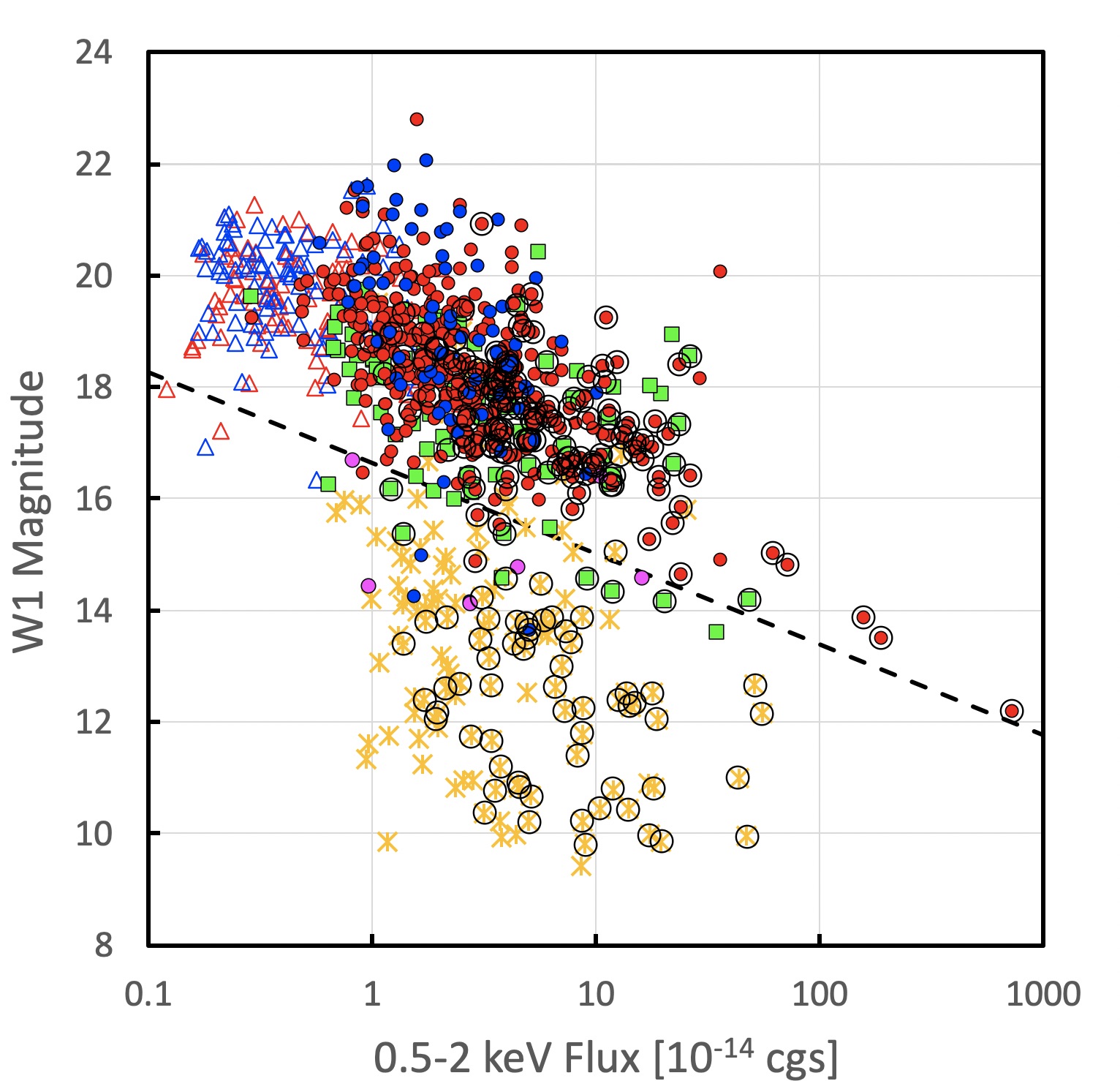}
    \caption{Left: AB magnitude versus X--ray flux. The symbols are the same as in Figure \ref{fig:solid_extent} (right), but in addition we show the X--ray upper limit fluxes for a sample of AGN candidates selected from mid-infrared colours (red triangles for AGN1 candidates and blue triangles for AGN2 candidates). The dashed line indicates an X--ray to optical flux ratio of $f_{XO}$=1. Right: The same as the left figure, but with the \textit{CatWISE2020} W1 magnitude on the x-axis. In this case the dashed line shows the empirical discrimination between AGN and other objects defined in \citet{2018MNRAS.473.4937S}.}
    \label{fig:fxabw1}
\end{figure*}

Figure \ref{fig:findingPN} gives a visual impression of the difficulty of optical identification in the complicated situation of faint optical counterparts with relatively large error circles and possibly confused settings. ROSAT X--ray contours are superimposed in a logarithmic scale on a false-colour image with the \textit{Spitzer} bands I2 (red) and I1 (green) and the \textit{HEROES} \textit{i}-band image (blue). This 18\arcmin$\times$15\arcmin~ image is centered on the planetary nebula NGC6543 close to the center of the \textit{HEROES} field (see also Figure \ref{fig:PN}). The maximum-likelihood X--ray positions are shown as cyan circles with 2$\sigma$ error radii, and the final optical counterparts are indicated with magnitudes in magenta. The planetary nebula (XID \#389) is clearly detected in X--rays, but there is another X--ray (XID \#377) source close to and confused with it. This source has two possible stellar counterparts, but the \textit{XMM-Newton} image clearly selects the brighter star. The faint X--ray source (XID \#388) in the lower center of the image has two credible AGN counterparts. Given these difficulties, we had to resort to the tedious task of visually inspecting every individual error circle and manually selecting and characterising the most likely optical counterpart, considering a number of prior expectations. Previously published deep X--ray surveys \citep[e.g.][]{2005ARA&A..43..827B} show that the largest fraction of X--ray sources at our flux limit are AGN, followed by stars, and clusters of galaxies. This is also the case for the H06 catalogue. Stars can usually be discriminated rather easily, because at the same X--ray flux they are about 5 magnitudes brighter than both AGN and cluster galaxies (see e.g. Figure \ref{fig:fxabw1}). However, given the rather large density of stars in our field, there is the possibility for misidentification. Brighter and lower-redshift clusters of galaxies can often be readily identified through their extended X--ray emission or the concentration of bright galaxies associated with the X--ray source. However, there are also fainter and higher-redshift cluster candidates without significant X--ray extent. For the objects with unclear identifications we first searched for a possible cluster or group by looking for photometric redshift concentrations in a circle with radius of 30--120\arcsec around the X--ray source, and indeed could identify a number of photometric cluster candidates this way. Higher-redshift (\textit{z}>0.8) clusters, where the optical magnitudes of even the brighter cluster members are very faint and thus do not stand out against the field galaxies, are easier to identify in the mid-infrared images. 

AGN with optical counterpart magnitudes \textit{i}$_{AB}$<19--20 are typically the brightest and often point-like objects in their X--ray error circle, and thus easy to identify. Problems arise, when the optical counterparts are fainter than \textit{i}$_{AB}$>20. Then the likelihood to have an unassociated field object with a magnitude brighter than the actual X--ray counterpart in the error circle increases substantially. Also, at fainter X--ray fluxes and optical magnitudes the fraction of unobscured type-1 AGN (AGN1 hereafter) decreases, and absorbed/obscured type-2 AGN (AGN2 hereafter), which are harder to discriminate from normal galaxies, become more abundant. In this situation mid-infrared imaging becomes crucial. In general, both AGN1 and AGN2 are brighter in the mid-infrared channels than normal galaxies, and very often the X--ray counterpart is the brightest \textit{WISE} or \textit{Spitzer} source in the error circle. Most AGN can be readily identified through their peculiar mid-infrared colours \citep[see e.g.][]{2012ApJ...753...30S,2013ApJ...772...26A}. We used the classical criterion W1--W2>0.8 to identify mid-infrared AGN candidates. The dotted black line in Figure \ref{fig:histmag} (right) shows the randomly expected number of mid-infrared selected AGN per X--ray error circle, which is below 1 for the relevant magnitude range, thus confirming the reliability of this selection. \citet{2013ApJ...772...26A} show that this selection contains some interlopers from normal galaxies with redshifts \textit{z}>1.5, which could in principle be discriminated using the \textit{Spitzer} [5.8]--[8.0] colours. Since we do not have longer wavelength \textit{WISE} or \textit{Spitzer} photometry of the \textit{HEROES} field, we resorted to looking at the spectral energy distribution of the photometric redshift fits and gave priority to candidates containing a significant AGN contribution in their model SED. Finally, at faint magnitudes (e.g. W1>20) even the mid-infrared colour selection runs into problems, first because the statistical errors in the detection hamper the proper colour determination, and secondly, because the \textit{WISE} data become significantly confusion limited. For the central 10 deg$^2$ part of our survey covered by \textit{Spitzer} we could, however, go a step further because these images resolve the \textit{WISE} source confusion and also go about a magnitude deeper. In a handful of cases we even found \textit{Spitzer} sources in the center of otherwise empty X--ray error circles, so called "infrared dropouts". We manually determined limiting optical magnitudes and detections at the corresponding positions and could determine photometric redshifts in the range 1<\textit{z}<6 for these objects.

For each of the potential X--ray counterparts we determined photometric redshifts as described in subsection \ref{subsec:photoz}. We visually inspected each photometric redshift fit and manually clipped outliers. For missing bands we manually determined the magnitude values or upper limits. For AGN, apart from the photometric redshift, we also determined a coarse characterization based on the best-fit model SED. Models with a clear type-1 broad line contribution to the SED (>10\% in case of hybrid galaxy/AGN models) with intrinsic extinction E(B-V)<0.2 were characterized as AGN1, SED models with type-2 character (Seyfert-2, QSO-2, star forming galaxies) and/or intrinsic extinction E(B-V)$\ge$0.2 were characterized as AGN2. Pure galaxy SED models with intrinsic extinction E(B-V)<0.2 were characterized as galaxies, if the corresponding X--ray luminosity was below log(L$_X$/(erg s$^{-1}$))<42, and as AGN1 for larger luminosities. It is clear, that this  characterization is only indicative and does not replace a true spectroscopic identification for individual objects. However, sample properties of object classes can still be assessed. 

In finally assessing an identification quality (IQ) for each possible counterpart, we also took into account the position information for 477 X--ray sources from the literature ( 431 matches with 2RXS, 83 with 4XMM-DR9, and 40 with XMMSL2; see subsection \ref{subsec:IDcatalog} above). The 2RXS catalogue gave better identification accuracy for 7 of the 431 common objects (1.6\%), while \textit{XMM-Newton} gave unique new positions for 10 out of 123 common sources (8.1\%). Overall, our identification quality can therefore be regarded a quite reliable. An identification quality of IQ=2 means a high likelihood single optical counterpart. There are 766 X--ray sources in this category. IQ=0 means a rejection of this counterpart in the presence of a better counterpart. 74 sources are in this category, i.e. about 10\% of the IQ=2 counterparts contain one or more less likely candidates, mainly unrelated galaxies or stars. An identification quality of IQ=1 means several possible counterparts without a clear priority ranking. There are 39 X--ray sources in this category (i.e. $\sim$5\% of the sample), with a total of 80 possible counterparts, which are dominated by AGN candidates. In this category there are also a number of possible dual AGN with two AGN candidates at similar redshift within the X--ray error circle. Table \ref{Table:IDSummary} shows a summary of classifications in the three quality classes. We compare this with the equivalent samples of soft X--ray selected objects in the COSMOS field. Table \ref{Table:ID} gives the catalog of all possible optical counterparts for the X--ray sources, including their identification quality.

\begin{table}
\caption{Summary of optical identifications}
\label{Table:IDSummary}
\centering
\begin{tabular}{lrrrr}
\hline\hline
Type            & IQ=2 & IQ=1 & IQ=0 & COSMOS\\
\hline
\\
AGN1            &  412  &  42   &    & 1118\\   
AGN2            &   79  &  25   &    & 1492\\ 
Galaxy          &    9  &   1   & 23 &  114\\
Cluster/Group   &  121  &   4   &  4 &   70\\
Star            &  145  &   8   & 47 &  113\\
\\
Total           &  766  &  80   & 74 & 2907\\
\hline
\end{tabular}
\end{table}

\setlength{\tabcolsep}{5.5pt}

\begin{table*}
\caption{ \textit{ROSAT} NEP Raster Candidate Optical ID Catalogue}
\vskip -0.5truecm
\label{Table:ID}
\begin{center}
\begin{tabular}{rllrrrrrrllll}
\hline\hline
(1) & (13)   & (14)    &(15)&(16)& (17)     &(18)&(19)&(20)&(21)&(22)&(23)&(24)\\ 
No & RA$_O$ & DEC$_O$ & AB & f$_{XO}$ & IR& I12 &ID & z & Qual & logL &IQ&Comment\\
\hline
1a	&	261.4750009	&	68.1612177	&	19.8i	&	0.03	&	19.4W	&	1.11	&	1	&	1.39	&	ph	&	44.8	&	2	&	AGN	\\
1b	&	261.465965	&	68.157951	&	13.4P	&	-2.56	&		&		&	5	&		&	ph	&		&	0	&	star	\\
2	&	261.768928	&	69.4465842	&	21.2P	&	0.89	&	19.2W	&	0.45	&	1	&	1.21	&	ph	&	44.9	&	2	&	BL 1700	\\
3	&	261.9797816	&	67.8124949	&	18.4i	&	-0.31	&	17.8W	&	1.22	&	1	&	1.33	&	ph	&	44.9	&	2	&	AGN	\\
4	&	262.1450026	&	67.5404091	&	19.4i	&	0.03	&	17.6W	&	0.76	&	1	&	0.649	&	NED	&	44.1	&	2	&	AGN1 1770	\\
5	&	262.414355	&	68.7944358	&	9.0G	&	-3.94	&	10.8W	&	0.01	&	5	&		&	NED	&		&	2	&	STAR G 1800	\\
6	&	262.4861218	&	66.8642273	&	13.9G	&	-2.28	&	13.6W	&	0.04	&	5	&		&	ph	&		&	2	&	star	\\
7	&	262.5535323	&	65.7389400	&	25.1i	&	1.98	&	21.0W	&	1.00	&	2	&	1.84	&	ph	&	44.9	&	2	&	AGN	\\
8	&	262.588411	&	68.1700626	&	19.0i	&	-0.30	&	17.6W	&	1.13	&	1	&	0.58	&	ph	&	43.9	&	2	&	AGN	\\
9a	&	263.0047282	&	65.3847411	&	13.8G	&	-2.41	&	14.5W	&	-0.06	&	5	&		&	ph	&		&	1	&	star	\\
9b	&	263.0120908	&	65.3914923	&	17.0i	&	-1.12	&	16.7W	&	0.20	&	4	&	0.51	&	ph	&	43.7	&	1	&	grp	\\
10	&	263.0122741	&	67.8054554	&	18.3i	&	-0.63	&	16.3W	&	1.10	&	1	&	0.34	&	ph	&	43.3	&	2	&	AGN	\\
...\\
796	&	277.4256734	&	67.8200237	&	17.0i	&	-0.56	&	17.4W	&	0.81	&	1	&	0.478	&	NED	&	44.2	&	2	&	AGN 5750	\\
797	&	277.4420624	&	64.5888202	&	11.7G	&	-3.27	&	13.3W	&	-0.02	&	5	&		&	NED	&		&	2	&	STAR 5760	\\
798	&	277.5087869	&	66.7564460	&	18.2i	&	-0.25	&	17.1W	&	0.53	&	1	&	0.289	&	NED	&	43.5	&	2	&	AGN1 5790	\\
799	&	278.0510896	&	68.5356066	&	16.4i	&	-1.01	&	16.2W	&	0.22	&	4	&	0.588	&	KCWI	&	44.2	&	2	&	CL 5920	\\
800	&	278.1256984	&	68.6144681	&	7.4G	&	-4.37	&	9.9W	&	0.76	&	5	&		&	NED	&		&	2	&	STAR G5 5950	\\
801	&	278.1476274	&	68.8015538	&	16.7z	&	0.32	&		&		&	4	&	0.205	&	KCWI	&	44.4	&	2	&	CL 5970	\\
802	&	278.1672663	&	69.0789133	&	10.5G	&	-3.21	&	10.9W	&	-0.02	&	5	&		&	ph	&		&	2	&	star	\\
803	&	278.4423412	&	69.3600977	&	20.8P	&	1.24	&	20.1W	&	0.67	&	1	&	4.33	&	ph	&	46.8	&	2	&	AGN	\\
804	&	278.6416759	&	69.5293133	&	11.4G	&	-2.33	&	12.1W	&	-0.02	&	5	&		&	NED	&		&	2	&	STAR 6051	\\
805	&	278.7231695	&	69.4008168	&	13.5i	&	-1.94	&		&		&	4	&		&	ph	&		&	2	&	grp	\\
\hline
\end{tabular}
\end{center}
\vskip -0.2truecm
Column explanation: (1) internal XID identification with letters indicating multiple possible counterparts; (13) and (14) optical source coordinates in J2000.0; (15) optical AB magnitude and corresponding source (G=Gaia, P=Pan-STARRS, S=SDSS, \textit{i},\textit{r},\textit{z}=HSC); (16) X--ray to optical flux ratio $f_{XO}$; (17) IR magnitude and corresponding source (I=IRAC I1, W=Wise W1), (18) IR colour W1-W2/I1-I2; (19) optical ID (1=AGN1, 2=AGN2, 3=Galaxy, 4=cluster/group, 5=star); (20) redshift; (21) redshift quality (NED,Keck,WIYN=spectroscopic, ph=photometric(; (22) observed 0.5--2 keV luminosity log(L$_X$) [erg s$^{-1}$]; (23) identification quality IQ; (24) comment.
\end{table*}

\subsection{The \textit{ROSAT} NEP Optical Identifications Catalogue}
\label{subsec:IDcatalog}

Table \ref{Table:ID} gives the final catalogue of optical identifications for the 805 X--ray sources in Table \ref{Table:XID}, again in abbreviated form. (The complete catalogue is available in the online publication.) For convenience we number the columns consecutively with Table \ref{Table:XID}. The majority of the bright star magnitudes is from the \textit{Gaia} DR2 catalogue \citep{2018A&A...616A...1G}, while most of the galaxy magnitudes are from the \textit{Subaru} HSC \textit{i}-band (or \textit{z}-band), but there are small additions from \textit{Pan-STARRS}, and \textit{SDSS}. Most spectroscopic redshifts are from the literature, the majority from H06, but others from the NASA Extragalactic Database (NED). For a small number of objects specific spectroscopic observing runs have been performed with the \textit{Keck} DEIMOS and KCWI instruments, as well as HYDRA on the \textit{WYIN} telescope; details for these will be published elsewhere. The photometric redshifts are from the \textit{LePhare} analysis described above. Column (21) gives the logarithm of the observed 0.5--2 keV luminosity (i.e. no rest-frame correction). Column (22) gives a comment for most sources. Objects from H06 are identified with their catalog numbers in the comment column.

\begin{figure*}[htbp]
    \includegraphics[width=.49\textwidth]{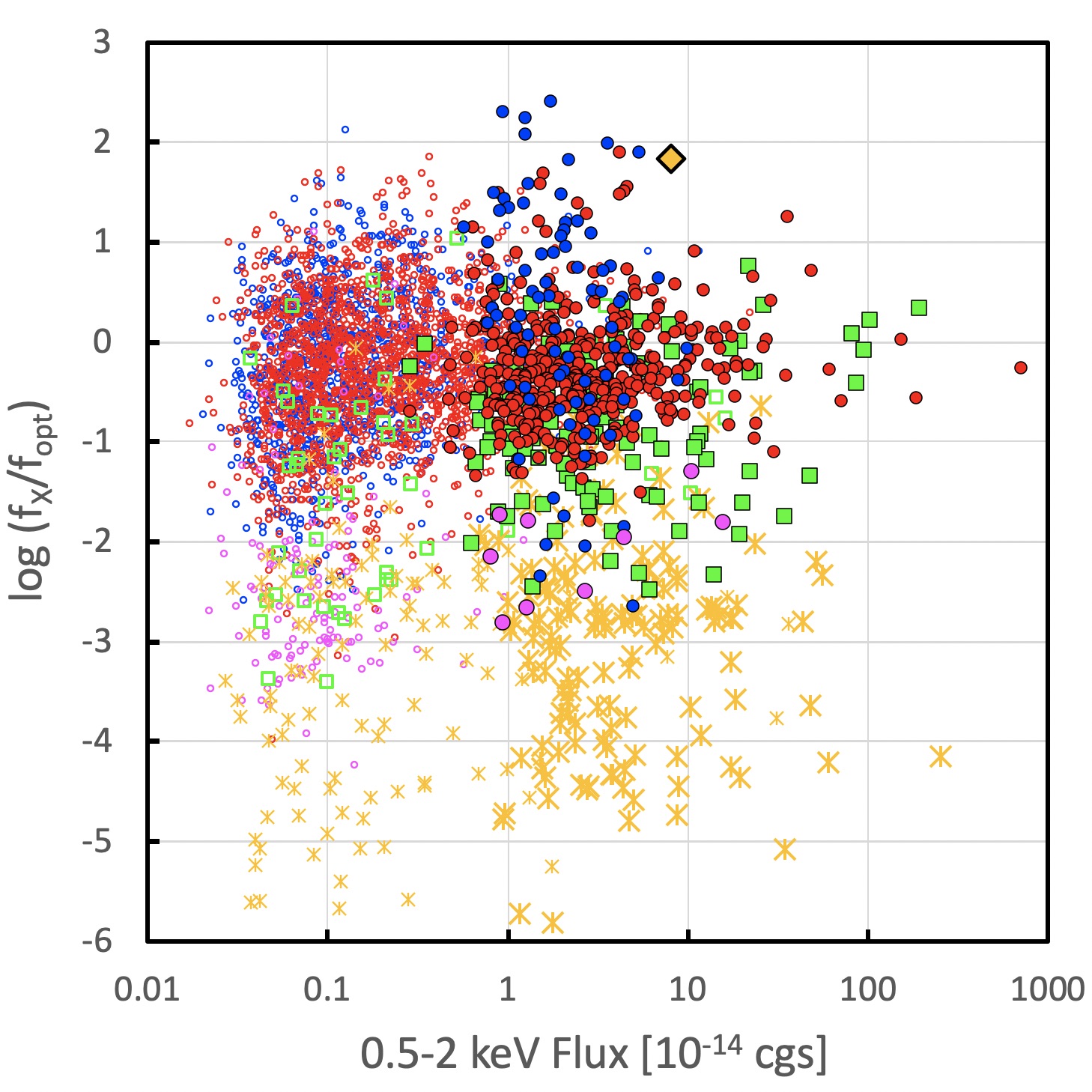}
    \includegraphics[width=.49\textwidth]{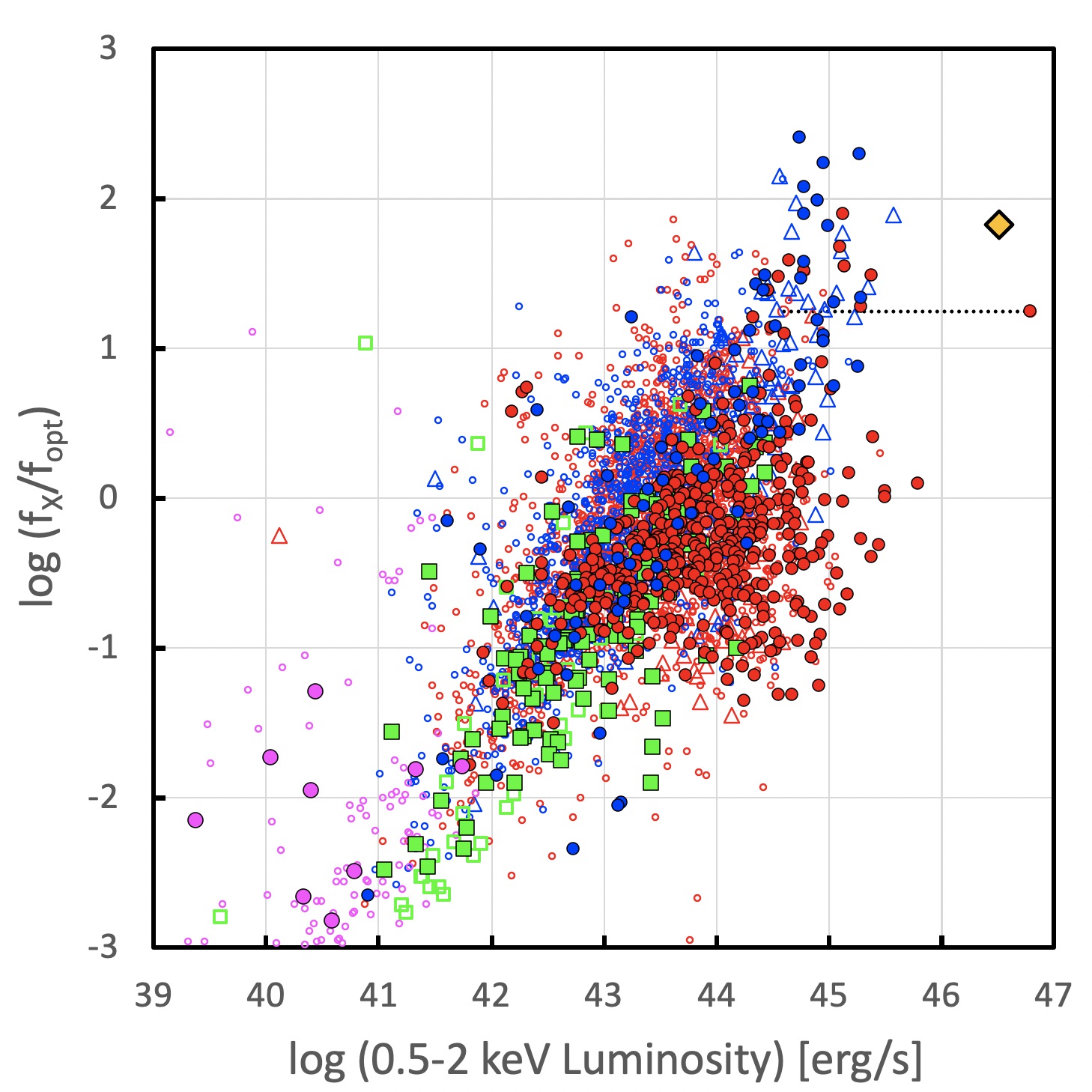}
    \caption{Left: X--ray to optical flux ratio $f_{XO}$ versus X--ray flux. In addition to the \textit{ROSAT} NEP raster scan catalogue objects (colours as in Figure \ref{fig:offset} without the H06 objects indicated), at fainter fluxes the objects from the largely spectroscopically identified \textit{Chandra} legacy survey of the COSMOS field \citep{2016ApJ...817...34M} have been added with small open symbols and the same colours. The large orange diamond shows the position of the highest-redshift radio loud quasar CFHQS J142952+544717 recently identified in the eROSITA all-sky survey \citep{2020MNRAS.497.1842M}. Right: X--ray to optical flux ratio versus X--ray luminosity. Same symbols as the left figure, but the upper limits from the mid-infrared AGN candidates are added as red and blue triangles. The dotted line refers to the highest-luminosity object XID803, which has a second photometric redshift solution at lower luminosity.}
    \label{fig:fxfxo}
\end{figure*}

\begin{figure*}[htbp]
    \includegraphics[width=.49\textwidth]{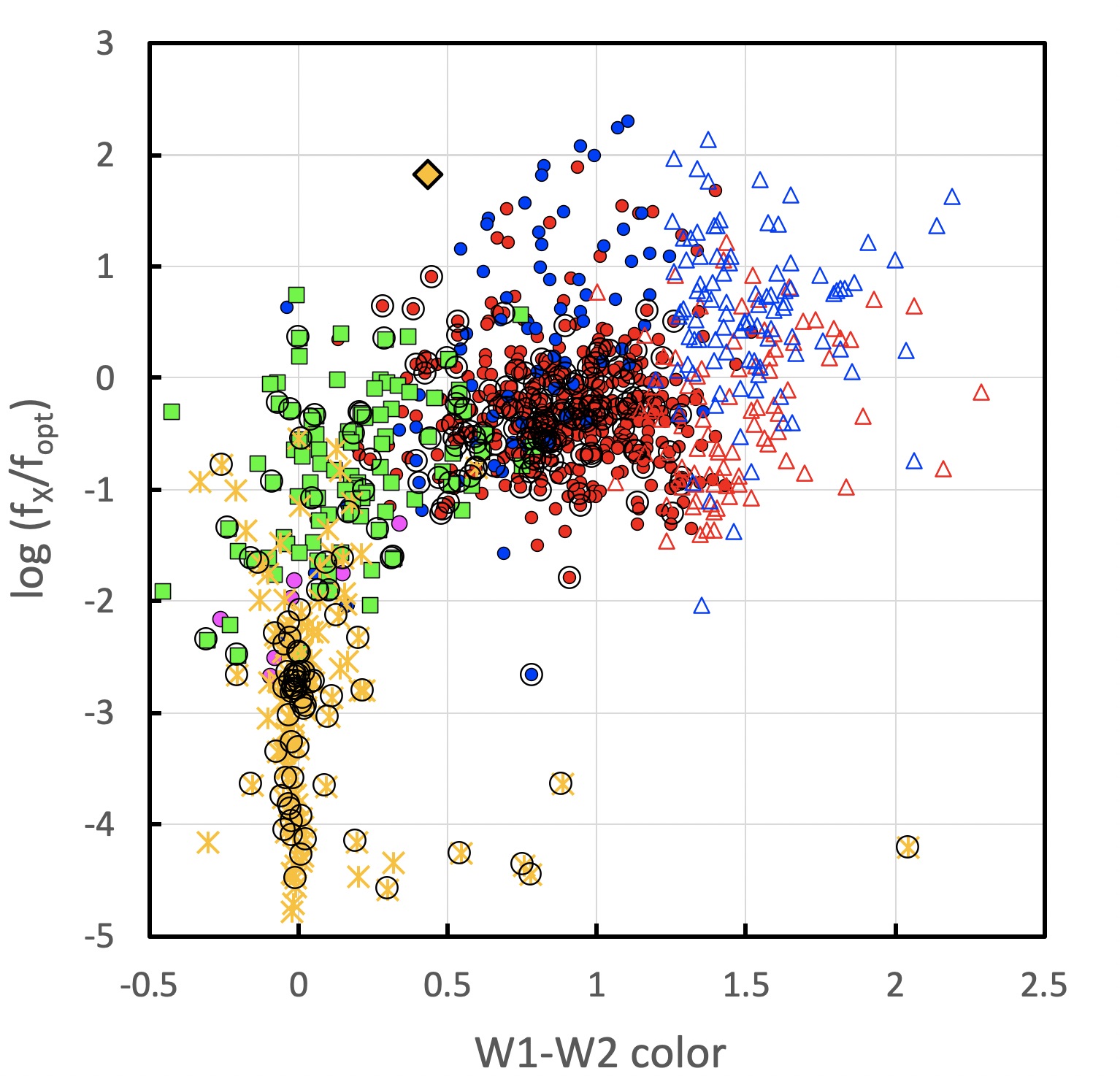}
    \includegraphics[width=.49\textwidth]{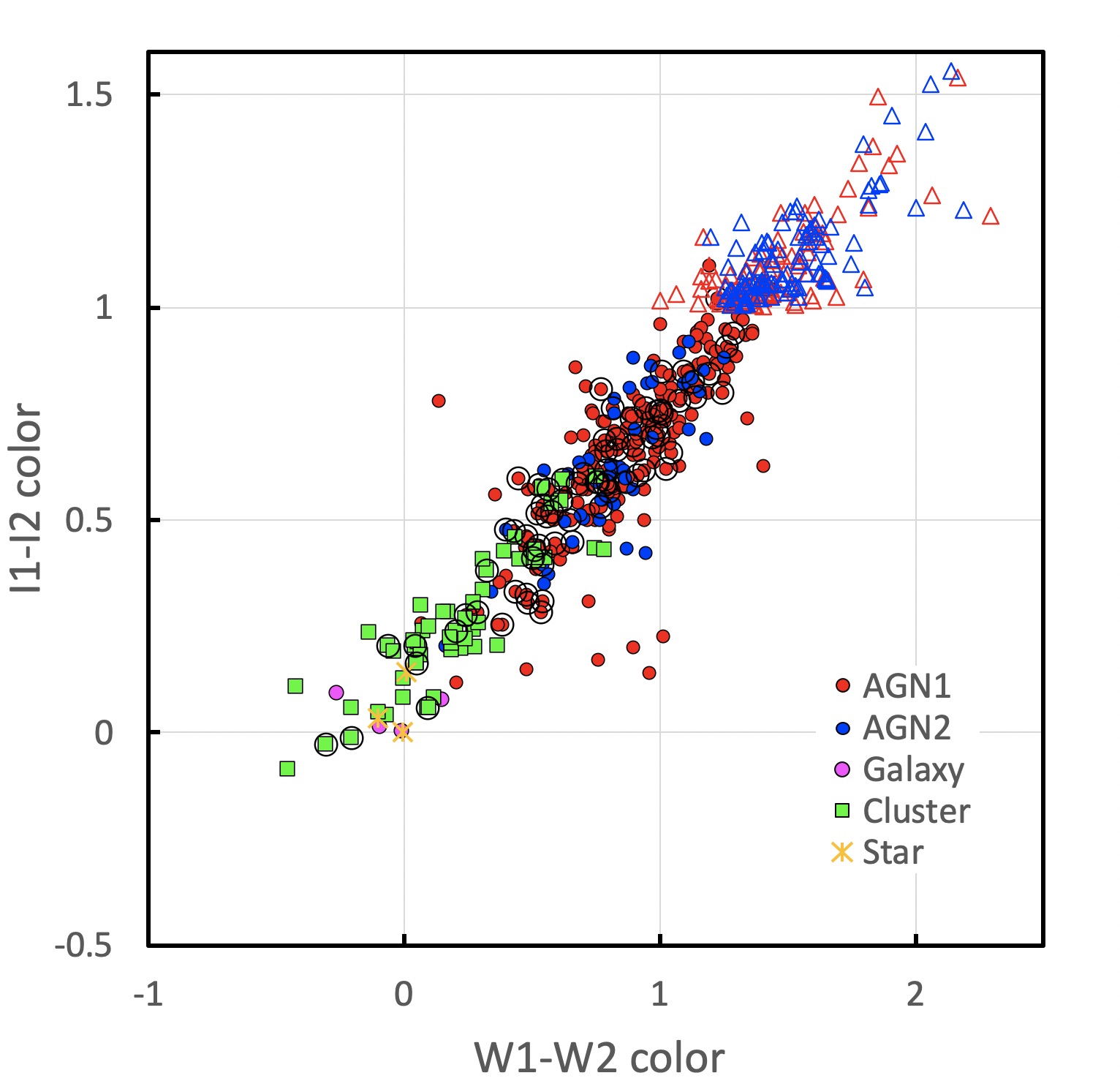}
    \caption{Left: X--ray to optical flux ratio $f_{XO}$ versus \textit{CatWISE2020} W1--W2 mid-infrared colours.  Right: \textit{Spitzer} IRAC I1--I2 versus \textit{CatWISE2020} W1--W2 mid-infrared colour. Symbols are the same as in Figure \ref{fig:fxabw1} and \ref{fig:fxfxo}.}
    \label{fig:W1fxo}
\end{figure*}

\begin{figure*}[htbp]
    \includegraphics[width=.49\textwidth]{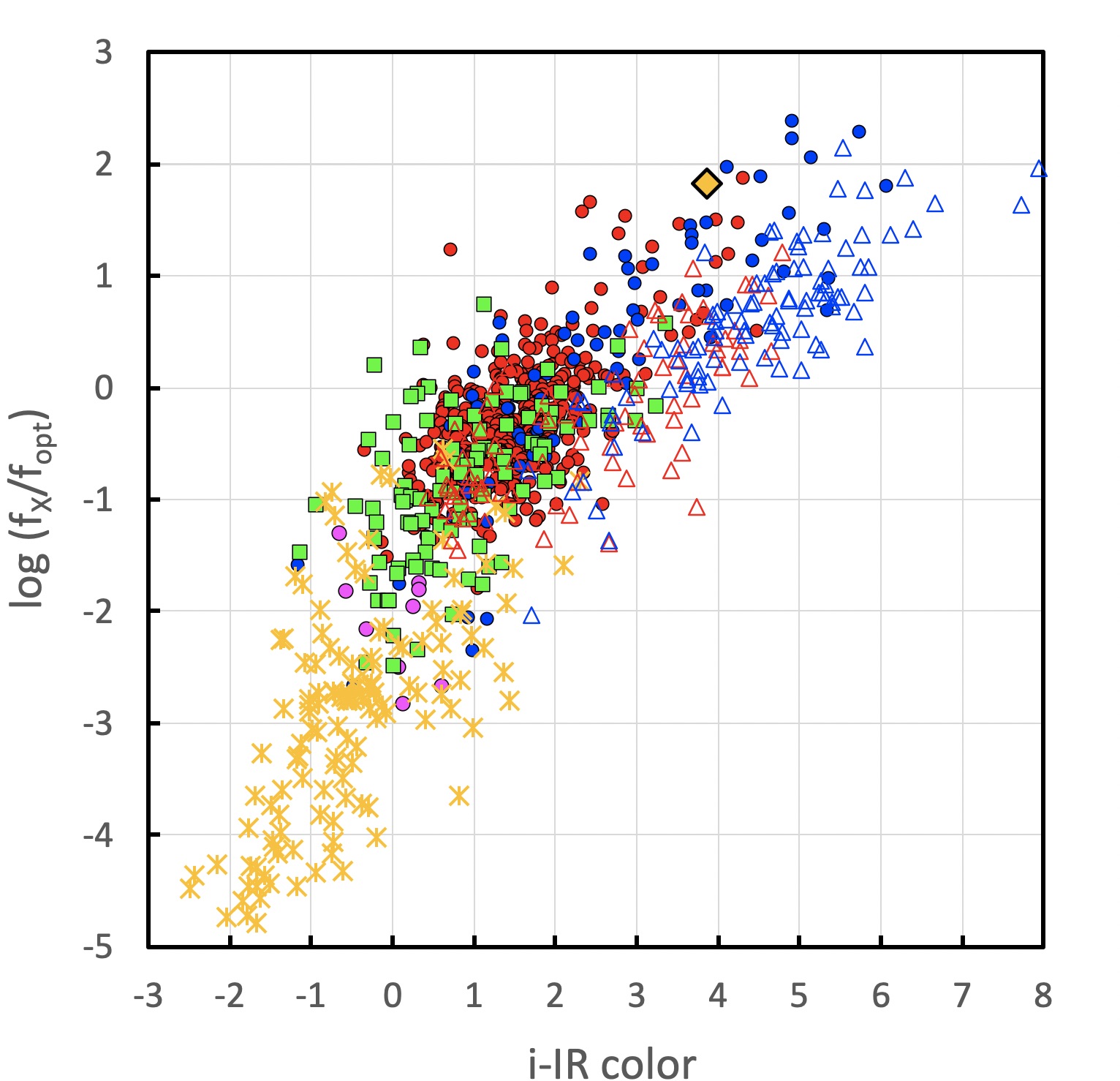}
    \includegraphics[width=.49\textwidth]{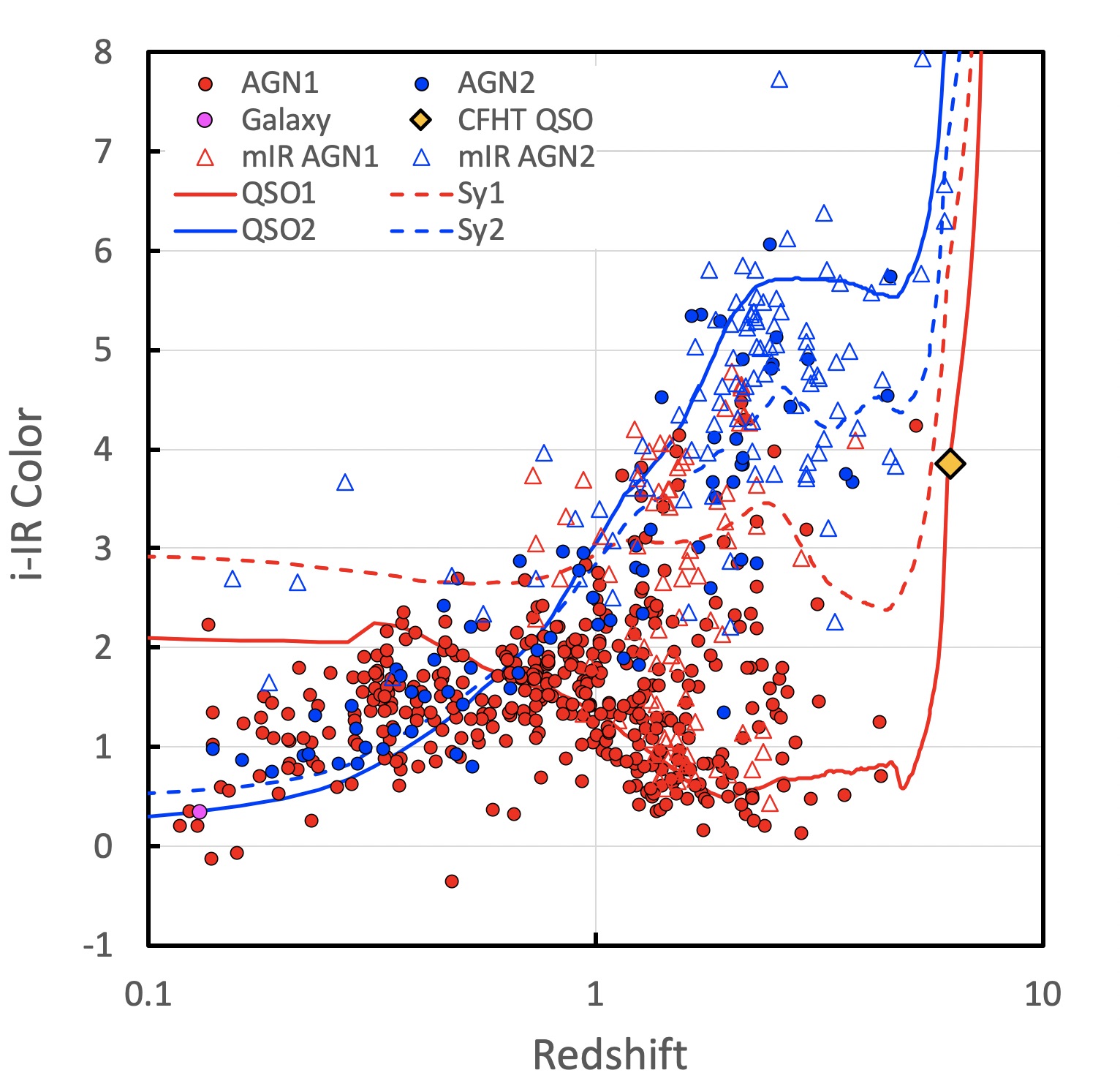}
    \caption{Left: X--ray to optical flux ratio $f_{XO}$ versus the optical--infrared colour \textit{i}--IR (symbols are the same as in Figure \ref{fig:offset}, but with the addition the of mid-infrared selected AGN candidates as in Figure \ref{fig:fxabw1}). Right: optical to infrared colour \textit{i}--IR as a function of redshift for AGN1 (red), AGN2 (blue) and galaxies (magenta). SED models for QSO1, QSO2, Sy1, and Sy2 from the photometric redshift fits are overplotted (see legend).}
    \label{fig:iIfxoz}
\end{figure*}

\begin{figure*}[htbp]
    \includegraphics[width=.49\textwidth]{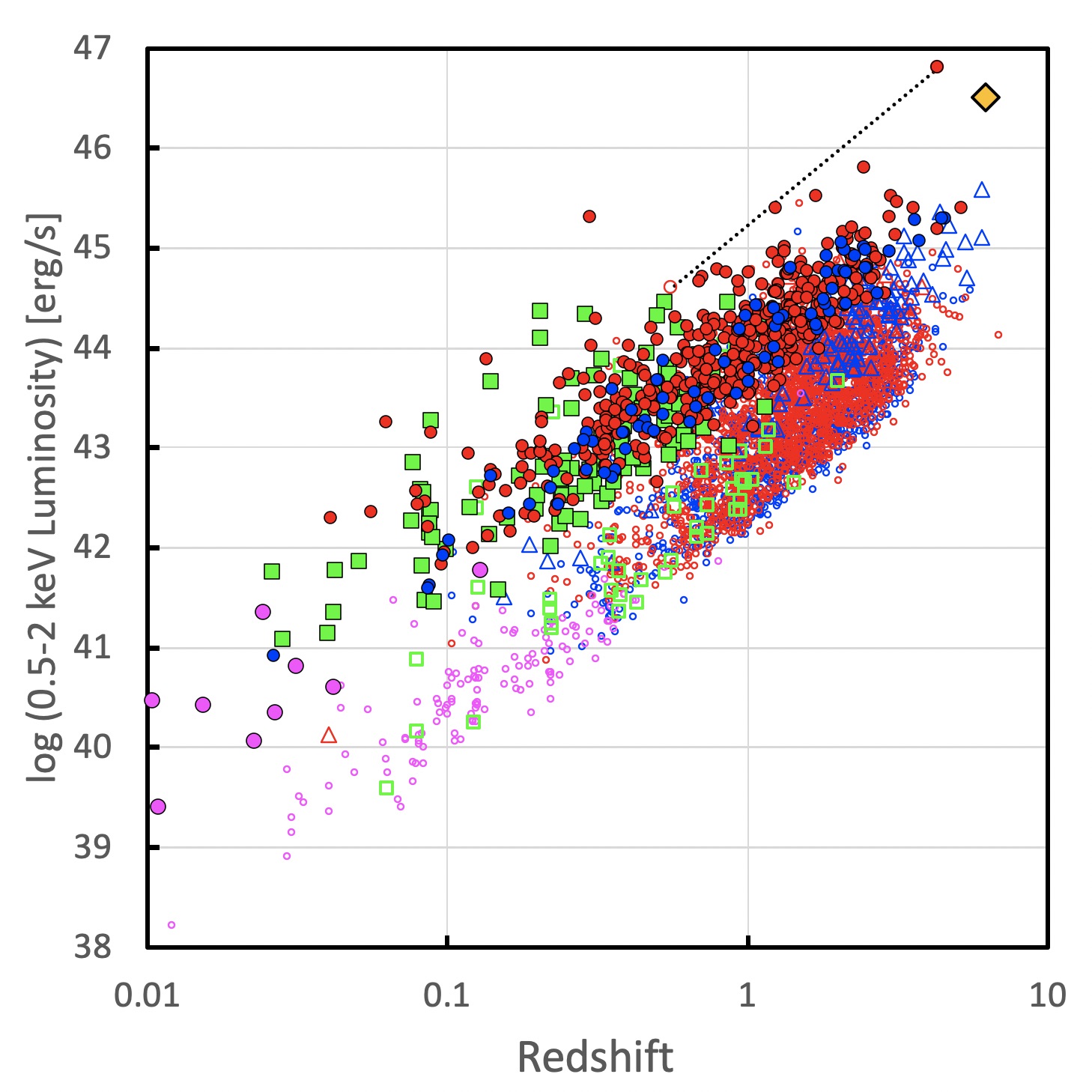}
    \includegraphics[width=.49\textwidth]{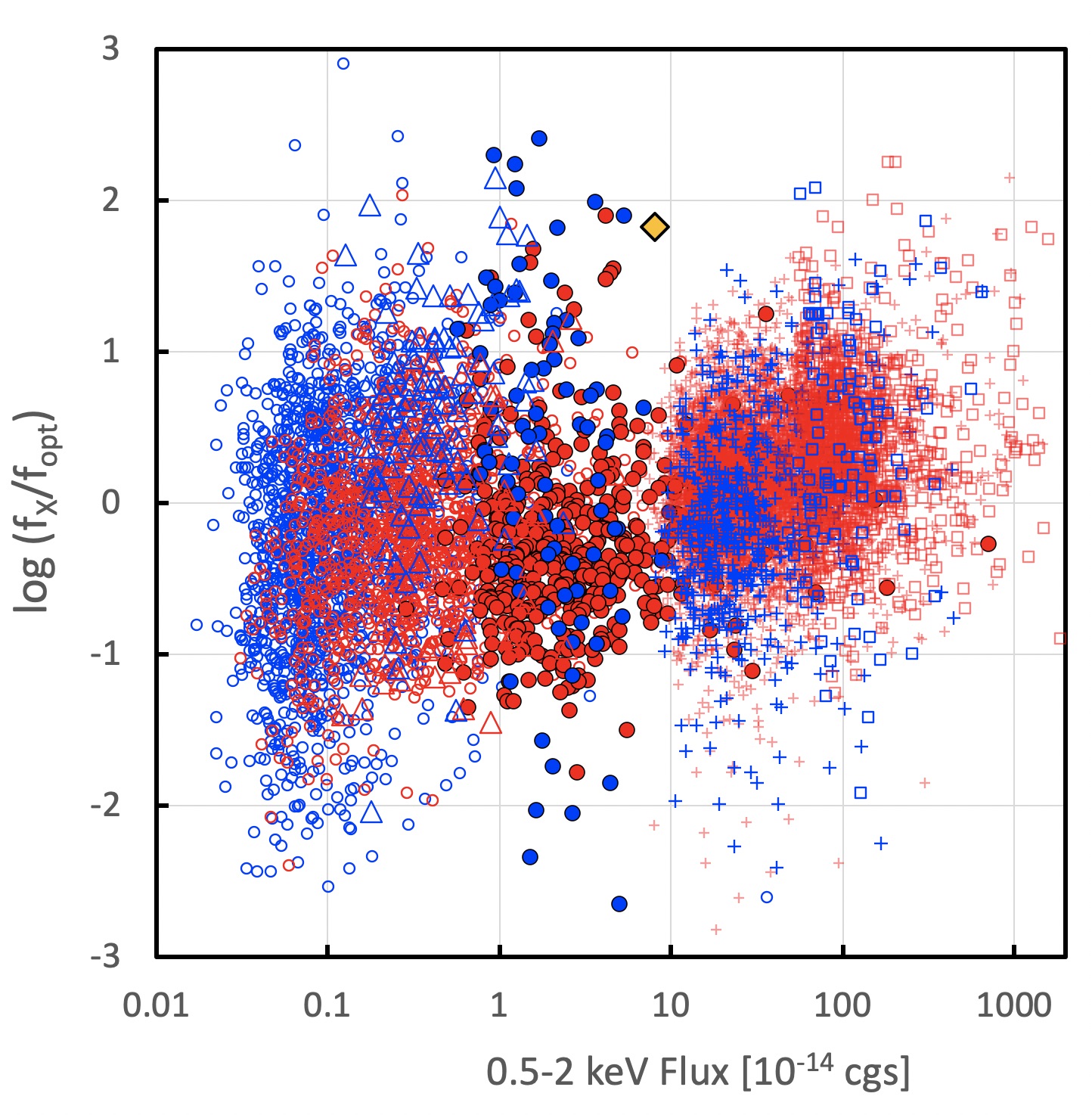}
    \caption{Left: The 0.5-2 keV X--ray luminosity versus redshift for the \textit{ROSAT} NEP and \textit{Chandra} COSMOS data (same symbols as in Figure \ref{fig:fxfxo}). Right the X--ray to optical flux ratio $f_{XO}$ versus X--ray flux for all AGN from the \textit{Chandra} COSMOS sources (small open circles), the mid-infrared sample upper limits (triangles), the ROSAT NEP raster scan (solid circles), the SPIDERS 2RXS catalog (open squares), and the SPIDERS XMMSL2 sources (plus signs). AGN1 are shown in red, AGN2 in blue symbols. CFHQS J142952+544717 is shown as orange diamond.}
    \label{fig:LXLfxo}
\end{figure*}

\subsection{Multiwavelength properties of X--ray counterparts}

Figures \ref{fig:fxabw1}--\ref{fig:LXLfxo} display the results of the optical identification process for the NEP raster scan catalogue. Figure \ref{fig:fxabw1} gives the classical correlation diagram between X--ray flux and optical magnitude (left) and W1 magnitude (right). This diagnostic diagram has originally been introduced by \citet{1988ApJ...326..680M} for the \textit{Einstein} Medium Sensitivity Survey, and has later been applied on numerous occasions, e.g. for the HELLAS2XMM Survey \citep{2003A&A...409...79F} and the \textit{XMM-Newton} survey of the COSMOS field \citep{2010ApJ...716..348B}. The different source classes segregate in a well-known fashion in this diagram, with AGN1 following a relatively tight correlation around a constant X--ray to optical flux ratio, which  is calculated as $f_{XO}$=$log(F_X/F_{opt})$=$log(F_X)$+0.4$\cdot$\textit{i}$_{AB}$+5.37. The dashed line in the left figure indicates an X--ray to optical flux ratio of unity, i.e. $f_{XO}$=0, while in the right diagram it gives the empirical discrimination line between AGN and stars/galaxies derived by \citet{2018MNRAS.473.4937S}. Clusters appear on average optically somewhat brighter, while stars are at significantly brighter magnitudes for the same X--ray flux. AGN2 are relatively rare at brighter X--ray fluxes, but their fraction increases towards lower fluxes. There is a significant new class of optically very faint AGN at lower X--ray fluxes, with magnitudes \textit{i}$_{AB}$>22 and X--ray to optical flux ratios $f_{XO}$=1--3. These are also dominant in the population of "infrared dropouts" with \textit{i}--IR>3 discussed below. Their photometric fits indicate spectral energy distributions with significant AGN contributions in the redshift range 1<\textit{z}<6. While there are 11 AGN1 in this category (about 3\% of the AGN1 sample), a total of 24 AGN2 fall in this group, which is almost 1/3 of the AGN2 population in our sample. In the right hand diagram, when plotted against mid-infrared magnitudes, these objects are less apparent, but still among those with the faintest mid-infrared magnitudes. Open triangles show the sample of mid-infrared selected AGN candidates (see below), at their 2$\sigma$ X--ray flux upper limits. A large number of those, too, fall in the category of optically very faint infrared dropouts.

\setlength{\tabcolsep}{7pt}

\begin{table*}
\caption{AGN candidates selected by mid-infrared colours}
\vskip -0.5truecm
\label{Table:mirAGN}
\begin{center}
\begin{tabular}{lllrrrrrrlllrr}
\hline\hline
(1) & (2)   & (3)    &(4)&(5)& (6)     &(7)&(8)&(9)&(10)&(11)&(12)&(13)&(14)\\ 
RA$_{IRAC}$	&	DEC$_{IRAC}$	&	AB	&	I1	&	I$_{12}$	&	eI$_{12}$	&	ID	&	ZPH	&	ML	&	uCTS	&	exp	&	uFX	&	ufxo	&	uL	\\
\hline
264.5209641	&	66.9128461	&	19.6 	&	18.0	&	1.53	&	0.09	&	2	&	0.19	&	0.0	&	2.46	&	2637	&	1.12	&	-0.74	&	42.0	\\
265.0664984	&	68.9359497	&	23.8 	&	18.6	&	1.30	&	0.17	&	2	&	2.18	&	0.0	&	2.18	&	2824	&	0.92	&	0.85	&	44.5	\\
265.6071114	&	65.7153652	&	19.5 	&	18.6	&	1.09	&	0.18	&	1	&	1.60	&	0.0	&	2.60	&	6042	&	0.52	&	-1.12	&	43.9	\\
265.6225486	&	65.9595430	&	22.8 	&	18.3	&	1.24	&	0.13	&	2	&	1.69	&	0.0	&	2.93	&	6115	&	0.57	&	0.27	&	44.0	\\
265.8146541	&	65.4302529	&	22.5 	&	18.7	&	1.02	&	0.23	&	2	&	2.51	&	0.0	&	2.63	&	4906	&	0.64	&	0.16	&	44.5	\\
265.8449932	&	66.8526275	&	24.8 	&	21.0	&	1.45	&	0.29	&	2	&	4.67	&	0.0	&	3.98	&	5912	&	0.81	&	1.21	&	45.2	\\
265.8618438	&	65.9624821	&	26.3 	&	20.0	&	1.08	&	0.24	&	2	&	6.00	&	2.8	&	5.74	&	6876	&	1.00	&	1.88	&	45.6	\\
265.9138475	&	66.4394202	&	21.0 	&	20.0	&	1.07	&	0.24	&	1	&	1.51	&	0.0	&	3.56	&	6006	&	0.71	&	-0.36	&	44.0	\\
265.9656502	&	66.6286979	&	24.2 	&	19.6	&	1.19	&	0.24	&	2	&	2.13	&	0.2	&	3.82	&	5884	&	0.78	&	0.94	&	44.4	\\
265.9777722	&	65.5437333	&	25.0 	&	20.0	&	1.03	&	0.29	&	2	&	3.68	&	0.0	&	3.67	&	5781	&	0.76	&	1.26	&	45.0	\\
...\\
272.9433227	&	65.1663909	&	22.5 	&	18.4	&	1.17	&	0.13	&	2	&	1.28	&	0.0	&	2.74	&	10941	&	0.30	&	-0.16	&	43.4	\\
272.9505745	&	65.0632444	&	21.6 	&	18.9	&	1.20	&	0.16	&	2	&	0.48	&	0.0	&	2.47	&	10321	&	0.29	&	-0.52	&	42.4	\\
272.9659244	&	65.8118536	&	21.7 	&	18.7	&	1.08	&	0.12	&	1	&	1.24	&	1.0	&	15.61	&	17647	&	1.06	&	0.07	&	44.0	\\
273.0271344	&	65.4764329	&	26.4 	&	20.8	&	1.20	&	0.29	&	2	&	4.12	&	0.0	&	2.83	&	15440	&	0.22	&	1.25	&	44.5	\\
273.0977434	&	66.1503762	&	21.4 	&	20.2	&	1.07	&	0.26	&	1	&	1.31	&	0.0	&	2.60	&	12639	&	0.25	&	-0.67	&	43.4	\\
273.1717517	&	68.3447007	&	20.3 	&	18.0	&	1.05	&	0.29	&	2	&	0.56	&	3.0	&	8.42	&	4255	&	2.37	&	-0.13	&	43.5	\\
273.207994	&	65.0189810	&	21.7 	&	18.1	&	1.08	&	0.17	&	2	&	0.28	&	0.0	&	2.77	&	9646	&	0.34	&	-0.40	&	41.9	\\
273.2409142	&	65.9689075	&	20.3 	&	19.2	&	1.09	&	0.16	&	1	&	2.13	&	0.0	&	2.41	&	13816	&	0.21	&	-1.19	&	43.8	\\
273.7493138	&	66.2812471	&	20.2 	&	18.7	&	1.12	&	0.16	&	1	&	1.33	&	0.0	&	2.82	&	8422	&	0.40	&	-0.95	&	43.6	\\
273.9294596	&	65.7268385	&	22.6 	&	18.6	&	1.05	&	0.22	&	2	&	1.46	&	0.3	&	5.97	&	12355	&	0.58	&	0.18	&	43.9	\\

\hline
\end{tabular}
\end{center}
\vskip -0.2truecm
Column explanation: (1) and (2) \textit{Spitzer} IRAC source coordinates in J2000.0; (3) \textit{Subaru} HSC \textit{i}-band (or \textit{z}-band) AB magnitude; (4) IRAC I1 magnitude; (5) and (6) I$_{12}$=I1--I2 mid-infrared colour and its error eI$_{12}$; (7) suggested AGN classification (1=AGN1, 2=AGN2); (8) photometric redshift; (9) maximum likelihood value of X--ray (non)detection; (9) 95.4\% upper limit counts; (11) ROSAT exposure time; (12) upper limit on the X--ray flux in units of 10$^{-14}$ erg cm$^{-2}$ s$^{-1}$; (13) upper limit X--ray to optical flux ratio $f_{XO}$; (14) logarithm of the upper limit observed 0.5--2 keV luminosity. 
\end{table*}

Figure \ref{fig:fxfxo} (left) shows the NEP sources with the X--ray to optical flux ratio $f_{XO}$ on the y-axis. In addition we plot the optical identifications from the Chandra legacy survey of the COSMOS field \citep{2016ApJ...819...62C,2016ApJ...817...34M,2018ApJ...858...77H,2019MNRAS.483.3545G} using the same colours, but smaller, open symbols. Despite the difference in X--ray flux limit of about a factor of 10, the optical to X--ray flux ratio distributions of the two samples are very similar, and overlap at the bright end. The fraction of stars is significantly smaller in COSMOS than in the NEP field. This is on one hand due to the lower Galactic latitude of the NEP ($\delta$$\sim$30$\degree$), compared to the COSMOS field ($\delta$$\sim$42$\degree$), but also due to the reduction of stars at fainter X--ray fluxes. The large orange diamond shows the position of the highest-redshift radio loud quasar CFHQS J142952+544717, the most X--ray luminous quasar ever observed at z > 6, which was recently identified in the eROSITA all-sky survey \citep{2020MNRAS.497.1842M}. Figure \ref{fig:fxfxo} (right) shows the same objects as in the left figure, but now with the (observed) 0.5--2 keV X--ray luminosity on the X--axis. In addition the X--ray upper limits are plotted as the open triangles as in Figure \ref{fig:fxabw1}. There is a general correlation between X--ray luminosity and X--ray to optical flux ratio, at least for galaxies, clusters and AGN2. This is likely caused by the optical counterparts with galaxy-type SEDs getting redder and thus relatively fainter with redshift. The more luminous AGN1, however, for which the SED is dominated by a power law spectrum, retain their optical to X--ray flux ratios in the range --1<$f_{XO}$<0.5 even at higher redshifts. There is one AGN1 at \textit{z}=4.43 with an X--ray luminosity log(L$_X$/(erg s$^{-1}$))=46.8 erg/s, even higher than that of CFHQS J142952+544717. However, the photometric redshift probability distribution for this object shows a second peak at lower redshifts (\textit{z}$\sim$0.56), which would bring the luminosity down to more ballpark values, indicated by the dotted line. The true redshift needs to be confirmed spectroscopically.

\begin{figure*}[htbp]
    \includegraphics[width=.49\textwidth]{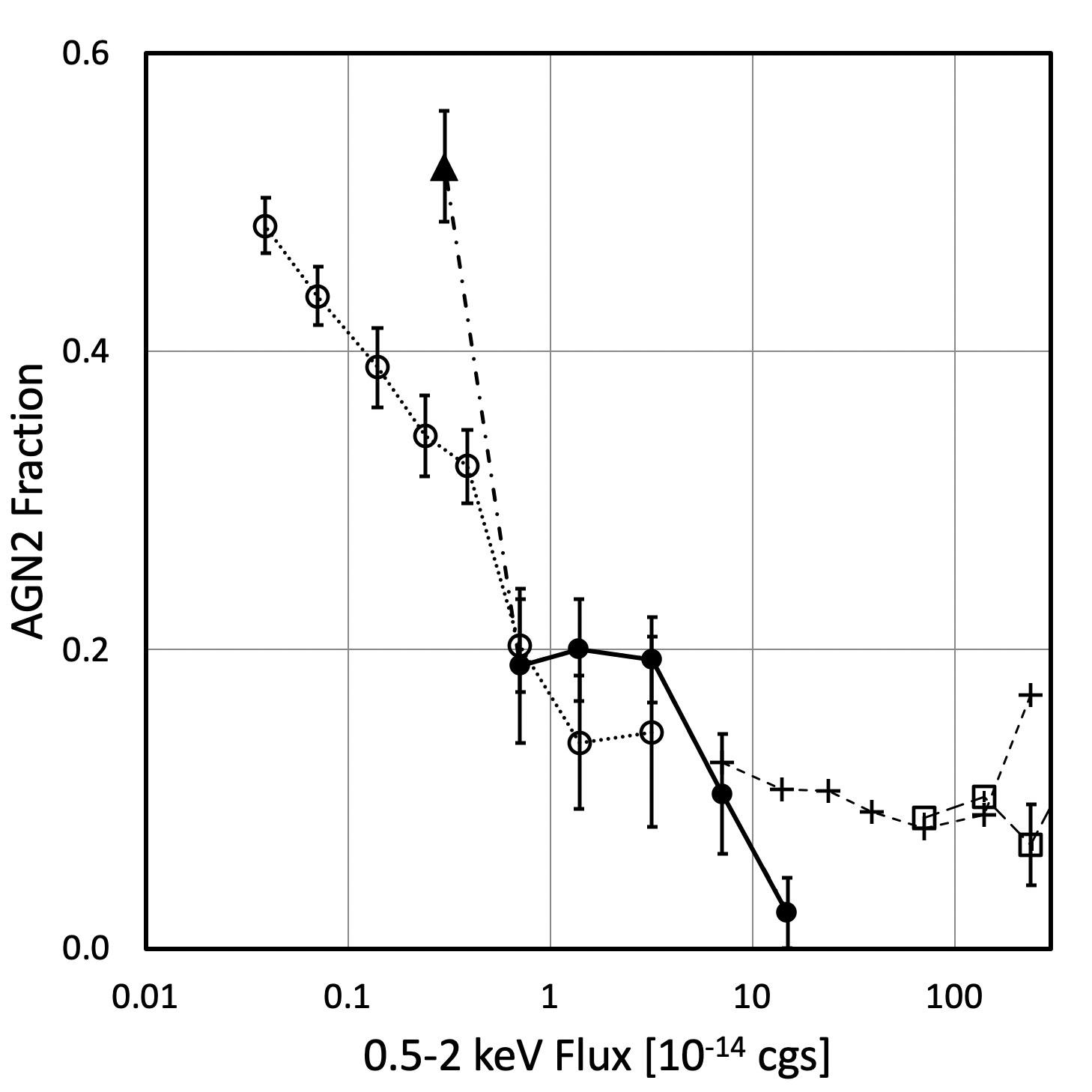}
    \includegraphics[width=.49\textwidth]{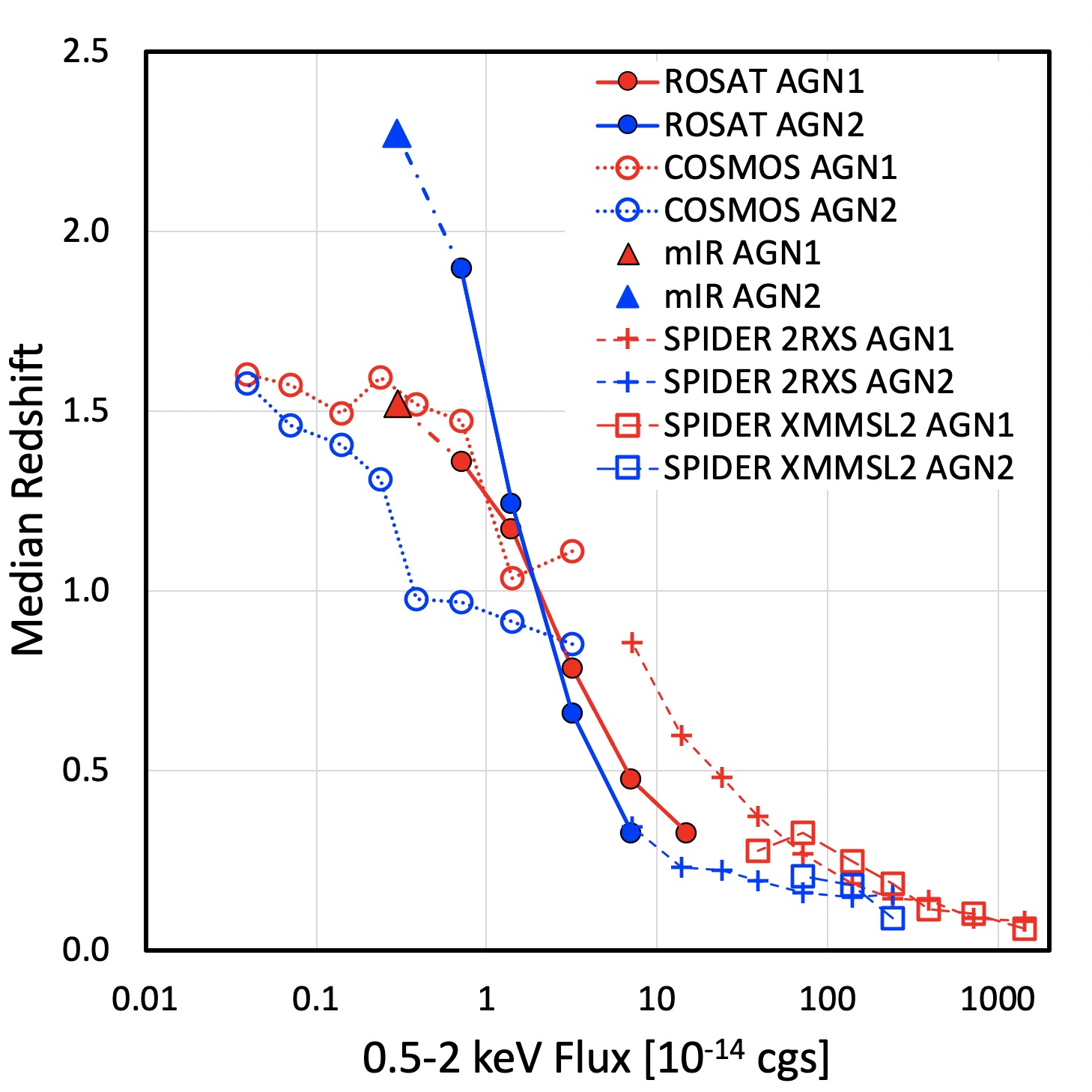}
    \includegraphics[width=.49\textwidth]{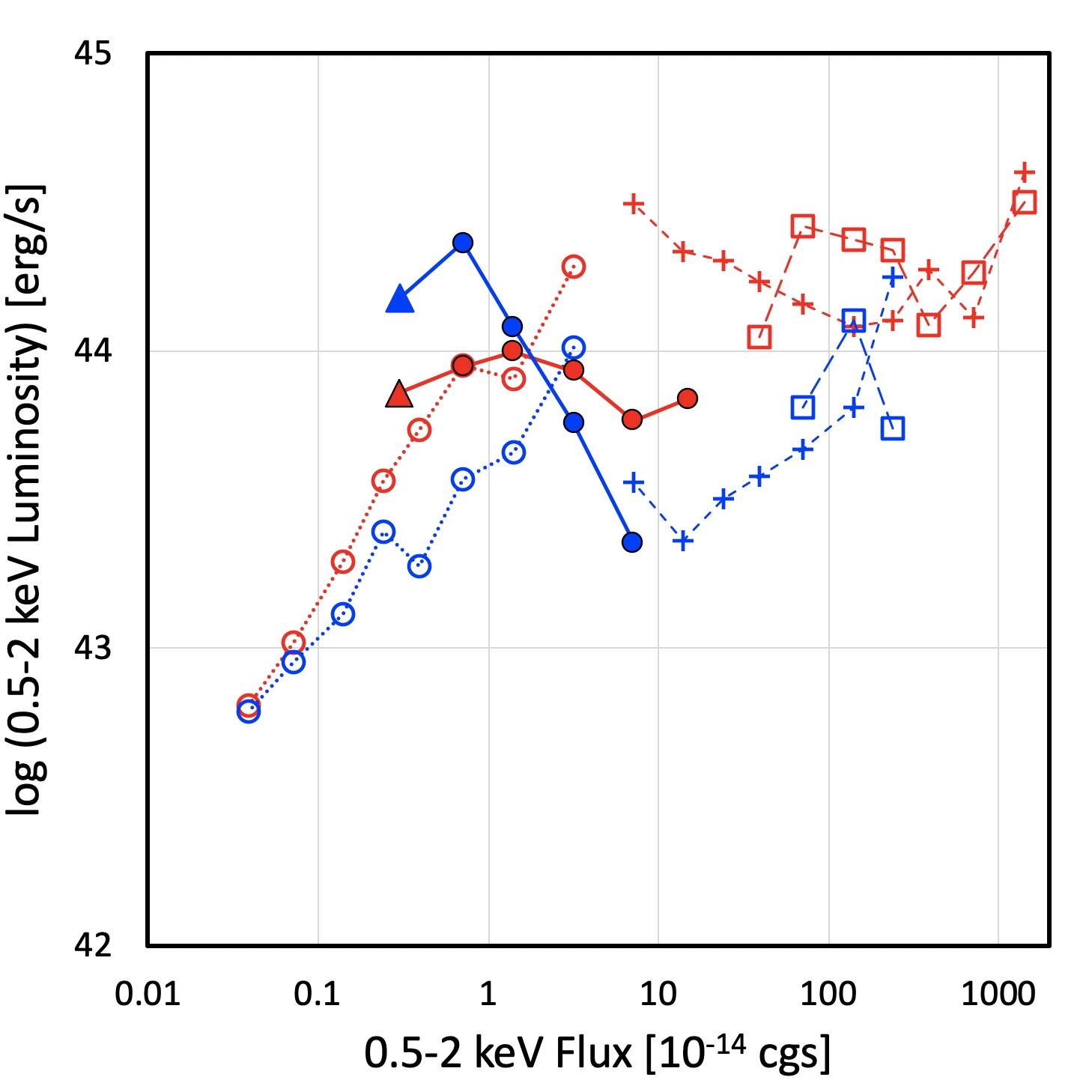}
    \includegraphics[width=.49\textwidth]{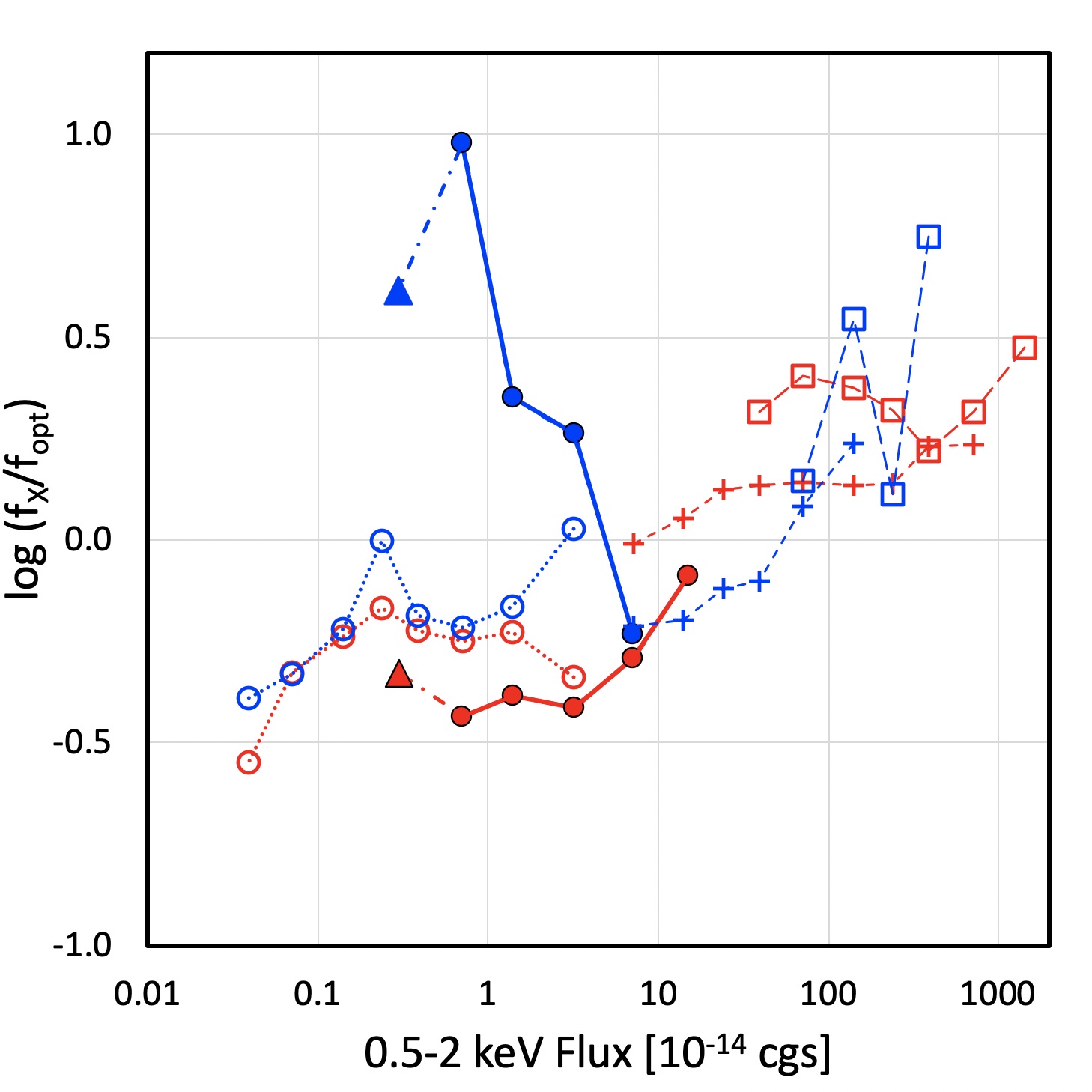}
    \caption{Ensemble properties of AGN1 (red) and AGN2 (blue) grouped by X--ray flux. Filled symbols show the \textit{ROSAT} NEP sample, open symbols the soft-band detected \textit{Chandra} COSMOS sample. The triangles show the upper limits for the mir-IR selected AGN sample. In the upper left the median redshifts are shown. The upper right panel gives median X--ray luminosities, and the lower left median X--ray to optical flux ratios. Finally, the lower right panel shows the AGN2 fraction (AGN/(AGN1+AGN2) as a function of soft X--ray flux.}
    \label{fig:Summary}
\end{figure*}

Figures \ref{fig:W1fxo} and \ref{fig:iIfxoz} show the power of mid-infrared colours for the X--ray source characterization and AGN selection. Figure \ref{fig:W1fxo} (left) displays the X--ray to optical flux ratio as a function of the \textit{CatWISE2020} W1--W2 colour. AGN dominate at W1--W2>0.7, but are also present down to W1--W2>0.3. (The classical mid-infrared AGN selection of \citet{2012ApJ...753...30S} assumes W1--W2>0.8). Clusters and AGN have overlapping ranges of $f_{XO}$, but clearly segregate in the mid-infrared colours. There is a number of bright stars with relatively large W1--W2 colors, which are likely spurious due to saturation in the \textit{WISE} images. This figure also contains the even redder selection of mid-infrared AGN candidates without X--ray detections (triangles, see below). Figure \ref{fig:W1fxo} (right) shows the comparison between the mid-infrared colours from \textit{CatWISE2020} (W1--W2) and the I1--I2 colours from the \textit{Spitzer} IRAC catalogue in the overlapping region. There is a clear correlation between the two mid-infrared colours, but in particular for the faintest mid-infrared objects the identification quality is significantly improved in the \textit{Spitzer} data. Figure \ref{fig:iIfxoz} (left) shows the correlation between the X--ray to optical flux ratio $f_{XO}$ and the optical to infrared color \textit{i}--IR. Here the infrared magnitude IR is either taken as the \textit{CatWISE2020} W1 or the \textit{Spitzer} IRAC I1 value. While the AGN1 form a rather compact clump centered around $f_{XO}$=--0.5$\pm$1 and \textit{i}--IR=1.5$\pm$1 in this diagram, the AGN2 are drawn out over a larger range of values up to $f_{XO}$<3 and \textit{i}--IR<8. We define infrared dropouts as objects with \textit{i}--IR>3. Figure \ref{fig:iIfxoz} (right) shows the run of the optical-infrared colours with redshift in comparison to some representative SED models. There is a clear segregation between AGN1, which maintain rather blue colour throughout all redshift ranges up to \textit{z}<5, and AGN2, which turn into infrared dropouts above redshifts \textit{z}>1. The most luminous QSO1 and QSO2 templates straddle the range of optical-infrared colours, while the lower luminosity Seyfert1 and Seyfert2 models lie in between. The mid-infrared selected AGN candidates with X--ray upper limits (triangles, see below) follow the same trends. Interestingly, the so far highest-redshift radio loud quasar CFHQS J142952+544717 from the eROSITA All-sky survey \citep{2020MNRAS.497.1842M} perfectly fits the QSO1 template for its redshift of \textit{z}=6.18.  Finally, Figure \ref{fig:LXLfxo} shows the coverage of the different samples in the X--ray luminosity versus redshift plane. The ROSAT NEP data reaches significantly higher X--ray luminosities of log(L$_X$/(erg s$^{-1}$))<46) than the both the COSMOS sample and the list of mid-infrared selected upper limits. This is only topped by the highest X--ray luminosity source CFHQS J142952+544717 discovered in the eROSITA all-sky survey \citep{2020MNRAS.497.1842M} and possibly one of the AGN1 in our sample (see above). The upper redshift cutoff of all samples is rather similar, with the bulk of objects at \textit{z}<4, but a significant number of AGN1 and AGN2 candidates at \textit{z}=4--7.

\subsection{X--ray properties of Mid-infrared selected AGN}
\label{subsec:mirAGN}

The richness and coverage of the \textit{HEROES} NEP dataset allows for the selection of a high-quality AGN sample based purely on mid-infrared colours. For this purpose we use the overlapping range of the \textit{WISE} and \textit{Spitzer} coverage in the field and select objects with red W1--W2 as well as I1-I2 colours. Following the correlation between these two infrared colours shown in Figure \ref{fig:W1fxo} (right) we chose objects with W1--W2>1, I1--I2>1 and the error $\delta$(I1--I2)<0.2. A total of 185 mid-infrared AGN candidates has been selected this way. We applied the same photometric redshift determination and classification procedure as for the X--ray selected AGN above, yielding 88 AGN1 and 97 AGN2 candidates with median redshifts around <\textit{z}$_{AGN1}$>=1.5 and <\textit{z}$_{AGN2}$>=2.3, respectively. We then offered the mid-infrared AGN candidate positions to the ROSAT ML detection algorithm, to determine X--ray count rates or upper limits for these objects at their fixed positions. We chose 95.4\% upper limit count rates, corresponding to a Gaussian 2$\sigma$ probability. Nine of the 186 mid-infrared AGN candidates have significant X--ray detections (with $L_{exi}$>12). Four of those are part of the X--ray catalogue in Table \ref{Table:XID}. The other five have not been detected, mainly because of confusion issues. X--ray fluxes and upper limits have been determined the same way as for the X--ray sources above. The relevant data is shown as triangles in Figures \ref{fig:fxabw1}--\ref{fig:LXLfxo}. In general, these correlations show that the mid-infrared selected AGN candidates are a continuation in parameter space of the X--ray selected AGN. They have similar X--ray to optical flux ratios and similar optical-infrared colours as the X--ray selected population, but are typically a factor of a few fainter than the X--ray sources. Their photometric redshift fits are much more dominated by large mid-infrared fraction SEDs (e.g. \texttt{MRK231} and \texttt{TORUS} (see section \ref{subsec:photoz}). 

It is interesting to note, that the median redshift of the mid-IR selected AGN2 is significantly higher than that of the AGN1 (see also Figure \ref{fig:Summary}). This is likely due to a selection effect in the mid-IR discussed by \citet{2013ApJ...772...26A}. The mid-infrared (W1–W2) color of Type 1 AGN is reddest in the redshift range 1<\textit{z}<2 , and progressively gets bluer at redshifts \textit{z}>2 (see Figure 1 of their paper). At redshifts \textit{z}>3, AGN1 completely fall out of the W1--W2>0.8 criterion. A similar effect can be seen in Figure \ref{fig:iIfxoz} (right). This effect causes the mid-IR selection to systematically miss higher-redshift AGN1.

We predict that most of the mid-IR selected AGN will be detected in the future deeper and harder \textit{eROSITA} coverage of the NEP. Table \ref{Table:mirAGN} gives the catalogue of 185 mid-infrared colour selected AGN candidates, again in abbreviated form. (The complete catalogue is available in the online publication.)

\section{Discussion and Conclusions}

We compare the ensemble properties of our AGN with other soft X--ray selected samples over a wide range of limiting fluxes. In addition the the \textit{Chandra} catalogue in the COSMOS field already displayed above, we include the Second ROSAT all-sky source catalogue 2RXS \citep{2016A&A...588A.103B}, and the Second \textit{XMM-Newton} Slew Survey Catalogue XMMSL2. 
The SPectroscopic IDentification of ERosita Sources (SPIDERS) survey, an SDSS-IV programme aimed at obtaining spectroscopic classification and redshift measurements for complete samples of sufficiently bright X-ray sources \citep{2017MNRAS.469.1065D,2018MNRAS.473.4937S,2020A&A...636A..97C} is the so far largest systematic spectroscopic observation of X-ray selected samples. It targeted the 2XRS and XMMSL2 sources in a sky area of 5129 square degrees covered in SDSS-IV by the eBOSS survey. We used the Value Added Catalogs from the MPE SPIDERS home page\footnote{https://www.mpe.mpg.de/XraySurveys/SPIDERS/}, which include cross-correlations with various galaxy cluster catalogues. The complete catalogues contain a total of 19821 and 2342 X--ray sources for 2RXS and XMMSL2, respectively. From the 2RXS SPIDERS catalogue we selected 10113 sources with detection likelihood $L_{exi}$>10 in the total (0.1--2.4 keV) \textit{ROSAT} band  and converted the X--ray fluxes to the 0.5--2 keV band, assuming an unabsorbed power law spectrum with photon slope --2. For XMMSL2 we selected 1860 sources with a detection likelihood $L_{exi}$>10 in the 0.5--2 keV band and used the corresponding soft fluxes given in the catalogue. We correlated the source lists with the SIMBAD and NED databases in order to obtain additional identifications beyond the original SPIDERS spectroscopic target selection. After this procedure, a total of 2607 and 227 objects remain unidentified in the 2RXS and XMMSL2 catalogues, respectively, corresponding to 26\% and 12\% incompleteness for the two selected subsamples. This allows a reasonably reliable assessment of ensemble properties of the different source classes, although the incompleteness can intruduce some systematic errors in the analysis (see below). Following \citet{2020A&A...636A..97C} we classified as unobscured AGN1 those catalogue entries with spectroscopic classes \texttt{'BALQSO'}, \texttt{'BLAGN'}, \texttt{'BLAZAR'}, \texttt{'QSO'}, \texttt{'QSO\_BAL'}, as well as sources from the literature with a QSO or Seyfert-1 (Sy1, Sy1.2) designation. As AGN2 we classified all sources with spectroscopic class \texttt{'NLAGN'}, all sources classified as \texttt{'GALAXY'}, which do not belong to a cluster of galaxies and have an X--ray luminosity log(L$_X$/(erg s$^{-1}$))>42, as well as sources from the literature with a Seyfert-2 (Sy1.5, Sy1.9, Sy2) designation. This way we classify a total of 5541 / 1046 objects as AGN1, and 603 / 101 as AGN2 for the 2RXS / XMMSL2 catalogues, respectively. The X--ray to optical flux ratio as a function of X--ray flux for these sources is compared to the ROSAT NEP and COSMOS samples in Figure \ref{fig:LXLfxo}  (right).

Figure \ref{fig:Summary} shows ensemble properties of the AGN1 and AGN2 samples grouped by 0.5--2 keV X--ray flux. The upper left graph clearly shows a strong increase of the fraction of AGN2 with decreasing X--ray flux. The mid-IR selected, non X--ray detected AGN (filled triangle), assuming that their ensemble X--ray flux is somewhat below their median upper limit (3$\times$10$^{-15}$ erg cm$^{-2}$ s$^{-1}$), show an even larger AGN2 fraction. This is likely due to the selection effect discussed in section \ref{subsec:mirAGN}.  
A dependence of the obscured AGN fraction on luminosity and redshift has first been discussed by  \citet{1991MNRAS.252..586L}. A similar trend of increasing obscuration with decreasing X--ray luminosity has been observed in hard X--ray selected samples of AGN over a very broad range of fluxes, including the \textit{Chandra} deep fields \citep{2005AJ....129..578B,2008A&A...490..905H}.

However, looking at the median redshift (upper right), X--ray luminosity (lower left), and X--ray to optical flux ratio (lower right) ensemble properties, there is a significant difference between the ROSAT NEP and the other samples. In general, for AGN1 of all samples the median redshift increases, and the median luminosity as well as the median X--ray to optical flux ratio decrease with decreasing flux across the whole range. In the overlapping flux ranges there are systematic differences between the samples, which could well be explained by the systematics of source identification and characterization procedures (e.g. the identification/classification incompleteness or the difference between photometric and spectroscopic clsssification). But these are relatively small compared to the overall trends. 
The XMMSL2, 2RXS and COSMOS AGN2 roughly follow the same trends in ensemble properties as the  AGN1. 

However, the \textit{ROSAT} NEP AGN2 behave differently: at the faintest fluxes they have the highest median $f_{XO}$ of all samples. This fact is independent from photometric redhift errors, which could in principle introduce systematic effects, when comparing median luminosities. As discussed above, photometric redshifts of AGN2 have smaller uncertainties than for AGN1. We can therefore conclude that the X--ray faintest {\it ROSAT} NEP AGN2 have somewhat larger redshifts and X--ray luminosities than the AGN1 at corresponding fluxes. The mid-IR selected AGN2 follow or even accentuate this trend.  

As already demonstrated in Figures \ref{fig:fxabw1}--\ref{fig:LXLfxo}, we  see the emergence of a new population of high-luminosity, high-redshift obscured AGN2, which have not been recognized in previous X--ray selected AGN samples. The main new observational phenomenon is the comparatively large X--ray to optical flux ratio $f_{XO}$ of this AGN2 population, accompanied by a large fraction of IR dropouts (see e.g. Figure \ref{fig:iIfxoz} left). 

We have investigated, why this population has not been recognized in previous X-ray surveys, in particular in the \textit{Chandra} COSMOS-Legacy survey. The soft X--ray selected sample of this survey contains 2969 objects \citep{2016ApJ...819...62C} and their original optical/IR identification catalogue \citep{2016ApJ...817...34M} contains 74 objects with ambiguous, sub-threshold or completely missing identifications. We looked at all these sources using the most recent publicly available deep optical (\textit{SUBARU} Hyper-Suprime-CAM), NIR (ULTRAVISTA) and \textit{Spitzer} (SPLASH) image cutouts\footnote{https:irsa.ipac.caltech.edu/data/COSMOS/index\_cutouts.html$/$} and found that 28 of these (i.e. 38\%) are indeed IR-dropouts (according to our above criterion \textit{i}--IR>3), and thus correspond to the newly identified AGN2 population.\footnote{The remaining unidentified sources typically are objects blinded by bright nearby stars or galaxies.} These objects have been included in the diagrams in Figures \ref{fig:fxfxo} and \ref{fig:LXLfxo}, but their contribution to the overall sample of AGN2 in the COSMOS survey is relatively small, so that they do not appear as a dominant new population. 

Similarly, the SPIDERS 2RXS and XMMSL2 catalogues contain a number of high $f_{XO}$ sources, but they are not dominant \citep[see also][]{2018MNRAS.473.4937S}. Due to the combination of larger solid angle, and sufficient X--ray sensitivity, together with the excellent HEROES multiwavelength coverage and in particular the \textit{Spitzer} mid-IR angular resolution and sensitivity the \textit{ROSAT} NEP sample reaches a dominant AGN2 population with optically very faint infrared X--ray counterparts (IR dropouts), which turn out to be predominantly higher luminosity  (log(L$_X$/(erg s$^{-1}$))>44), higher redshift (\textit{z}>1.5) AGN2. Their photometric redshift fits often require mid-IR dominated SEDs (e.g. \texttt{MRK231} and \texttt{TORUS} models) reminiscent of the \textit{Spitzer} power law AGN detected in the \textit{Chandra} deep fields \citep{2007ApJ...660..167D}, albeit  at higher redshifts.  The same is true for the sample of mid-IR selected, but not X--ray detected AGN candidates. 

Here it is worth emphasizing, that the \textit{Spitzer} data are crucial in our analysis. At the faintest infrared magnitudes, where many of our X--ray counterparts reside, the WISE data is significantly confusion limited, which can easily produce spurious red IR colours, and also does not allow to uniquely identify the often very faint optical counterpart. Only the higher angular resolution of the \textit{Spitzer} data allows to obtain secure IR colors and optical identifications at the faintest IR magnitudes.    

Figures \ref{fig:fxfxo} (right) and \ref{fig:LXLfxo} show that due to the combination of larger solid angle and sufficient sensitivity the \textit{ROSAT} NEP sample reaches X--ray luminosities around log(L$_X$/(erg s$^{-1}$))<46 for all AGN, which is larger than previous X--ray surveys, which are typically limited at log(L$_X$/(erg s$^{-1}$))<45. The discovery of the so far highest X--ray luminosity quasar in the \textit{eROSITA} all-sky survey \citep{2020MNRAS.497.1842M} confirms this trend.

Our unique HEROES multiwavelength dataset is important as a reference sample for future deep surveys in the NEP region, in particular for \textit{eROSITA}, but also for \textit{Euclid} and \textit{SPHEREX}. We predict that most of the absorbed distant AGN should be readily picked up by \textit{eROSITA}, but they require sensitive mid-IR imaging to be recognized as X--ray counterparts.

\begin{acknowledgements}
We would like to thank an anonymous referee for very helpful suggestions to improve the manuscript. GH would like to thank Olivier Ilbert and Mara Salvato for important help with the \textit{LePhare} photometric redshift code, and  Mara as well for very helpful comments on the paper draft. We would like to thank Len Cowie for designing the HEROES survey, and the former IfA REU students Cameron White \& Victoria Jones who have accompanied the HEROES observations and analysis, as well as taken a few HYDRA spectra of X--ray sources at the WIYN telescope. E.M.H. would like to acknowledge the NSF grant AST-1716093, partly funding the HEROES activities. We would like to acknowledge the excellent \textit{ROSAT} data archive provided by the Max-Planck-Institute for Extraterrestrial Physics, Garching, Germany. This research has made use of data obtained from XMMSL2, the Second \textit{XMM-Newton} Slew Survey Catalogue, produced by members of the XMM SOC, the EPIC consortium, and using work carried out in the context of the EXTraS project ("Exploring the X-ray Transient and variable Sky", funded from the EU's Seventh Framework Programme under grant agreement no. 607452). This research has made use of data obtained from the 4XMM \textit{XMM-Newton} serendipitous source catalogue compiled by the 10 institutes of the \textit{XMM-Newton} Survey Science Centre selected by ESA. This work is based in part on observations made with the \textit{Spitzer} Space Telescope, which was operated by the Jet Propulsion Laboratory, California Institute of Technology under a contract with NASA and it also makes use of data products from the \textit{Wide-field Infrared Survey Explorer}, which is a joint project of the University of California, Los Angeles, and the Jet Propulsion Laboratory/California Institute of Technology, funded by NASA. It is also based in part on data collected at \textit{Subaru} Telescope, which is operated by the National Astronomical Observatory of Japan, as well  as on observations obtained with MegaPrime/MegaCam, a joint project of \textit{CFHT} and CEA/DAPNIA, at the Canada-France-Hawaii Telescope (\textit{CFHT}) which is operated by the National Research Council (NRC) of Canada, the Institut National des Science de l'Univers of the Centre National de la Recherche Scientifique (CNRS) of France, and the University of Hawaii. The authors wish to recognize and acknowledge the very significant cultural role and reverence that the summit of Maunakea has always had within the indigenous Hawaiian community. We are most fortunate to have the opportunity to conduct observations from this mountain.  
\end{acknowledgements}

\bibliographystyle{aa} 
\bibliography{ROSATNEP} 

\begin{table*}
\caption{Appendix: \textit{ROSAT} Observing Log.}
\label{Table:log}
\centering
\begin{tabular}{lrrrlll}
\hline\hline
ROS. ID & UT0 & UT1 & Exp. [s] & PSPC & PI & Target\\
\hline
100378p & 900622.043711 & 900624.132107 & 50406 & C & MPE                & WFC BACKGROUND\\
120011p & 900626.221911 & 900629.070323 &  6249 & C & MPE                & DUMMY NEP 1   \\
120012p & 900626.234713 & 900628.070129 &  5601 & C & MPE                & DUMMY NEP 2   \\
120013p & 900627.012443 & 900629.081407 &  4796 & C & MPE                & DUMMY NEP 3   \\
120014p & 900627.030735 & 900628.203358 &  5090 & C & MPE                & DUMMY NEP 4   \\
120111p & 900708.111319 & 900710.151343 &  2340 & C & MPE                & DUMMY NEP 1   \\
120113p & 900708.125011 & 900709.070750 &  1437 & C & MPE                & DUMMY NEP 3   \\
120114p & 900708.161244 & 900708.212559 &  2334 & C & MPE                & DUMMY NEP 4   \\
120112p & 900708.174134 & 900708.195629 &  1963 & C & MPE                & DUMMY NEP 2   \\
930521s & 900711.132922 & 910813.075954 &  & C & MPE & RASS \\
150068p & 900716.213258 & 900718.085233 &   771 & C & BARSTOW & WD1821+643    \\
160030p & 910305.162834 & 910324.772892 & 3327 & B & MPE     & PSPC NEP      \\
000015p & 910423.415322 & 910423.424304 & 776 & B & MPE     & NEP\\
000016p & 910423.427973 & 910423.490276 & 5145 & B & MPE     & NEP\\
000051p & 910521.084841 & 910521.085957 &   292 & B & MPE     & DUMMY POINTING\\
170075p & 910802.062431 & 910803.080717 & 41204 & B & MPE     & NEP           \\
000026p & 920221.075604 & 920224.043053 & 44989 & B & MPE     & Dummy Pointing\\
700585p & 920325.212613 & 920325.215133 &  1465 & B & HASINGER & NEP RASTER    \\
700581p & 920327.020115 & 920330.194818 &  3658 & B & HASINGER & NEP RASTER    \\
700640p & 920404.120308 & 920404.125707 &  2995 & B & HASINGER & NEP RASTER    \\
700630p & 920404.233739 & 920407.012843 &  1691 & B & HASINGER & NEP RASTER    \\
700629p & 920405.085351 & 920405.091132 &   993 & B & HASINGER & NEP RASTER    \\
700650p & 920406.151548 & 920406.155633 &  2391 & B & HASINGER & NEP RASTER    \\
700561p & 920406.170112 & 920406.173158 &  1792 & B & HASINGER & NEP RASTER    \\
700569p & 920406.201446 & 920406.204318 &  1335 & B & HASINGER & NEP RASTER    \\
700607p & 920408.145304 & 920408.154337 &  2979 & B & HASINGER & NEP RASTER    \\
700571p & 920410.131643 & 920410.135651 &  2355 & B & HASINGER & NEP RASTER    \\
700575p & 920410.144243 & 920410.153206 &  2909 & B & HASINGER & NEP RASTER    \\
700576p & 920411.095126 & 920411.103924 &   634 & B & HASINGER & NEP RASTER    \\
700586p & 920411.131624 & 920411.135046 &  2005 & B & HASINGER & NEP RASTER    \\
700578p & 920411.143220 & 920411.144952 &   991 & B & HASINGER & NEP RASTER    \\
700413p & 920411.174304 & 920411.180248 &  1128 & B & EDELSON & 1821+64       \\
700584p & 920412.002912 & 920412.020855 &  2100 & B & HASINGER & NEP RASTER    \\
700560p & 920412.045431 & 920412.052054 &  1516 & B & HASINGER & NEP RASTER    \\
700580p & 920412.130931 & 920412.134533 &  2058 & B & HASINGER & NEP RASTER    \\
700582p & 920412.145054 & 920412.152056 &  1642 & B & HASINGER & NEP RASTER    \\
700562p & 920412.162739 & 920412.165623 &  1202 & B & HASINGER & NEP RASTER    \\
700511p & 920412.191800 & 920412.214403 &  1606 & B & HASINGER & NEP RASTER    \\
700564p & 920412.222920 & 920413.020359 &  2680 & B & HASINGER & NEP RASTER    \\
700565p & 920413.000542 & 920413.002659 &  1209 & B & HASINGER & NEP RASTER    \\
700568p & 920413.075908 & 920413.083528 &  1832 & B & HASINGER & NEP RASTER    \\
700567p & 920413.093920 & 920413.102801 &  2621 & B & HASINGER & NEP RASTER    \\
700573p & 920413.112404 & 920413.120426 &  2347 & B & HASINGER & NEP RASTER    \\
700577p & 920413.124724 & 920517.055852 &  4531 & B & HASINGER & NEP RASTER    \\
700592p & 920413.144558 & 920413.151548 &  1557 & B & HASINGER & NEP RASTER    \\
700572p & 920413.155822 & 920413.231327 &  1473 & B & HASINGER & NEP RASTER    \\
700570p & 920413.162236 & 920413.165122 &  1249 & B & HASINGER & NEP RASTER    \\
700566p & 920413.173606 & 920413.213728 &  1491 & B & HASINGER & NEP RASTER    \\
700596p & 920414.081434 & 920414.083057 &   921 & B & HASINGER & NEP RASTER    \\
700583p & 920414.093313 & 920414.102312 &  2665 & B & HASINGER & NEP RASTER    \\
700600p & 920414.112001 & 920414.115820 &  2244 & B & HASINGER & NEP RASTER    \\
700608p & 920414.124215 & 920414.133421 &  3048 & B & HASINGER & NEP RASTER    \\
700587p & 920414.143823 & 920414.150949 &  1292 & B & HASINGER & NEP RASTER    \\
700594p & 920520.020722 & 920520.023215 &  1303 & B & HASINGER & NEP RASTER    \\
700589p & 920520.051748 & 920520.054317 &  1375 & B & HASINGER & NEP RASTER    \\
700949p & 930307.130709 & 930307.134010 &  1858 & B & EDELSON & 1821+64       \\
700948p & 930412.105649 & 930412.131058 &  1980 & B & EDELSON & 1821+64       \\
701523p & 930808.025343 & 930810.024153 & 24741 & B & BOLLER & MRK 507       \\
800498p & 930923.101851 & 930923.183213 &  5408 & B & BURNS & ABELL 2304    \\
000071p & 931126.428402 & 931126.462187 & 2919 & B & MPE     & Dummy Pointing\\
999995p & 940227.012316 & 970302.064742 &  11695 & B & MPE     & Idle Point    \\
999989p & 980219.203418 & 980219.212610 &  3061 & B & MPE     & IDLE POINT N2 \\
\hline
\end{tabular}
\end{table*}

\end{document}